\documentclass{article}

\usepackage{arxiv}
\usepackage{algorithm}
\usepackage{algorithmicx}
\usepackage{algpseudocode}
\floatname{algorithm}{Algorithm} 
\usepackage[utf8]{inputenc} 
\usepackage[T1]{fontenc}    
\usepackage{hyperref}       
\usepackage{url}            
\usepackage{booktabs}       
\usepackage{amsfonts}       
\usepackage{nicefrac}       
\usepackage{microtype}      
\usepackage{lipsum}
\usepackage{graphicx}
\usepackage{amsmath} 
\usepackage{bm}      

\usepackage{tikz}
\usepackage[edges]{forest}

\definecolor{paired-light-blue}{RGB}{198, 219, 239}
\definecolor{paired-dark-blue}{RGB}{49, 130, 188}
\definecolor{paired-light-orange}{RGB}{251, 208, 162}
\definecolor{paired-dark-orange}{RGB}{230, 85, 12}
\definecolor{paired-light-green}{RGB}{199, 233, 193}
\definecolor{paired-dark-green}{RGB}{49, 163, 83}
\definecolor{paired-light-purple}{RGB}{218, 218, 235}
\definecolor{paired-dark-purple}{RGB}{117, 107, 176}
\definecolor{paired-light-gray}{RGB}{217, 217, 217}
\definecolor{paired-dark-gray}{RGB}{99, 99, 99}
\definecolor{paired-light-pink}{RGB}{222, 158, 214}
\definecolor{paired-dark-pink}{RGB}{123, 65, 115}
\definecolor{paired-light-red}{RGB}{231, 150, 156}
\definecolor{paired-dark-red}{RGB}{131, 60, 56}
\definecolor{paired-light-yellow}{RGB}{231, 204, 149}
\definecolor{paired-dark-yellow}{RGB}{141, 109, 49}
\tikzset{%
    parent/.style =          {align=center,text width=1.5cm,rounded corners=3pt, line width=0.3mm, fill=gray!10,draw=gray!80},
    child/.style =           {align=center,text width=2.3cm,rounded corners=3pt, fill=blue!10,draw=blue!80,line width=0.3mm},
    grandchild/.style =      {align=center,text width=2cm,rounded corners=3pt},
    greatgrandchild/.style = {align=center,text width=1.5cm,rounded corners=3pt},
    greatgrandchild2/.style = {align=center,text width=1.5cm,rounded corners=3pt},    
    referenceblock/.style =  {align=center,text width=1.5cm,rounded corners=2pt},
    top_class/.style =           {align=center,text width=2cm,rounded corners=3pt, fill=paired-light-gray!50,draw=paired-dark-gray!65,line width=0.3mm},
     topclass_wide/.style =           {align=center,text width=2.5cm,rounded corners=3pt, fill= paired-light-gray!50,draw=cyan!0,line width=0.3mm}, 
    generation/.style =           {align=center,text width=2.5cm,rounded corners=3pt, fill= paired-light-orange!50,draw=paired-dark-orange!75,line width=0.3mm}, 
    generation_wide/.style =           {align=center,text width=2.5cm,rounded corners=3pt, fill= paired-light-orange!50,draw=paired-dark-orange!75,line width=0.3mm}, 
    generation_more/.style =           {align=center,text width=2.5cm,rounded corners=3pt, fill= paired-light-orange!50,draw=paired-dark-orange!75,line width=0.3mm},   
    generation_work/.style =           {align=center,text width=2.5cm,rounded corners=3pt, fill= paired-light-orange!50,draw= cyan!0,line width=0.3mm},
    encoder/.style =           {align=center,text width=2cm,rounded corners=3pt, fill=paired-light-orange!50,draw=paired-dark-orange!65,line width=0.3mm},  
    encoder_more/.style =           {align=center,text width=4cm,rounded corners=3pt, fill=paired-light-orange!50,draw=paired-dark-orange!65,line width=0.3mm}, 
    encoder_work/.style =           {align=center,text width=4.0cm,rounded corners=3pt, fill=paired-light-orange!50,draw=red!0,line width=0.3mm},    
    gpa/.style =           {align=center,text width=2.5cm,rounded corners=3pt, fill=paired-light-blue!50,draw=paired-dark-blue!65,line width=0.3mm},
    gpa_wide/.style =           {align=center,text width=2.5cm,rounded corners=3pt, fill=paired-light-blue!50,draw=paired-dark-blue!65,line width=0.3mm},   
    gpa_work/.style =           {align=center, text width=2.5cm,rounded corners=3pt, fill=paired-light-blue!50,draw=blue!0,line width=0.3mm},
    data/.style =           {align=center,text width=2cm,rounded corners=3pt, fill=paired-light-blue!50,draw=paired-dark-blue!65,line width=0.3mm},
    data_wide/.style =           {align=center,text width=3cm,rounded corners=3pt, fill=paired-light-blue!50,draw=paired-dark-blue!65,line width=0.3mm},   
    data_work/.style =           {align=center, text width=4.5cm,rounded corners=3pt, fill=paired-light-blue!50,draw=blue!0,line width=0.3mm},  
    model/.style =           {align=center,text width=2cm,rounded corners=3pt, fill=paired-light-orange!50,draw=paired-dark-orange!65,line width=0.3mm},  
    model_more/.style =           {align=center,text width=4cm,rounded corners=3pt, fill=paired-light-orange!50,draw=paired-dark-orange!65,line width=0.3mm}, 
    model_work/.style =           {align=center,text width=4.5cm,rounded corners=3pt, fill=paired-light-orange!50,draw=red!0,line width=0.3mm},    
    pretraining/.style =           {align=center,text width=2cm,rounded corners=3pt, fill= paired-light-green!50,draw=paired-dark-green!75,line width=0.3mm}, 
    pretraining_wide/.style =           {align=center,text width=2.5cm,rounded corners=3pt, fill= paired-light-green!50,draw=paired-dark-green!75,line width=0.3mm}, 
    pretraining_more/.style =           {align=center,text width=2.5cm,rounded corners=3pt, fill= paired-light-green!50,draw=paired-dark-green!75,line width=0.3mm},   
    pretraining_work/.style =           {align=center,text width=2.5cm,rounded corners=3pt, fill= paired-light-green!50,draw= cyan!0,line width=0.3mm},      
    finetuning/.style =           {align=center,text width=2cm,rounded corners=3pt, fill= paired-light-purple!50,draw=paired-dark-purple!75,line width=0.3mm},   
    finetuning_work/.style =           {align=center,text width=4.5cm,rounded corners=3pt, fill= paired-light-purple!50,draw= orange!0,line width=0.3mm},        
    inference/.style =           {align=center,text width=2cm,rounded corners=3pt, fill= paired-light-red!35,draw=paired-light-red!90,line width=0.3mm},           
    inference_more/.style =           {align=center,text width=4cm,rounded corners=3pt, fill= paired-light-red!35,draw=paired-light-red!90,line width=0.3mm},
    inference_work/.style =           {align=center,text width=4.5cm,rounded corners=3pt, fill= paired-light-red!35,draw= magenta!0,line width=0.3mm},         
}
\title{Deep Learning in Single-Cell and Spatial Transcriptomics Data Analysis: Advances and Challenges from a Data Science Perspective}

\author{
  Shuang Ge \\
  Tsinghua University\\
  Peng Cheng Laboratory\\
  \texttt{ges23@mails.tsinghua.edu.cn} \\
  \And
  Shuqing Sun \\
  Tsinghua University\\
  \And
  Huan Xu\\
  Anhui University of Science and Technology\\
  \And
  Qiang Cheng \thanks{Corresponding author} \\
  University of Kentucky\\
  \texttt{Qiang.Cheng@uky.edu}
  \And
  Zhixiang Ren \thanks{Corresponding author} \\
  Peng Cheng Laboratory\\
  \texttt{jason.zhixiang.ren@outlook.com} \\
}

\begin{document}
\maketitle

\begin{abstract}
The development of single-cell and spatial transcriptomics has revolutionized our capacity to investigate cellular properties, functions, and interactions in both cellular and spatial contexts. Despite this progress, the analysis of single-cell and spatial omics data remains challenging. First, single-cell sequencing data are high-dimensional and sparse, often contaminated by noise and uncertainty, obscuring the underlying biological signals. Second, these data often encompass multiple modalities, including gene expression, epigenetic modifications, metabolite levels, and spatial locations. Integrating these diverse data modalities is crucial for enhancing prediction accuracy and biological interpretability. Third, while the scale of single-cell sequencing has expanded to millions of cells, high-quality annotated datasets are still limited. Fourth, the complex correlations of biological tissues make it difficult to accurately reconstruct cellular states and spatial contexts. Traditional feature engineering-based analysis methods struggle to deal with the various challenges presented by intricate biological networks. Deep learning has emerged as a powerful tool capable of handling high-dimensional complex data and automatically identifying meaningful patterns, offering significant promise in addressing these challenges. This review systematically analyzes these challenges and discusses related deep learning approaches. Moreover, we have curated 21 datasets from 9 benchmarks, encompassing 58 computational methods, and evaluated their performance on the respective modeling tasks. Finally, we highlight three areas for future development from a technical, dataset, and application perspective. This work will serve as a valuable resource for understanding how deep learning can be effectively utilized in single-cell and spatial transcriptomics analyses, while inspiring novel approaches to address emerging challenges.
\end{abstract}



\section{Introduction}
The advancement of single-cell and spatial transcriptomics techniques has facilitated in-depth investigations of cellular characteristics, functions, and interactions, considering both cellular activity and spatial context within tissues. Single-cell RNA sequencing (scRNA-seq) quantifies gene expression at the cellular level, thereby elucidating cellular composition, gene expression patterns, and molecular characteristics\cite{sun2024single,fan2020single}. Recognized as the Method of the Year by Nature Methods in 2013\cite{nature_methods_2013}, scRNA-seq has significantly advanced research into complex biological questions, including mechanisms of disease resistance\cite{li2023single,wang2020single}, tissue heterogeneity\cite{keller2019unravelling,zeng2023understanding}, targeted therapies\cite{zhang2023single}, and embryonic development\cite{wang2024single}. However, tissue dissociation disrupts spatial cell distribution and intercellular interactions, thereby constraining our understanding of the intricate processes occurring within multicellular organisms.

Spatial transcriptomics (ST) generates spatially resolved transcriptomic data to create detailed tissue maps at the subcellular level. This technique represents a significant advance in the field of transcriptomics, transitioning from cellular resolution to spatially sub-cellular resolution. In recognition of its importance in biomedical research, Nature Methods named spatially resolved transcriptomics as the Method of the Year in 2020\cite{marx2021method}.

Single-cell and spatial transcriptomics are crucial for studying the microenvironment at cellular and spatial resolutions, respectively. However, the complexity of biological tissues and the limitations of current sequencing techniques present significant analytical challenges. In this review, we discuss four major challenges in single-cell and spatial transcriptomics from a data science perspective: data sparsity, diversity, scarcity, and correlation. Our aim is to elucidate the origins of these challenges, explore potential solutions, and provide insights into the underlying mechanisms of the methodology. In terms of data sparsity, we examine issues such as the curse of dimensionality, noise, and uncertainty. Concerning data diversity, we categorize the integration of single-cell and spatial transcriptomics data into two primary types: multimodal integration and multi-source integration. When dealing with data scarcity, we focus on missing data annotations and missing modalities. Finally, from a data correlation perspective, we analyze methods for modeling spatiotemporal dependencies and incorporating prior knowledge.

With the increasing volume and diversity of data, traditional analysis techniques for bulk RNA sequencing are becoming increasingly inadequate for single-cell and spatial transcriptomics\cite{fan2020single}. Deep learning (DL), a powerful tool for modelling large-scale, high-dimensional complex data, has demonstrated its versatility across numerous scientific domains, including small molecule modeling\cite{myung2024deep,bennett2020predicting}, protein structure prediction\cite{chowdhury2022single,pearce2021deep}, and drug development\cite{catacutan2024machine}, etc. 

\begin{figure*}[ht]
  \centering
  \includegraphics[width=\linewidth]{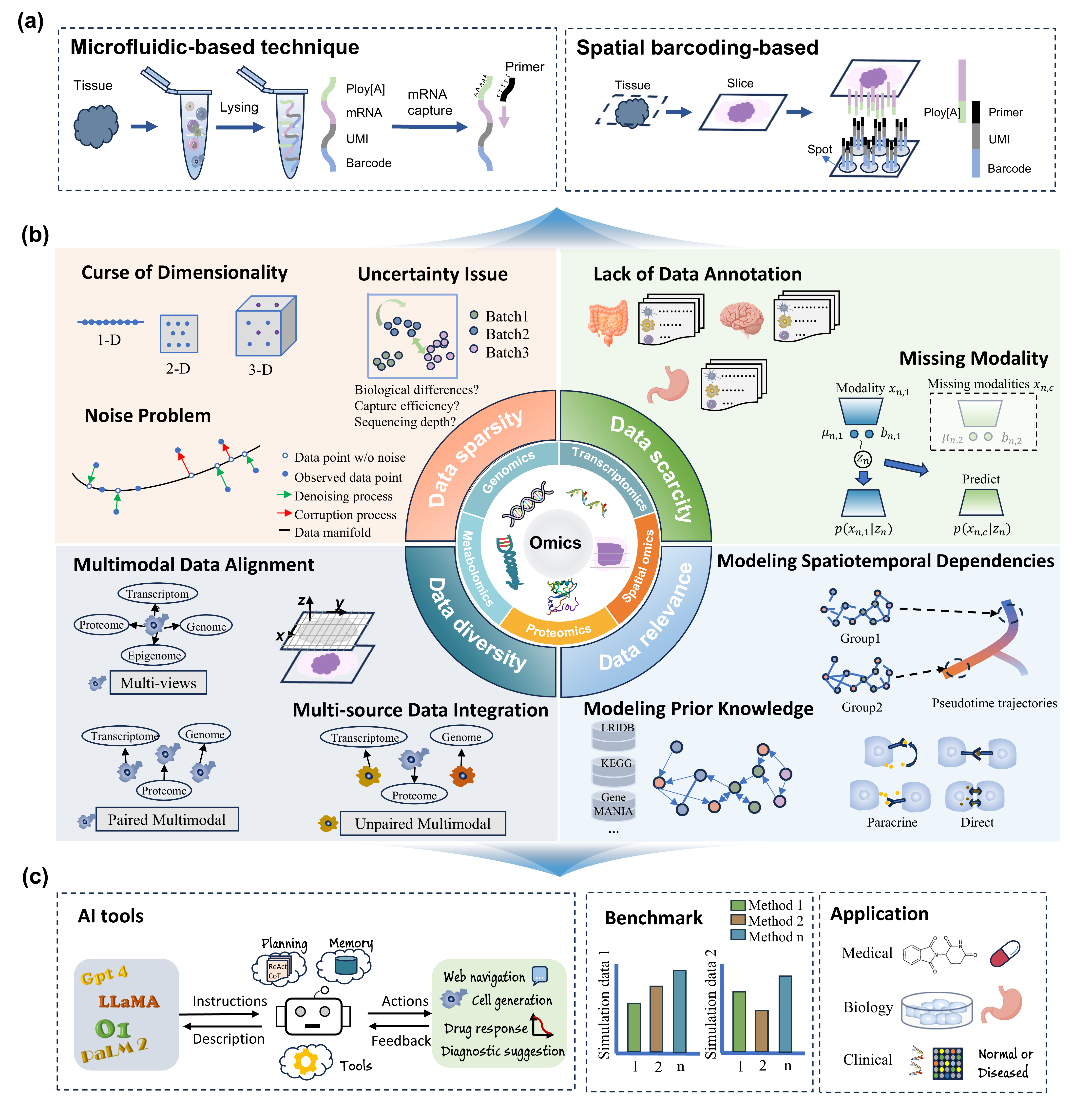}
  \caption{\textbf{The overall structure of the article is organized into three main sections.} (a) An overview of key sequencing technologies in single-cell and spatial transcriptomics; (b) A discussion of four significant scientific and technical challenges within the field from a data science perspective, namely: data sparsity, data diversity, data scarcity, and data correlation; (c) An exploration of potential future perspectives that includes innovative AI methodologies, benchmark datasets and evaluation metrics, as well as applications of DL in practical scenarios. Some components of this figure are drawn by Figdraw.}
  \label{fig1}
\end{figure*}

Recent reviews\cite{erfanian2023deep,bao2022deep,flores2022deep,brendel2022application} of deep learning (DL) applications in single-cell data have introduced methods such as multilayer perceptrons (MLP)\cite{taud2018multilayer}, autoencoder (AE)\cite{hinton2006reducing}, generative adversarial network (GAN)\cite{goodfellow2020generative}, convolutional neural network (CNN)\cite{lecun1989backpropagation}, and graph neural network (GNN)\cite{scarselli2008graph}. These reviews explored both traditional and DL methods across various stages of the scRNA-seq analysis pipeline. But they do not summarize current technological advances and challenges from a data science perspective. Moreover, existing data analysis techniques may not always be effective for addressing novel problems as the number of modalities continues to grow. This review aims to discuss four major data science challenges and exploring relevant methods within these contexts. We highlight DL techniques by comparing them with traditional machine learning approaches and emphasize their advantages, particularly when integrated with statistical frameworks. Each algorithm is discussed alongside its mathematical foundations, focusing on both similarities and differences. Finally, we outline future directions in three key areas: the application of novel AI methodologies, the development of fair and robust benchmark datasets with biologically interpretable evaluation metrics, and the exploration of DL applications in practical scenarios. This review provides a comprehensive overview of DL applications in single-cell and spatial transcriptomics data analysis from a data science perspective, offering insights that could inspire innovative solutions to emerging challenges in biological and medical research. The overall structure of the article is shown in Fig. \ref{fig1}.

\section{Transcriptomic data}\label{sec2}

\begin{figure*}[ht]
  \centering
  \includegraphics[width=\linewidth]{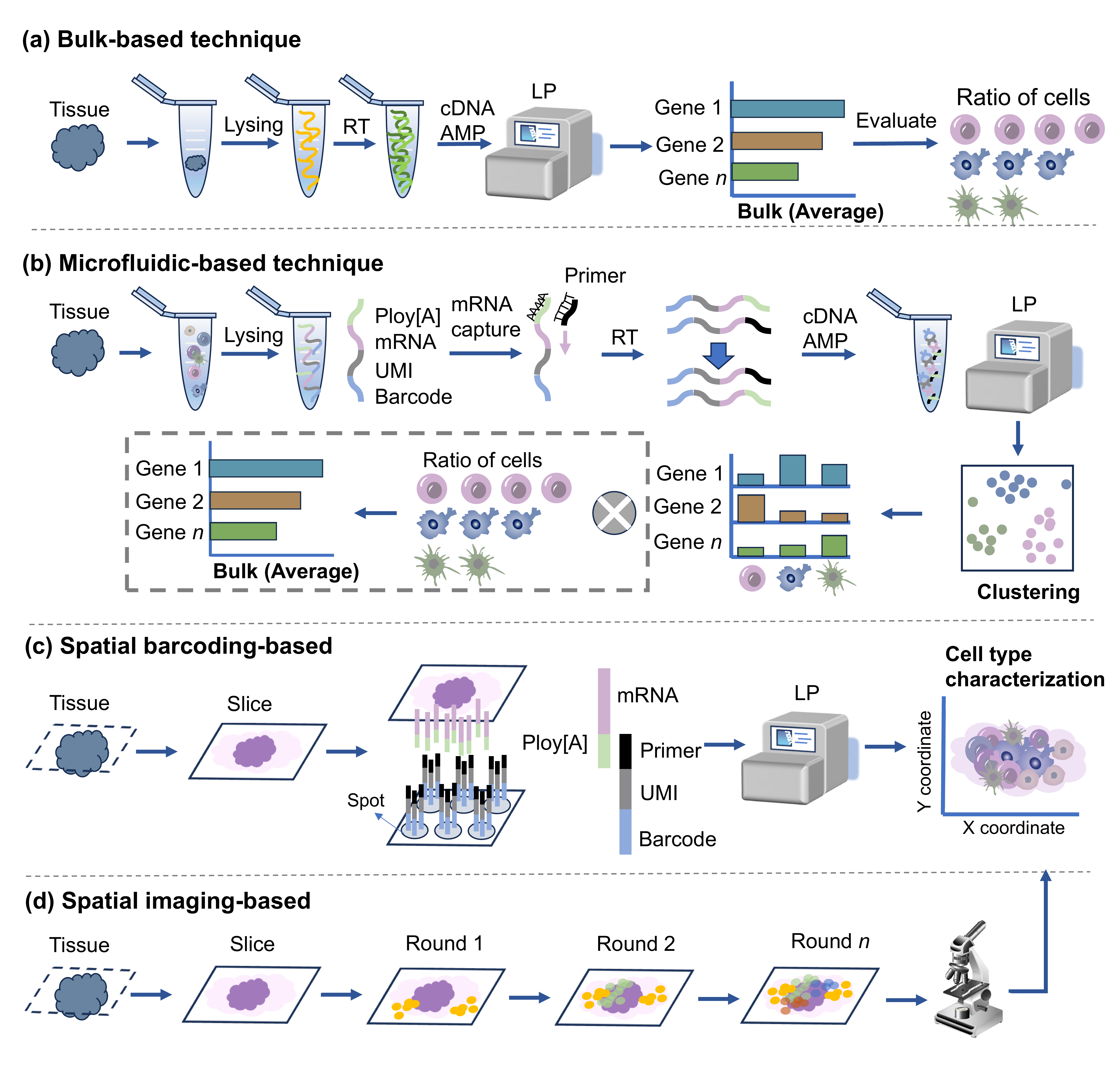}
  \caption{\textbf{The sequencing pipeline for single cell and spatial transcriptomics data.} (a) Bulk-based technique provides average gene expression profiles at the tissue level, with cell proportions estimated through deconvolution methods. (b) Microfluidic-based techniques isolate individual cells into droplets or wells, followed by barcoding and sequencing. (c) Spatial barcode-based techniques utilize cell barcodes to capture poly-adenylated RNA molecules in situ before reverse transcription. (d) Targeted in situ sequencing employs specifically designed probes to bind RNA or cDNA targets, leveraging in situ spatial information.}
  \label{fig1.1}
\end{figure*}

Bulk RNA sequencing (RNA-seq) provides average gene expression profiles at the tissue level, limiting its ability to accurately represent cellular heterogeneity. Consequently, it becomes challenging to discern whether the observed differences are due to changes in cellular composition or variations in gene expression (Fig. \ref{fig1.1}). The scRNA-seq addresses this limitation by profiling gene expression at the single-cell level. Additionally, spatial transcriptomics integrates sequencing data with spatial context, providing a more comprehensive understanding of tissue construction and function.

\subsection{Single-cell transcriptomes}\label{subsec1}
In 2009, scRNA-seq technology emerged, making it possible to study the transcriptomes of individual cells. Single-cell sequencing technology requires four main steps: (1) isolation of single cells (2) reverse transcription (3) cDNA amplification (4) sequencing library preparation and sequencing.

\textbf{Isolation of single cells.} Isolation of single cells refers to the process of separating individual cells from a complex tissue or cell population. Accurate and reliable capture is essential for single-cell sequencing. The dissociation methods mainly include mechanical dissociation, enzymatic dissociation and chemical dissociation. Target cells are selected from single-cell suspensions based on specific characteristics such as size, fluorescence, or surface labeling.

Fluorescence-activated cell sorting (FACS)\cite{herzenberg1976fluorescence} is a widely used high-throughput technique that labels cells with fluorescent dyes or antibodies targeting specific molecules on or within the cell. In flow cytometry, a laser excites a fluorescence marker, which emits a signal that is measured to quantify the molecular content. However, this method is less effective for cells with low marker expression and struggles to distinguish a subset of cells with similar fluorescence markers. Typical sequencing methods include Smart-seq\cite{hagemann2020single}, VASA-seq\cite{salmen2022high}, FLASH-seq\cite{hahaut2022fast}. 

Magnetic-activated cell sorting (MACS)\cite{miltenyi1990high} is another high-throughput isolation technique that separates and enriches specific cell types by binding magnetic beads to target cell proteins. The beads are conjugated with antibodies or other ligands, but unlike FACS, MACS isolates cells based on surface protein expression rather than gene expression, sorting them into positive and negative populations. MACS is primarily employed for the initial enrichment of cells and does not facilitate precise single-cell sorting as FACS does.


Microfluidic-based techniques exploit the inherent physical properties of cells for separation. These properties encompass cell size, shape, electrical polarizability, electrical impedance, density, deformability, magnetic susceptibility, and hydrodynamic characteristics \cite{gossett2010label}. In droplet-based microfluidics, individual cells are encapsulated within small droplets that are suspended in an immiscible fluid. Techniques such as InDrop\cite{klein2015droplet}, Drop-seq\cite{macosko2015highly}, and 10x Chromium\cite{zheng2017massively} build upon this methodology. Microwell-based scRNA-seq methods—such as CytoSeq\cite{fan2015combinatorial}, Seq-Well\cite{gierahn2017seq}, and Microwell-seq\cite{han2018mapping}—involve placing cells into discrete wells to ensure that each well contains either a single cell or none at all.

\textbf{Reverse transcription.} RNA cannot be directly sequenced within the cell. After cell lysis, the released RNA must be reverse transcribed to generate complementary DNA (cDNA). The poly(A) tailing method employs an oligo-dT primer that binds to the 3'-poly(A) tail of mRNA, thereby facilitating its reverse transcription into cDNA. During this process, additional nucleotide sequences, such as cell-specific barcodes and uniform molecular identifiers (UMIs) for mRNA, are incorporated to uniquely label each cell and distinguish individual mRNA molecules.

\textbf{cDNA amplification.} Since mRNA is typically present in very low quantities within individual cells, it often proves inadequate for sequencing purposes. Therefore, cDNA amplification is necessary to produce sufficient amounts for subsequent library preparation. The most widely utilized method for this process is PCR-based amplification.

\textbf{Sequencing library construction.} The first step in library preparation involves converting nucleic acids into a sequencing library, where DNA or RNA molecules are ligated to platform-specific adapters.

\subsection{Spatial transcriptomes}
Spatial transcriptomic techniques can be broadly categorized into two main types based on whether positional information is encoded before sequencing: (1) next-generation sequencing-based methods and (2) imaging-based methods.

Next-generation sequencing-based approaches encompass both the earlier microdissection-based techniques and the more widely adopted barcode-based approaches. Microdissection techniques isolate regions of interest through physical segmentation or optical selection, followed by collection for library preparation and sequencing. Microdissection-based techniques include tomo-seq\cite{junker2014genome}, STRP-seq\cite{schede2021spatial}, Geo-seq\cite{chen2017spatial}, PIC-seq\cite{giladi2020dissecting}, TIVA\cite{lovatt2014transcriptome}, NICHE-seq\cite{medaglia2017spatial}. Of these, PIC-seq, TIVA, and NICHE-seq achieve cellular-level resolution\cite{sun2024single}. However, physical segmentation is often performed manually, making it time-consuming. Additionally, optical selection requires the insertion of specialized markers into living cells or model organisms, which limits its application to FFPE human samples. Accurately locating spatial locations remains a significant challenge, often resulting in relatively low spatial resolution.

The barcode-based sequencing technique, inspired by scRNA-seq, utilizes cell barcodes to capture poly-adenylated RNA molecules in situ before reverse transcription. This process is facilitated by a capture probe that incorporates a spatial barcode, a unique molecular identifier (UMI), and poly-T oligonucleotides, followed by the synthesis of complementary DNA (cDNA). Spatial barcodes operate analogously to cellular barcodes, ensuring an accurate mapping of transcriptomes obtained from tissue slices back to their original locations. The spatial resolution of this method depends on the distance between adjacent spots, achieving a maximum resolution of approximately $0.5\sim0.7 \mu{m}$, which facilitates subcellular analysis. However, enhancing spatial resolution often leads to compromises in detection sensitivity and gene coverage. Examples of barcode-based sequencing technique include 10x Visium\cite{staahl2016visualization}, Slide-seq (V2)\cite{stickels2021highly}, HDST\cite{vickovic2019high}, Stereo-seq\cite{chen2022spatiotemporal}, Seq-scope\cite{cho2021microscopic, kim2024seq}, Decoder-seq\cite{cao2024decoder}. Moreover, Open-ST\cite{schott2024open} can generate ST in 3D. 

In contrast to barcode-based techniques, image-based methods directly leverage in-situ spatial information without the need for spatial barcodes. Techniques such as in situ hybridization (ISH) and in situ sequencing (ISS) utilize gene-specific complementary DNA or RNA probes that bind to target sequences within fixed cells or tissues. Subsequently, spatial mapping of gene expression is accomplished by imaging, typically employing fluorescence or other markers. However, the number of detectable transcripts is constrained by optical limitations, allowing to detect only a few hundred targets. Ongoing technological advances aim to enhance the multiplexing capability. For example, ISH-based MERFISH\cite{xia2019spatial} utilizes 3-color imaging and can analyze approximately 10,000 genes with only 23 rounds of imaging. Other techniques include seqFISH+\cite{eng2019transcriptome}.

RNA techniques based on in situ sequencing (ISS) can be divided into targeted and untargeted approaches. Targeted ISS involves the binding of RNA or cDNA targets with specifically designed probes, such as padlock probes, followed by rolling-circle amplification (RCA) to replicate these targets for sequencing. In contrast, the untargeted ISS transcribes the transcript into cDNA using standard reverse transcription, which is followed by DNA amplification and sequencing. This approach does not require pre-selection of target genes, but may exhibit lower detection efficiency. Examples of untargeted ISS include STARmap\cite{wang2018three}, and ExSeq\cite{alon2021expansion}.

\subsection{Database}

\begin{table*}
\centering
  \caption{Commonly used database of million-level single cells. The table summarises the name, type, number of species (abbreviated as "Species"), number of datasets (abbreviated as "Datasets"), number of tissues (abbreviated as "Tissues"), number of cells (in millions, abbreviated as "Cells (M)"), and number of cell types (abbreviated as "Cell types") for each database. H: Human, M: Mus musculus.}
  \label{tab1}
  \begin{tabular}{cccccccc}
  
    \toprule
        & Name & Type & Species & Datasets & Tissues & Cells (M) & Cell types \\
    \midrule
	
1&  GEO\cite{barrett2012ncbi}  &   Comprehensive   &  \textbackslash  & 4348 &	 \textbackslash &  \textbackslash &  \textbackslash \\
2&	HCA\cite{regev2017human} &   Comprehensive  & 1 (H) & 200 & 80	 &	19.9 & 200 \\
3&  SCP\cite{tarhan2023single} &   Disease       & 16     & 780 &	106   & 55.1 &	640  \\
4&	CELLxGENE\cite{megill2021cellxgene} & Comprehensive &	1 (H) & 1634	& \textbackslash &	98.6 & 942 \\
5&	PanglaoDB\cite{franzen2019panglaodb} &	Comprehensive & 2 (H\&M)  & \textbackslash & 258 &	5.6 &	\textbackslash\\
6&	ABC Atlas & Brain & 23	& 166 &	\textbackslash & 4.0 &	34   \\
7&	CSEM\cite{zeng2022cancerscem} & Cancer & 1 (H) & 1466 & 74 &	7.3 &	80\\
8&	EA\cite{moreno2022expression}   &	Comprehensive  & 66 &	4451 &	\textbackslash  & 5.9 &	\textbackslash \\  
9&  HUSCH\cite{shi2023husch} & Comprehensive & 1 (H) &  185 & 45 & 3 & 270 \\
10& DISCO\cite{li2022disco} & Comprehensive & 1 (H) &  4593 & 107 &	18 & \textbackslash\\
11& EMBL-EBI\cite{li2015embl} & Comprehensive &	12 & 123 &\textbackslash &\textbackslash &\textbackslash\\
12& hECA\cite{chen2022heca} & Comprehensive &  1 (H) &	116 &	38	 & 1.1 &	146 \\
    \bottomrule
  \end{tabular}
\end{table*}
The volume of sequencing data has grown exponentially with the rapid advances in single-cell and spatial transcriptomics technologies, highlighting the need for curated databases, robust analysis pipelines, and effective visualization tools. This review collects 12 large-scale single-cell sequencing databases (Table \ref{tab1}) and 7 spatial transcriptomics databases (Table \ref{tab2}). A concise overview is provided in Table \ref{tab3}.
\begin{table*}
\centering
  \caption{Commonly used database of spatial transcriptomics data. The table summarises the name, type, number of species
(abbreviated as "Species"), number of datasets (abbreviated as "Datasets"), number of tissues (abbreviated as "Tissues"), number of samples (abbreviated as "Samples"), and number of publications (abbreviated as "Publications") for each database. H denotes Human. K is an abbreviation for thousand.}
  \label{tab2}
  \begin{tabular}{ccccccccc}
  
    \toprule
        & Name & Type & Species & Datasets & Tissues & Samples (k) & Publications \\
    \midrule

1&    SpatialDB\cite{fan2020spatialdb}  &   Comprehensive  &5  & 305  & \textbackslash &	 \textbackslash &  5  \\
2&	  Aquila\cite{zheng2023aquila}     &   Disease  & 5 & 110 & 26	 &	6.5 & 81 \\
3&    SOAR\cite{li2022soar}       &   Disease  & 11 & 304 &	40   & 2.8 & \textbackslash	  \\
4&	STOmicsDB\cite{xu2024stomicsdb} & Comprehensive &	17 & 231 & 128   &	7.7 & 7339 \\
5&	SPASCER\cite{fan2023spascer}  &	Comprehensive & 4  & 1082 & 16   &	\textbackslash & 43\\
6&	SODB\cite{yuan2023sodb} & Comprehensive & 12	& \texttt{>}2000 &	76 & \textbackslash &	\textbackslash   \\
7&	SORC\cite{zhou2024sorc} & Cancer & 1 (H) & 82 & 17 &	0.3	& \textbackslash\\
    \bottomrule
  \end{tabular}
\end{table*}

Single-cell omics highlights the significance of spatial context, which will increasingly be incorporated to develop multi-omics databases. The establishment of such a database not only emphasizes the integration of data sets from diverse sources, but also necessitates data preprocessing, analysis, visualization, user interaction, and other critical components.

\begin{table*}
\centering
  \caption{A list of 12 large-scale single-cell sequencing databases and 7 spatial transcriptomics databases. For single-cell sequencing, only data sets with million-level cells were included.}
  \label{tab3}
  \begin{tabular}{p{2.5cm}p{14cm}}
  
    \toprule
        Name & Brief introduction \\
    \midrule
GEO\cite{barrett2012ncbi}  &   A widely used repository for gene expression data, encompassing profiles from diverse species and experimental conditions. It includes 26,712 platforms and 4,348 datasets, each assigned a unique identifier (GEO Accession ID), and provides standardized formats and annotations. \\
HCA\cite{regev2017human} &   Delivers a comprehensive atlas of human cells, detailing their molecular and spatial characteristics across various organs, tissues, developmental stages, and disease states.\\
SCP\cite{tarhan2023single} &   Facilitates the sharing and exploration of single-cell genomics data. It enables researchers to contribute datasets and create visualizations without additional development effort. \\
CELLxGENE\cite{megill2021cellxgene} & Offers real-time tools for analyzing large-scale single-cell data. Its scalable and flexible framework allows users to adapt the code to specific analytical requirements.\\
PanglaoDB\cite{franzen2019panglaodb} &	Provides pre-processed and pre-computed analyses, simplifying data exploration. Its online interface supports queries on cell types, genetic pathways, and regulatory networks, removing the need for extensive preprocessing.\\
ABC Atlas & provides a platform for visualizingdata from multiple different cells in the mammalian brain and empower researchers to simultaneously explore and analyze multiple brain datasets.\\
CSEM\cite{zeng2022cancerscem} & Integrates scRNA-seq data from diverse human cancers, enabling researchers to explore immune profiles, gene expression dynamics, and metabolic reprogramming within tumor microenvironments.\\
EA\cite{moreno2022expression}   &	An open-access platform that provides comprehensive information on gene and protein expression across species, tissues, cell types, and biological conditions, with tools for data exploration, visualization, and analysis.\\  
HUSCH\cite{shi2023husch} & A comprehensive scRNA-seq database offering detailed cell-type annotations, gene expression visualizations, and functional analyses \\
DISCO\cite{li2022disco} & A comprehensive single-cell omics database that integrates over 18 million cells with harmonized metadata, offering tools like FastIntegration, CELLiD, and CellMapper for data integration, annotation, and projection onto global and tissue-specific atlases.\\
EMBL-EBI\cite{li2015embl} & Manages a comprehensive suite of open data resources and tools for life sciences, including the Pathogens Portal, which provides extensive biomolecular data on over 200,000 pathogen species to support infection biology, pathogen surveillance, and public health research.\\
hECA\cite{chen2022heca} &  Integrates over 1 million human cells from diverse datasets, providing advanced tools for data retrieval, multi-view biological representations, and customizable reference creation for applications in various biological studies.\\
SpatialDB\cite{fan2020spatialdb} & The first manually curated ST database. It includes 24 datasets (305 sub-datasets) spanning five species, generated using eight ST technologies. It features 6,000 gene-cell-type associations and supports automatic cell type annotation.\\
Aquila\cite{zheng2023aquila} & Facilitates transcriptomics and proteomics analyses for both 2D and 3D experiments. and provides diverse visualizations, such as spatial cell distributions, expression patterns, and marker co-localization. Researchers can securely upload and analyze their spatial transcriptomics data, enabling personalized exploration.\\
SOAR\cite{li2022soar} & Hosts large-scale ST datasets with systematic management and analysis. It ensures consistency through uniform processing and annotation, enhancing reliability for comparative studies and benchmarking.\\
STOmicsDB\cite{xu2024stomicsdb}  & Offers comprehensive analyses, including cell type annotation, spatial region and gene identification, and cell-cell interaction insights, facilitating deeper biological understanding.\\
SPASCER\cite{fan2023spascer} & Specializes in advanced analyses such as spatial transcriptomic deconvolution, spatial cell-cell interactions, gene pattern detection, and pathway enrichment, supporting more detailed spatial investigations.\\
SODB\cite{yuan2023sodb} & Accommodates a broad range of spatial transcriptomics technologies with a user-friendly interface. It enables users to generate molecular markers for specific regions, enhancing flexibility in data exploration.\\
SORC\cite{zhou2024sorc} & The first spatial transcriptomics database dedicated to cancer research, SORC includes 269 tissue slices from seven cancer types and integrates 46 single-cell data types. It provides a detailed spatial cell atlas, facilitating insights into tumor microenvironment interactions by uncovering specific genes and pathways.\\
    \bottomrule
  \end{tabular}
\end{table*}

\section{Challenges and DL Methods in Single-Cell Data Analysis}
\subsection{Data Sparsity}
Single-cell transcriptomes typically encompass tens of thousands of genes and exhibit considerable variability in expression across individual cells. Many genes remain inactive within a particular cell type, and even within the same cell type, certain genes may be transiently unexpressed due to the dynamic nature of the transcriptome and fluctuations in the cell cycle state. Consequently, gene expression matrices tend to be highly sparse, which poses challenges in modeling feature spaces, including issues related to the curse of dimensionality, noise, and uncertainty (Fig. \ref{fig2}). The overall structure of this section is shown in Fig. \ref{fig:Sparsity_structure}.

\subsubsection{Curse of Dimensionality}
The curse of dimensionality arises from the exponential increase in the volume of the space as the number of dimensions grows, which leads to sparsity and makes it more difficult to cover the space effectively with a limited number of observations. Consequently, the similarity between data points diminishes. To address this challenge, feature selection and dimensionality reduction techniques are frequently employed. Traditional methods for dimensionality reduction include parameterized approaches such as principal component analysis (PCA) \cite{mackiewicz1993principal}, along with variants like scPCA \cite{boileau2020exploring} and FastRNA \cite{lee2022fastrna}. Additionally, non-parametric methods such as t-distributed stochastic neighbor embedding (t-SNE) \cite{wang2021joint,linderman2019fast,kobak2019art} and uniform manifold approximation and projection (UMAP) \cite{becht2019dimensionality,kim2023similarity} are also widely utilized.

Nonparametric methods aim to map high-dimensional data into a lower-dimensional space while preserving local structure, typically described in terms of probabilities or metric learning. In contrast, parametric methods model the relationships between data points through explicit mathematical formulations with parameters estimated from training data. DL is a powerful parametric approach that employs neural networks for automated modeling, enabling end-to-end parameter learning directly from data. Compared to traditional methods, DL has demonstrated superior effectiveness in managing complex and high-dimensional feature spaces \cite{molho2024deep,hwang2024big,ma2022deep}. Scvis\cite{ding2018interpretable}, a nonlinear dimensionality reduction method based on the variational autoencoder (VAE) framework, integrates generative modeling with variational inference (Fig. \ref{fig2} (a)). By using two distinct neural network structures, it facilitates bidirectional mapping from high-dimensional data to low-dimensional (i.e., cell embedding) space, thereby preserving the global structure of the data. Consider a high-dimensional scRNA-seq dataset $\mathcal{D}=\left\{\mathbf{x}_n\right\}_{n=1}^N$, which comprises $N$ cells, where ${x}_n$ denotes the gene expression vector for each cell. It is assumed that the observed data is generated from a low-dimensional prior distribution given by $p\left(x_n \mid z_n, \theta\right)=\operatorname{Distribution}(\left(\mu_{\theta}\left(z_n\right)), \sigma_{\theta}\left(z_n\right)\right)$; this is typically modeled as a factorized standard normal distribution expressed as $p\left(\mathbf{z}_n \mid \boldsymbol{\theta}\right)=\prod_{i=1}^d \mathcal{N}\left(z_{n, i} \mid 0, \mathbf{I}\right)$, through a transformation parameterized by $\theta$. This parameter is difficult to compute directly and is instead approximated using a neural network. For each cell, the generative distribution can be expressed as the following integral:
\begin{equation}
    p\left(\mathbf{x}_n \mid \boldsymbol{\theta}\right)=\int p\left(\mathbf{z}_n \mid \boldsymbol{\theta}\right) p\left(\mathbf{x}_n \mid \mathbf{z}_n, \boldsymbol{\theta}\right) \mathrm{d} \mathbf{z}_{\mathrm{n}}
\end{equation}
However, computing the posterior distribution $\mathrm{p}(\mathrm{z_n} \mid \mathrm{x_n}, \theta)$ based on the observed data is intractable. To address this issue, a variational distribution $\mathrm{q}(\mathrm{z_n} \mid \mathrm{x_n}, \phi)$ is introduced as an approximation. It is assumed that the variational distribution follows a multivariate Gaussian distribution characterized by mean $\mu_\phi(\mathrm{x_n})$ and standard deviation $\sigma_\phi(\mathrm{x_n})$, both of which are functions of $x_n$, parameterized by a neural network. The model is then optimized to ensure that similar cells exhibit analogous posterior distributions. Consequently, the low-dimensional latent space effectively preserves the distance relations of the high-dimensional data, leading to efficient dimensionality reduction.

\begin{figure}
  \centering
  \includegraphics[width=\linewidth]{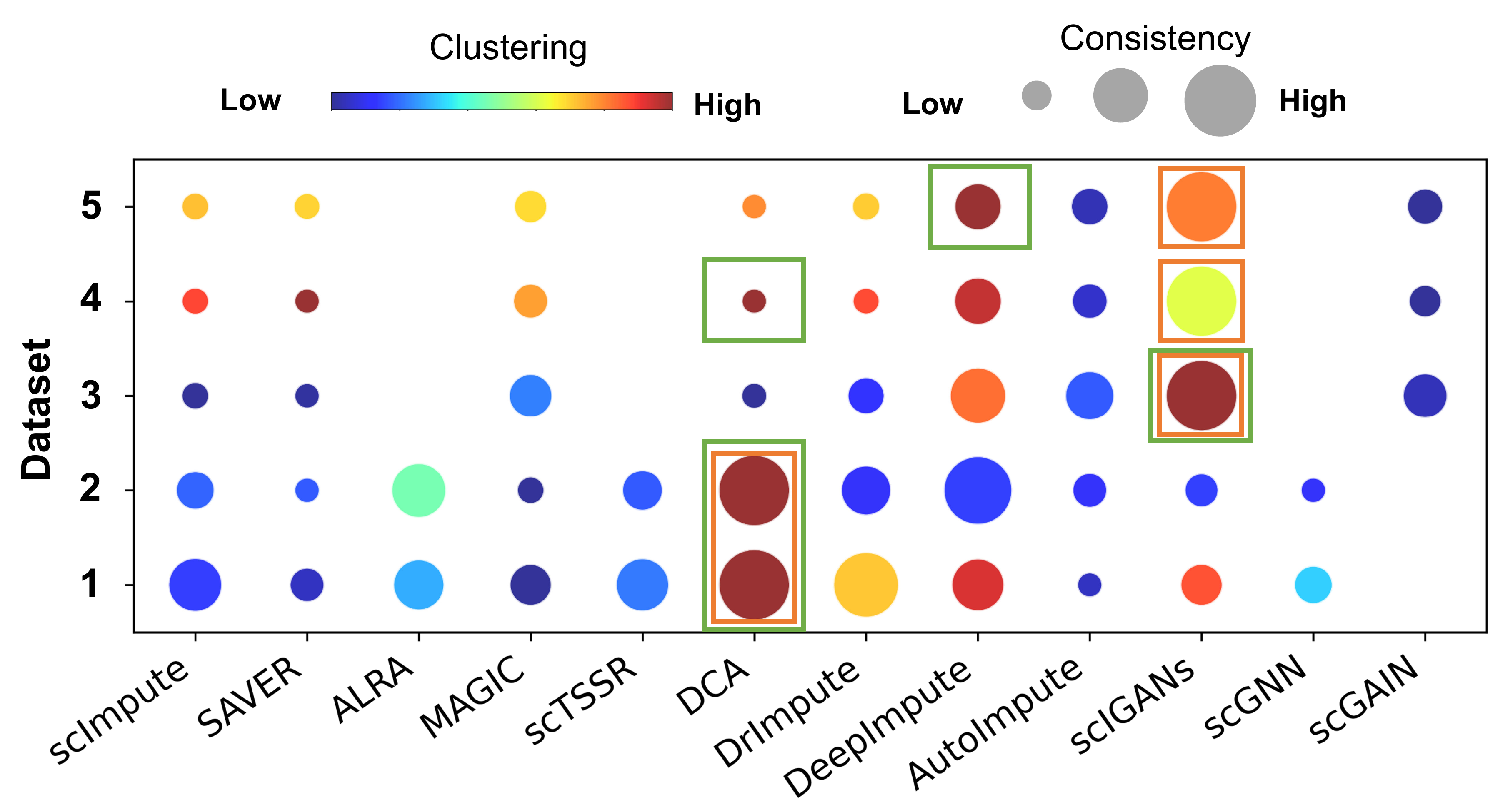}
  \caption{\textbf{Revisualize the benchmark results for data imputation from five benchmark datasets\cite{dai2022scimc, bai2024sae}.} In benchmark 1 (dataset 1 and 2), 'clustering' represents the average value of clustering evaluation metrics, including NMI and ARI, while 'consistency' includes PCC. In benchmark 2 (dataset 3-5), 'clustering' represents the mean of the NMI and ARI, and 'consistancy' refers to the mean metrics of F1, AUC and ACC. The green rectangle indicates the largest point size (imputation consistency), while the orange rectangle represents the highest color value (clustering performance). }
  \label{benchmark_sparsity}
\end{figure}

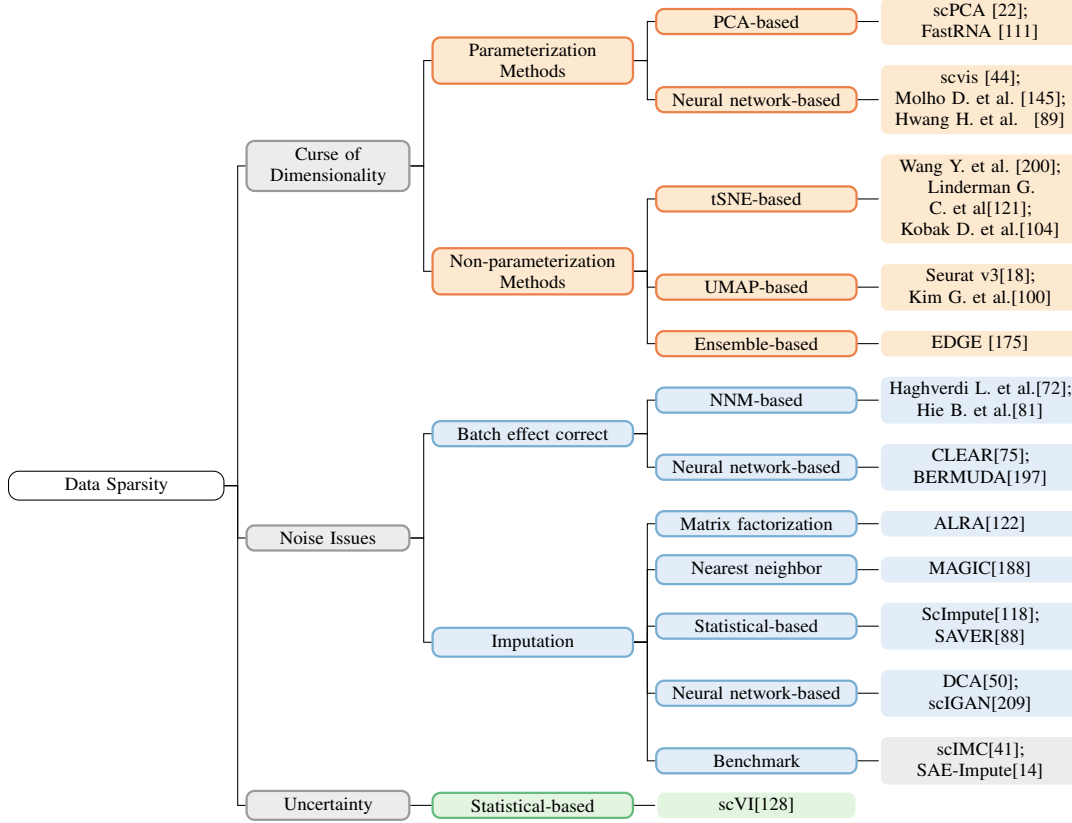
\begin{figure*}
    \scriptsize
    \hspace*{-30pt}
    \centering
    \begin{forest}
        for tree={
        forked edges,
        grow'=0,
        draw,
        rounded corners,
        node options={align=center,},
        text width=2.7cm,
        s sep=6pt,
        calign=edge midpoint,
        text=black,
        },
        [
        Data Sparsity,
        [
        Curse of~\\ Dimensionality, for tree={ top_class}
        [
        Parameterization~\\  Methods,
        for tree={fill=red!45,generation}
        [
        PCA-based,
        generation_more
        [
        scPCA~\cite{boileau2020exploring};\\
        FastRNA~\cite{lee2022fastrna},
        generation_work
        ]
        ]
        [
        Neural network-based,
        generation_more
        [
        scvis~\cite{ding2018interpretable};\\
        Molho D. et al.~\cite{molho2024deep};\\
        Hwang H. et al. ~\cite{hwang2024big},
        generation_work
        ]
        ]
        ]
        [
        Non-parameterization~\\ Methods,
        for tree={fill=red!45,generation}
        [
        tSNE-based,
        generation_more
        [
        Wang Y. et al.~\cite{wang2021joint};\\
        Linderman G. C. et al\cite{linderman2019fast};\\
        Kobak D. et al.\cite{kobak2019art},
        generation_work
        ]
        ]
        [
        UMAP-based,
        generation_more
        [
        Seurat v3\cite{becht2019dimensionality};\\
        Kim G. et al.\cite{kim2023similarity},
        generation_work
        ]
        ]
        [
        Ensemble-based,
        generation_more
        [
        EDGE~\cite{sun2020ensemble},
        generation_work
        ]
        ]
        ]
        ]
        [
        Noise Issues,
        for tree={top_class}
        [
        Batch effect correct,
        for tree={fill=green!45,gpa}
        [
        NNM-based,
        gpa_wide
        [
        Haghverdi L. et al.\cite{haghverdi2018batch};\\
        Hie B. et al.\cite{hie2019efficient},
        gpa_work
        ]
        ]
        [
        Neural network-based,
        gpa_wide
        [
        CLEAR\cite{han2022self};\\
        BERMUDA\cite{wang2019bermuda},
        gpa_work
        ]
        ]
        ]
        [
        Imputation,
        for tree={fill=blue!45,gpa}
        [
        Matrix factorization,
        gpa_wide
        [
        ALRA\cite{linderman2022zero},
        gpa_work
        ]
        ]
        [
        Nearest neighbor,
        gpa_wide
        [
        MAGIC\cite{van2018recovering},
        gpa_work
        ]
        ]
        [
        Statistical-based,
        gpa_wide
        [
        ScImpute\cite{li2018accurate};\\
        SAVER\cite{huang2018saver},
        gpa_work
        ]
        ]
        [
        Neural network-based,
        gpa_wide
        [
        DCA\cite{eraslan2019single};\\
        scIGAN\cite{xu2020scigans},
        gpa_work
        ]
        ]
        [
        Benchmark={fill=gray!45,top_class}
        topclass_wide
        [
        scIMC\cite{dai2022scimc};\\
        SAE-Impute\cite{bai2024sae},
        topclass_wide
        ]
        ]
        ]
        ]
        [
        Uncertainty,
        for tree={ top_class}
        [
        Statistical-based={fill=green!45, pretraining_more},
        pretraining_more
        [
        scVI\cite{lopez2018deep},
        pretraining_work
        ]
        ]
        ]
        ]
    \end{forest}
    \caption{\textbf{The structure of section "Data Sparsity" and related methods.} The tree chart outlines the challenges associated with processing sparse single-cell data, focusing on issues including the curse of dimensionality, noise, and uncertainty.}
    \label{fig:Sparsity_structure}
\end{figure*}

\subsubsection{Noise Issues}
Biological noise in scRNA-seq data arises from the intrinsic randomness of biological systems and variations in cellular states. In contrast, experimental noise reflects non-biological fluctuations due to technical limitations or random errors. Systematic biases, commonly referred to as batch effects, occur due to discrepancies in experimental conditions, instruments, reagents, and procedures across data batches. Furthermore, technical constraints or low capture efficiency often lead to missing values, resulting in a large number of zeros in the gene expression matrix. These zeros can obscure the true biological signal, a phenomenon known as dropout events. Batch effects and dropout events always co-occur in real-world datasets. Research has increasingly focused on batch effect correction and imputation methods to address these challenges.

\begin{figure*}[ht]
  \centering
  \includegraphics[width=\linewidth]{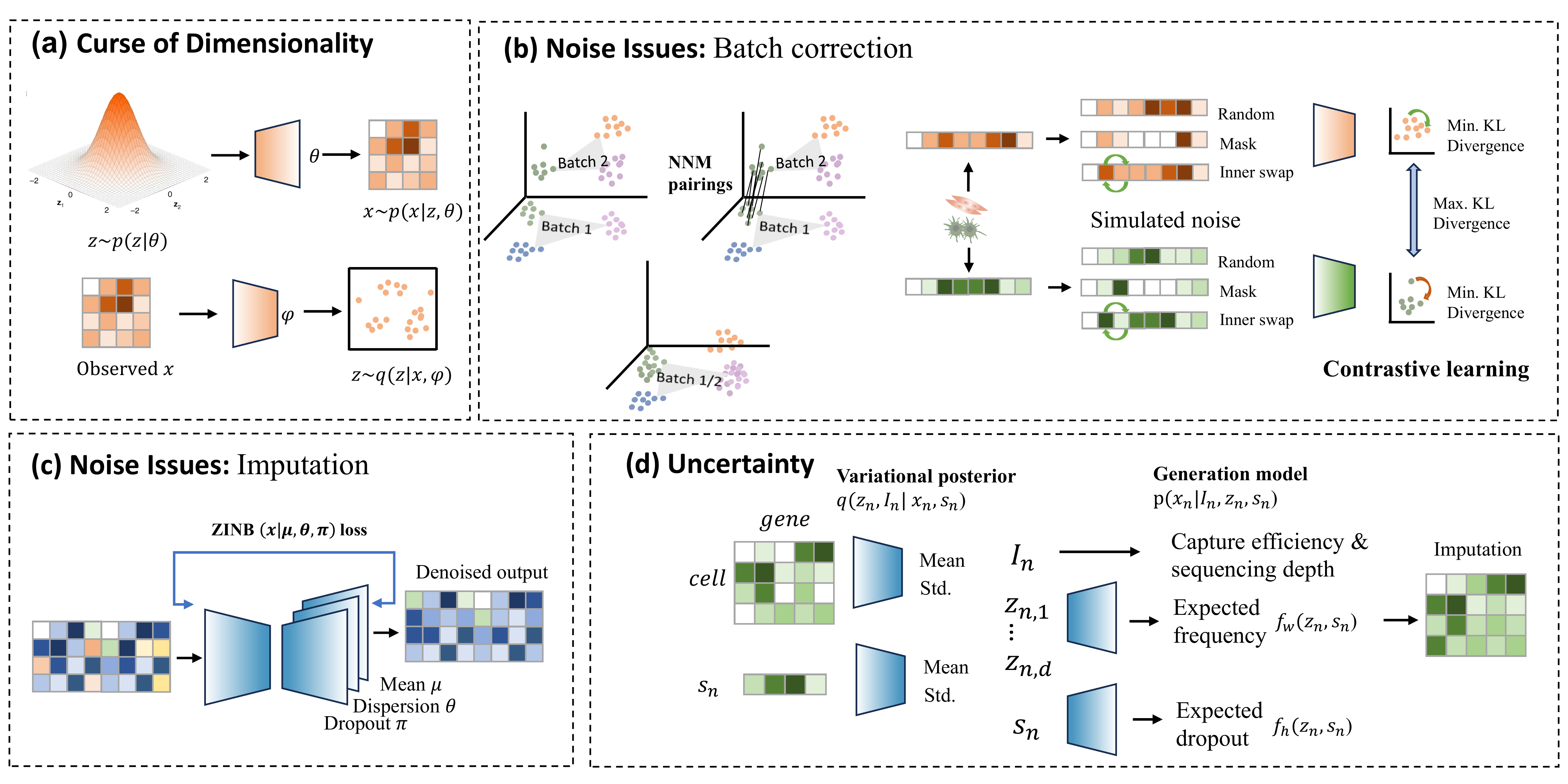}
  \caption{\textbf{The challenges and typical approaches for data sparsity.} Neural networks often modeling data representations in complex latent spaces, particularly in scenarios with increased factors of variability, such as uncertainties in experimental processes. (a) For curse of dimensionality, we plotted the framework of scvis\cite{ding2018interpretable}. (b) For batch correction, the neural networks (CLEAR\cite{han2022self}) shares the same objective as the nearest neighbors matching (NNM\cite{haghverdi2018batch}). Both approaches use distance to measure the similarity between samples, facilitating the clustering of samples of the same type across different batches. (c) For imputation, the VAE-based method (DCA\cite{eraslan2019single}) is used for noise separation through data reconstruction. (d) For modeling uncertainty, scVI incorporates stochastic factors inherent in the sequencing process, providing a framework to better capture and account for variability in the data.}
  \label{fig2}
\end{figure*}

Batch effect correction improves the comparability of datasets derived from different batches, ensuring that observed differences reflect genuine biological variation. This process involves mapping the identical cell types from various experiments to a common region within a latent space. Traditional methods, such as nearest neighbor matching (NNM)\cite{haghverdi2018batch,hie2019efficient}, address this issue by aligning representations across batches. DL-based approaches focus on the hidden space where samples are mapped to semantic representations, facilitating better learning of the underlying patterns. These methods preserve essential biological signals while filtering out irrelevant features through data reconstruction. The CLEAR algorithm\cite{han2022self}, which is based on contrastive learning, improves latent representations by constructing positive and negative sample pairs during training (Fig. \ref{fig2} (b)). Similarly, BERMUDA\cite{wang2019bermuda} aligns batch distributions using the maximum mean discrepancy (MMD) loss, facilitating the integration of data in the latent space.

Imputation methods are designed to reconstruct missing gene expression values by distinguishing technical noise from real biological zeros, relying on observed data patterns. Various approaches have been developed to address this issue, including traditional matrix factorization techniques, similarity modeling, statistical approaches, and DL strategies. Matrix factorization methods mitigate noise by preserving the dominant low-rank signal while discarding extraneous components. Traditional matrix factorization techniques are based on singular value decomposition (SVD) and non-negative matrix factorization (NMF), such as ALRA\cite{linderman2022zero}. Similarity-based approaches leverage relationships between cells to infer missing values. Actually, they leverage gene expression profiles from other cells. An example is MAGIC\cite{van2018recovering}, which employs a k-nearest neighbor algorithm to smooth and impute data based on local similarities. ScImpute\cite{li2018accurate} also relies on local structure for imputation. Statistical modeling methods employ predefined or probabilistic models to fit observed data. For example, SAVER\cite{huang2018saver} assumes gene expression follows a negative binomial distribution modeled through a Poisson-gamma mixture. It estimates the parameters using an empirical Bayesian approach with Poisson LASSO regression, and outputs the posterior means as imputed values. DL methods reframe traditional matrix operations as neural network layers, transforming parameter estimation into an optimization problem. Common frameworks include autoencoder-based models and generative architectures that adaptively learn complex patterns to enhance imputation accuracy.

Autoencoders (AEs) are widely used for imputation, effectively integrating dimensionality reduction and denoising within a unified framework. As unsupervised learning models, AEs transform high-dimensional input data into a lower-dimensional latent space, preserving essential features while eliminating redundant information. This latent representation provides contextual insights that facilitate the imputation process. The decoder reconstructs the original input from this compact representation with the objective of generating an output that closely resembles the initial data. For example, DCA\cite{eraslan2019single} employs an autoencoder with a ZINB noise model to infer key parameters such as the mean, dispersion, and dropout probabilities associated with gene expression data (Fig. \ref{fig2}(c)). The decoder produces a denoised reconstruction that is well aligned with the modeled data distribution. Variants of AEs, such as variational autoencoders (VAEs)\cite{kingma2013auto}, conditional autoencoders\cite{sohn2015learning}, and sparse autoencoders\cite{ng2011sparse}, offer additional flexibility tailored to specific applications.

Generative models, such as GANs, address the limitations of similarity-based methods that often lead to over-smoothed imputations. GAN-based models are designed to learn the underlying data distribution and generate new samples that closely resemble the denoised data. ScIGAN\cite{xu2020scigans} generates synthetic single-cell profiles instead of directly estimating missing values from observed data. This strategy minimizes overfitting to dominant cell types while improving imputation for rare cell populations. The distinctive design of scIGAN involves transforming real gene expression data into a two-dimensional image representation. The generator synthesizes gene expression profiles from latent variables, whereas the discriminator distinguishes between real and synthetic images. Both networks are trained adversarially and their performance is refined by iterative competition.

We have collected 12 methods, including scImpute\cite{li2018accurate}, SAVER\cite{huang2018saver}, ALRA\cite{linderman2022zero}, MAGIC\cite{van2018recovering}, scTSSR\cite{jin2020sctssr}, DCA\cite{eraslan2019single}, DrImpute\cite{gong2018drimpute}, DeepImpute\cite{arisdakessian2019deepimpute}, AutoImpute\cite{talwar2018autoimpute}, scIGANs\cite{xu2020scigans}, scGAIN\cite{gunady2019scgain}, to evaluate their imputation performance on five benchmark datasets\cite{dai2022scimc, bai2024sae}. The results demonstrate that DCA and scIGANs each achieved the highest imputation consistency across the two benchmarks, with both methods displaying robust clustering performance in four out of five datasets (Fig. \ref{benchmark_sparsity}).

\subsubsection{Uncertainty} 
Uncertainty issues typically arise from factors that contribute to ambiguity during analysis, decision-making, or prediction. This is caused by insufficient information, measurement errors, inaccurate model assumptions, or other sources of variability. In addition to the aforementioned approaches for addressing batch effects and dropout events, uncertainty quantification can enhance model selection and performance evaluation. This procedure facilitates the mitigation of uncertainties arising from experimental data and model assumptions\cite{gawlikowski2023survey}. An example is scVI\cite{lopez2018deep}, which explicitly incorporates batch annotations and addresses batch effects through conditional independence assumptions (Fig. \ref{fig2} (d)). This approach effectively isolates batch-related factors from the data, thereby reducing the uncertainties associated with batch differences and improving gene expression analysis. scVI models the observed expression $x_{ng}$ of each gene $g$ in each cell $n$ as a random sample from a zero-inflated negative binomial distribution (ZINB) denoted as $p(x_{ng}\mathrm{|}z_n, s_n, l_n)$. Here, $z_n$ represents a low-dimensional Gaussian vector that captures biological differences between cells; $l_n$ is a one-dimensional Gaussian variable that accounts for variation due to differences in capture efficiency and sequencing depth, serving as a cell-specific scaling factor; and $s_n$ denotes the batch annotation of the cell (if available). Employing variational inference and reparameterization techniques, scVI optimizes the posterior distribution via a variational lower bound. By incorporating sources of uncertainty, including cell-specific and batch-dependent features, this approach effectively preserves the information inherent in the original data. In contrast, posterior correction methods may rely on fixed assumptions, which can lead to the loss of critical information or introduce bias.

\subsection{Data diversity}

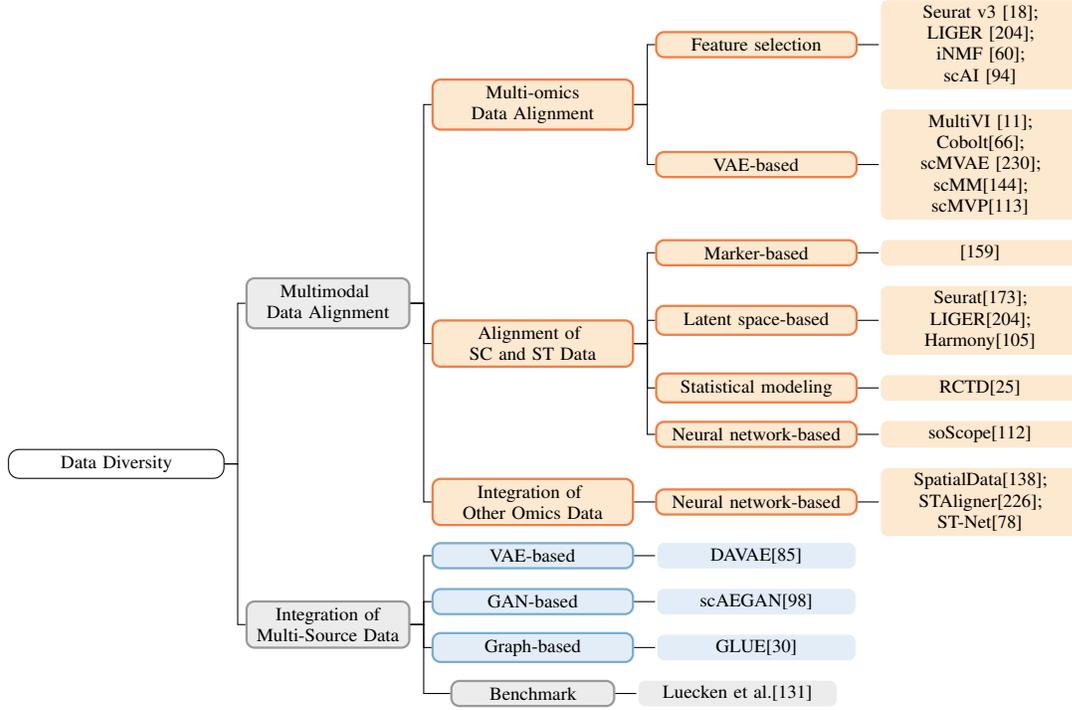
\begin{figure*}[ht]
    \scriptsize
    \hspace*{-30pt}
    \centering
    \begin{forest}
        for tree={
        forked edges,
        grow'=0,
        draw,
        rounded corners,
        node options={align=center,},
        text width=2.7cm,
        s sep=6pt,
        calign=edge midpoint,
        text=black,
        },
        [
        Data Diversity,
        [
        Multimodal ~\\ Data Alignment, for tree={ top_class}
        [
        Multi-omics~\\ Data Alignment,
        for tree={fill=red!45,generation}
        [
        Feature selection,
        generation_more
        [
        Seurat v3~\cite{becht2019dimensionality};\\
        LIGER~\cite{welch2019single};\\
        iNMF~\cite{gao2021iterative};\\
        scAI~\cite{jin2020scai},
        generation_work
        ]
        ]
        [
        VAE-based,
        generation_more
        [
        MultiVI~\cite{ashuach2023multivi};\\
        Cobolt\cite{gong2021cobolt};\\
        scMVAE~\cite{zuo2021deep};\\
        scMM\cite{minoura2021mixture};\\
        scMVP\cite{li2022deep},
        generation_work
        ]
        ]
        ]
        [
        Alignment of ~\\SC and ST Data,
        for tree={fill=red!45,generation}
        [
        Marker-based,
        generation_more
        [
        \cite{satija2015spatial},
        generation_work
        ]
        ]
        [
        Latent space-based,
        generation_more
        [
        Seurat\cite{stuart2019comprehensive};\\
        LIGER\cite{welch2019single};\\
        Harmony\cite{korsunsky2019fast},
        generation_work
        ]
        ]
        [
        Statistical modeling,
        generation_more
        [
        RCTD\cite{cable2022robust},
        generation_work
        ]
        ]
        [
        Neural network-based,
        generation_more
        [
        soScope\cite{li2024tissue},
        generation_work
        ]
        ]
        ]
        [
        Integration of  ~\\Other Omics Data,
        for tree={fill=red!45,generation}
        [
        Neural network-based,
        generation_more
        [
        SpatialData\cite{marconato2024spatialdata};\\
        STAligner\cite{zhou2023integrating};\\
        ST-Net\cite{he2020integrating},
        generation_work
        ]
        ]
        ]
        ]
        [
        Integration of Multi-Source Data,
        for tree={ top_class}
        [
        VAE-based={fill=green!45,gpa},
        gpa_wide
        [
        DAVAE\cite{hu2022versatile},
        gpa_work
        ]
        ]
        [
        GAN-based={fill=green!45,gpa},
        gpa_wide
        [
        scAEGAN\cite{khan2023scaegan},
        gpa_work
        ]
        ]
        [
        Graph-based={fill=green!45,gpa},
        gpa_wide
        [
        GLUE\cite{cao2022multi},
        gpa_work
        ]
        ]
        [
        Benchmark={fill=gray!45,top_class}
        topclass_wide
        [
        Luecken et al.\cite{luecken2022benchmarking},
        topclass_wide
        ]
        ]
        ]
        ]
    \end{forest}
    \caption{\textbf{The structure of section "Data Diversity" and related methods.} The tree chart outlines the challenges associated with processing multi-view single-cell data, focusing on issues including multimodal data alignment, and the integration of multi-source data.}
    \label{fig:diversity_structure}
\end{figure*}

\begin{figure*}[ht]
  \centering
  \includegraphics[width=\linewidth]{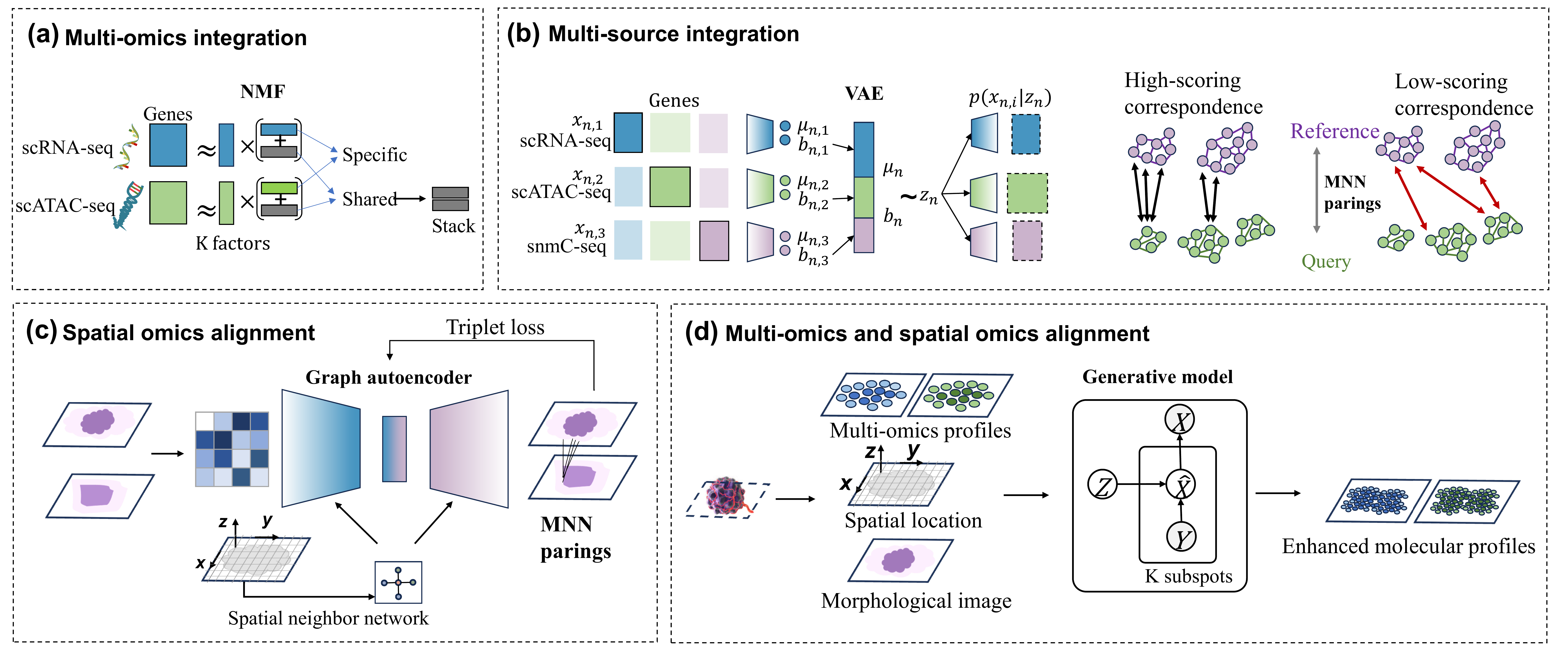}
  \caption{\textbf{The challenges and typical approaches for data diversity.} The integration of multi-modal and multi-omics data via DL is a trend in the study of data diversity. (a) For multi-omics integration, we plotted the framework of LIGER\cite{welch2019single}. Multi-omics data alignment aims to align similar features while preserving the unique characteristics of each modality. (b) For multi-source integration, Seurat v3\cite{stuart2019comprehensive} integrates scRNA-seq experiments with scATAC-seq based on anchors. GLUE\cite{cao2022multi} integrates omics-specific autoencoders to model regulatory interactions between omics layers. (c) For spatial alignment, STAligner\cite{zhou2023integrating} enables coordinate registration of stacked consecutive slices and 3D tissue reconstruction based on MNN and graph autoencoder. (d) For multi-omics and spatial transcriptomics alignment, soScope\cite{li2024tissue} jointly infers high-resolution spatial maps using a variational Bayesian inference network.}
  \label{fig3}
\end{figure*}

The "central dogma" delineates the flow of genetic information from DNA to RNA and subsequently to proteins,  establishing a foundational framework for understanding how gene expression influences cellular functions. Omics data, obtained through high-throughput techniques, provide a systematic characterization of the various molecular components within an organism, including the genome, transcriptome, proteome, and metabolome. Recent advances in multichannel sequencing now allow simultaneous measurement of multiple types of omics data. Current transcriptome-focused multimodal techniques include combinations such as gDNA-mRNA\cite{rodriguez2019unravelling, dey2015integrated}, mRNA-methylation\cite{angermueller2016parallel, hou2016single}, mRNA-ATAC\cite{cao2018joint,chen2019high,ma2020chromatin}, mRNA-proteome\cite{gerlach2019combined,stoeckius2017simultaneous}, and mRNA-methylation-ATAC\cite{wang2021single,clark2018scnmt}. The observed heterogeneity encompasses intra-sample heterogeneity (resulting from differences in sequencing depth, coverage, and data type), inter-sample heterogeneity (caused by variations in experimental design, sample handling, and sequencing protocols), and variability across species and individuals. The diversity and complexity inherent in single-cell data present significant challenges, particularly when it comes to aligning and integrating paired and unpaired datasets. "Paired" data refers to multimodal datasets derived from the same sample, whereas "unpaired" data consists of multimodal datasets from different samples or platforms. Multi-omics analysis integrates these diverse data types to facilitate a comprehensive understanding of organismal heterogeneity and regulatory mechanisms. The overall structure of this section is shown in Fig. \ref{fig:diversity_structure}.

\subsubsection{Multimodal Data Alignment}
This section focuses on the alignment of multimodal data, including multi-omics data as well as paired single-cell and spatial transcriptomics data, with the goal of uncovering intrinsic patterns in the alignment of homologous data. 

\paragraph{Multi-omics Data Alignment}
Multi-omics data alignment aims to align similar features while preserving the unique characteristics of each modality. LIGER\cite{welch2019single} employs non-negative matrix factorization (NMF) to uncover latent structures in the data, extracting both shared and modality-specific gene expression patterns while minimizing distances between datasets (Fig. \ref{fig3} (a)). iNMF\cite{gao2021iterative} builds on LIGER by enabling online learning for enhanced data integration. Other comparable methods include scAI\cite{jin2020scai}. MultiVI\cite{ashuach2023multivi} adopts a VAE framework where the encoder processes different molecular attributes, such as protein abundance, to generate modality-specific latent representations, denoted as $q\left(z_R \mid x_R, s\right)$ and $q\left(z_A \mid x_A, s\right)$. The cell state is estimated as the average of these two representations: $q\left(z \mid x_R, x_A, s\right)$, thereby forming a joint latent space of multiple modalities. Modality-specific decoders then generate the observed values using a negative binomial distribution for transcriptomic data and a Bernoulli distribution for chromatin accessibility data. A constraint is imposed on the latent space to minimize the distance between these two representations. Additional VAE-based models include Cobolt\cite{gong2021cobolt}, scMVAE\cite{zuo2021deep}, scMM\cite{minoura2021mixture}, and scMVP\cite{li2022deep}. scMVP maximizes the likelihood of jointly generated probabilities across multi-omics data by introducing a Gaussian mixture model (GMM) prior to obtain a shared latent embedding. Each modality is encoded separately using an asymmetric GMM-VAE model that incorporates two clustering consistency modules to align each imputed dataset while preserving the shared semantic information.

\paragraph{Alignment of Single-Cell and Spatial Transcriptomics Data}
Spatial mapping in ST involves aligning scRNA-seq data with physical spatial domains, matching the geometry of the spatial data. Traditional methods have aimed to reconstruct key marker genes by assuming continuity in gene expression or using local alignment information\cite{satija2015spatial}. However, these methods are hindered by limited capture rates, significant dropout events, and sparse gene distribution, making them error-prone and unable to generalize across different experimental settings. More recent approaches, such as Seurat\cite{stuart2019comprehensive}, LIGER\cite{welch2019single}, and Harmony\cite{korsunsky2019fast}, integrate scRNA-seq with ST data through shared latent spaces and mutual nearest neighbors (MNN). This integration facilitates the transfer of cell-type labels while enhancing weak ST signals. RCTD\cite{cable2022robust} integrates reference scRNA-seq data to model the average gene expression profiles of different cell types. It utilizes a hierarchical model to estimate the proportion of each cell type within individual spatial spots. The method applies a Poisson distribution to estimate gene expression counts and employs maximum likelihood estimation (MLE) for parameter estimation. soScope\cite{li2024tissue} adopts a multimodal DL framework that integrates spot-level omics maps (transcript, histone, DNA, protein), spatial relationships, and high-resolution morphological images. It jointly infers high-resolution spatial maps using a variational Bayesian inference network (Fig. \ref{fig3} (d)).

\paragraph{Integration of Other Omics Data}
Many open-source frameworks have been developed to facilitate the alignment and integration of multimodal spatial omics data, including SpatialData\cite{marconato2024spatialdata}, SSGATE\cite{lv2024multi}, STAligner\cite{zhou2023integrating} and SpatialGlue\cite{long2024deciphering}. For example, SpatialGlue\cite{long2024deciphering} can be used to integrate three modalities, including spatial epigenome–transcriptome and transcriptome–proteome modalities. For aligning 2D slices, STAligner\cite{zhou2023integrating} employs a triplet-based approach to identify mutual nearest neighbors that exhibit similar gene expression patterns across different slices. This method facilitates coordinate registration of stacked consecutive slices, enabling 3D tissue reconstruction (Fig. \ref{fig3} (c)). ST-Net\cite{he2020integrating} integrates paired H\&E-stained pathology images with ST data to train an end-to-end neural network for predicting spatially resolved transcriptomics from pathology images.

Recent studies have highlighted advancements in ST techniques \cite{hu2024benchmarking}, although the field is still challenged by a trade-off between spatial resolution and measurement throughput. spatial transcriptomics, bridging imaging and sequencing, holds great potential for integrating diverse modalities from histopathology and single-cell data, offering deeper insights into spatial organization, microenvironmental interactions, and histopathology. Moving forward, the integration of multimodal data remains a key challenge in single-cell and spatial transcriptomics analysis.

\begin{figure}
  \centering
  \includegraphics[width=\linewidth]{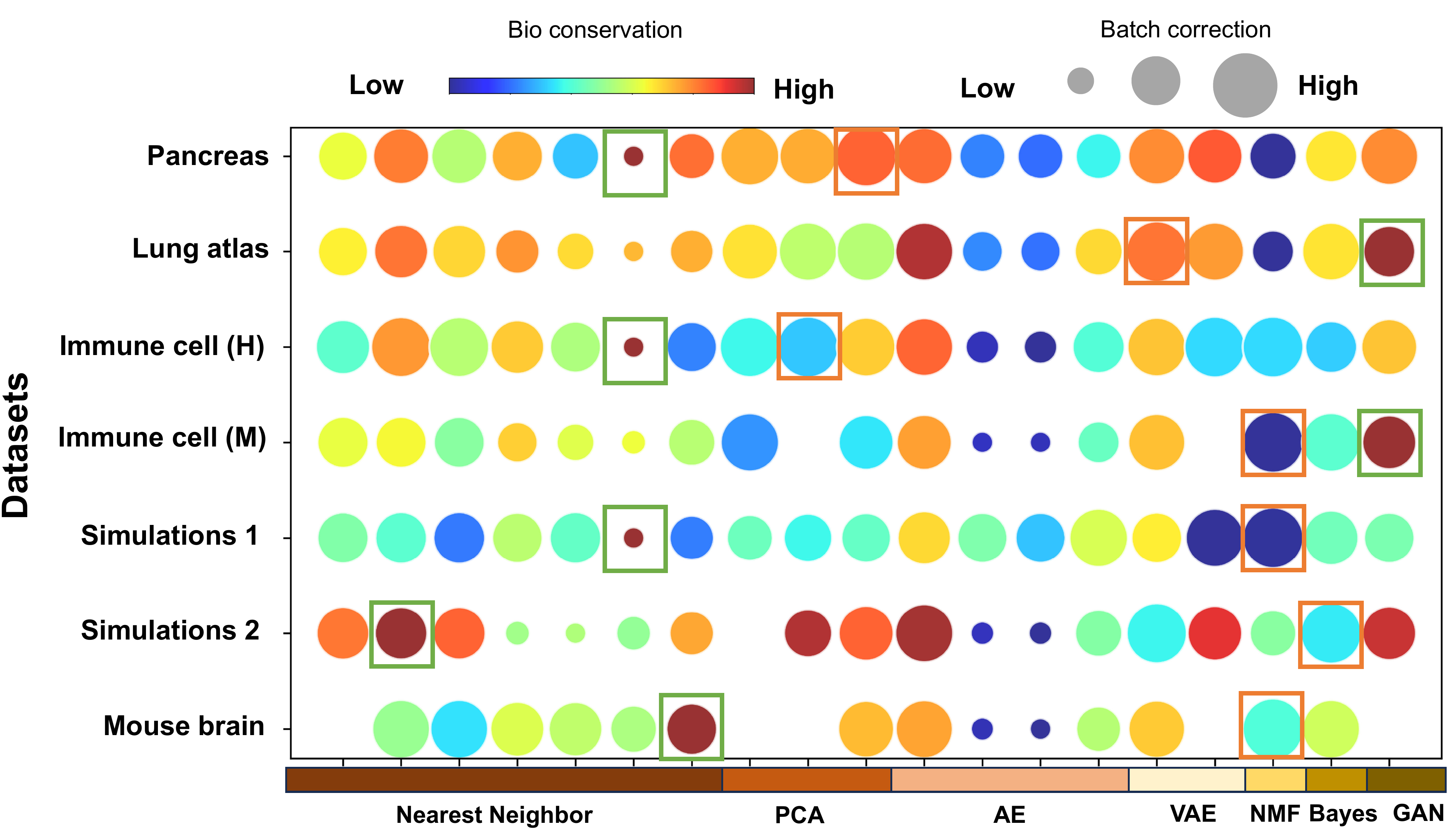}
  \caption{\textbf{Revisualize the benchmark results for multi-source data integration from seven benchmark datasets\cite{luecken2022benchmarking}.} This analysis includes 19 reported methods, including nearest neighbor (NN), PCA, AE, VAE, NMF, Bayes, and GAN. The orange rectangle represents the largest point size (batch correction), while the green rectangle indicates the points with the highest color value (bio-conservation), closest to red on the color bar.}
  \label{benchmark_diversity}
\end{figure}
\subsubsection{Integration of Multi-Source Data}

Integration of unpaired datasets, such as those derived from different samples or sequencing platforms, usually requires alignment of independent feature spaces. In this context, cross-modal integration aims to mitigate discrepancies in embeddings of the same cell type across heterogeneous modalities. Seurat v3\cite{stuart2019comprehensive} identifies common anchors (cell pairs) across datasets based on features (such as genes) that exhibit high variability among cells, integrating data using these anchors (Fig. \ref{fig3} (b)). It can integrate scRNA-seq with scATAC-seq, allowing for an investigation into chromatin differences.

Generative DL models have been widely employed to capture complex semantic relationships in multi-source datasets. For example, DAVAE\cite{hu2022versatile} integrates large-scale unpaired data through three essential components: a variational approximation network, a generative Bayesian neural network, and a domain adversarial classifier. This setup benefits the learning of latent representations, enhances data fitting, and mitigates batch effects to accurately represent cellular biological states across diverse datasets. 
Similarly, scAEGAN\cite{khan2023scaegan} combines autoencoders (AE) with conditional generative adversarial networks (cGAN) to facilitate the translation between different single-cell datasets. It effectively transforms the dataset by using AE to remove random noise, while employing cGAN with recurrent consistency regularization. GLUE integrates omics-specific autoencoders with graph-based coupling and adversarial alignment to model regulatory interactions between omics layers, supporting integrated regulatory inference for unpaired multi-omics datasets (Fig. \ref{fig3} (b)).

Recent benchmark for multi-source data integration has collected 19 methods for data integration on seven benchmark datasets, as detailed by Luecken et al.\cite{luecken2022benchmarking}, indicates that NMF-based and nearest neighbor approaches exhibit superior performance in average bio-conservation and batch correction across all datasets, respectively. We also observed the effectiveness of the VAE-based and GAN-based model in preserving biological consistency (Fig. \ref{benchmark_diversity}).

\subsection{Data scarcity}
\begin{figure*}
    \scriptsize
    \hspace*{-30pt}
    \centering
    \begin{forest}
        for tree={
        forked edges,
        grow'=0,
        draw,
        rounded corners,
        node options={align=center,},
        text width=2.7cm,
        s sep=6pt,
        calign=edge midpoint,
        text=black,
        },
        [
        Data Scarcity,
        [
        Missing Data Annotation, for tree={ top_class}
        [
        Statistical modeling={fill=green!45,gpa},
        gpa_wide
        [
        scDesign3\cite{song2024scdesign3},
        gpa_work
        ]
        ]
        [
        GAN-based={fill=green!45,gpa},
        gpa_wide
        [
        GRouNdGAN\cite{zinati2024groundgan},
        gpa_work
        ]
        ]
        [
        Benchmark={fill=gray!45,top_class}
        topclass_wide
        [
        Pratapa A. et al.\cite{pratapa2020benchmarking};\\
        Cao Y. et al.\cite{cao2021benchmark},
        topclass_wide
        ]
        ]
        ]
        [
        Missing Modalities,
        for tree={ top_class}
        [
        VAE-based={fill=green!45,pretraining},
        pretraining_wide
        [
        TotalVI\cite{gayoso2021joint};\\
        UniPort\cite{cao2022unified};\\
        POE\cite{wu2018multimodal};\\
        MOE\cite{shi2019variational};\\
        CGVAE\cite{liu2018constrained},
        pretraining_work
        ]
        ]
        [
        Benchmark={fill=gray!45,top_class}
        topclass_wide
        [
        Hu Y. et al.\cite{hu2024benchmarking};\\
        Makrodimitris S. et al.\cite{makrodimitris2024depth},
        topclass_wide
        ]
        ]
        ]
        ]
    \end{forest}
    \caption{\textbf{The structure of section "Data Scarcity" and related methods.} The tree chart outlines the challenges related to the scarcity of high-quality single-cell data, emphasizing issues such as missing data annotation and missing modalities.}
    \label{fig:sparcity_structure}
\end{figure*}
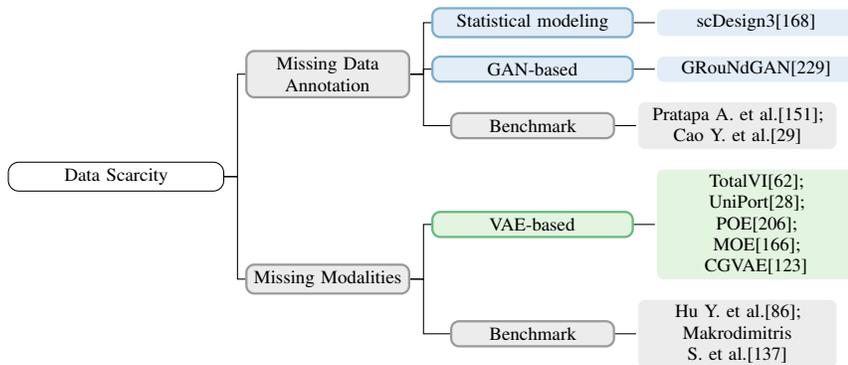

The overall structure of this section is shown in Fig. \ref{fig:sparcity_structure}.

\subsubsection{Missing Data Annotation}
Single cell data has reached sequencing scales of hundreds of millions. Due to the significant labor and time required in laboratory settings, existing datasets often present challenges in obtaining large-scale biological annotations. In addition, single-cell data contains many complex biological factors, making it difficult to obtain a reasonable and reliable ground truth. For instance, benchmark datasets for experimental analysis of single-cell population evolution require follow-up samples with known evolutionary trajectories and developmental timelines, which are difficult to obtain under experimental conditions\cite{lahnemann2020eleven}. Moreover, the reliability of model evaluation often depends on high-quality data annotations. For example, since regulatory interactions in databases are aggregated from broad datasets and lack specificity to particular biological systems, it is unreliable to evaluate the performance of gene regulatory network (GRN) inference algorithms. A key technical solution to this problem is the construction of simulation datasets.

It has been extensively employed to evaluate and compare computational methods, concentrate on specific biological features, and establish more precise benchmarks\cite{pratapa2020benchmarking}. Cao et al.\cite{cao2021benchmark} conducted a comprehensive review of various simulation approaches and proposed a framework for systematic benchmarking, highlighting their ability to capture biological signals and higher-order interactions, such as the mean-variance relationship among genes. However, most existing methods generate simulated data tailored to specific evaluation objectives, such as clustering or differential gene expression analysis. There are few tools specifically designed to create datasets that are applicable across diverse scenarios.

scDesign3\cite{song2024scdesign3} employs statistical modeling methods to generate single-cell multi-omics data and spatial transcriptomics data with known cell proportions. It standardizes generative modeling approaches across various data modalities, rather than focusing on only one modality. Given a cell state covariate $x_i$ (factors such as cell type, cell pseudotime, and cell spatial locations) and experimental design covariates $z_i$ (such as batch effects and experimental conditions), the measurement values $Y_{ij}$ are modeled according to a specific distribution $F_j\left(\cdot \mid \mathbf{X}_i, \mathbf{z}_i ; \mu_{i j}, \sigma_{i j}, p_{i j}\right)$. This is formulated as a generalized additive model for location scale and shape (GAMLSS) , characterized by its position, proportion, and shape parameters.

The model is parametrically represented, incorporating specific link functions for each feature $j's$ (distribution functions $\theta_j\left(\mu_{i j}\right)$, such as Gaussian (Normal), Bernoulli, Poisson, ZINB) These link functions correspond to the mean parameter $\mu_{i j}$, the scale parameter $\sigma_{i j}$(for example, standard deviation or dispersion), and zero-inflation proportion parameter $p_{ij}$. For instance, the specific link functions $\theta_j\left(\mu_{i j}\right)$ for features $j$ maps the mean parameter $\mu_{i j}$ to the model's linear predictor. This mapping depends on the chosen distribution function and consists of four key components:
\begin{equation}
    \theta_j\left(\mu_{i j}\right)=\alpha_{j 0}+\alpha_{j b_i}+\alpha_{j c_i}+f_{j c_i}\left(\mathbf{x}_i\right)
\end{equation}
The specific intercept $\alpha_{j 0}$ for feature $j$,  represents the mean of feature $j$ in the absence of other influencing factors. The batch effect $\alpha_{j b_i}$,  captures the influence of batch $b_i$ on feature $j$. The conditional effect $\alpha_{j c_i}$ denotes the impact of condition $c_i$ on feature $j$. The cell state covariates $x_i$, such as the effects associated with different cell types on feature $j$, are also considered. In scDesign3, the joint distribution of cellular features is constructed using a marginal cumulative distribution function and a copula model with parameters estimated by a maximum likelihood method. These parameters can be adjusted to generate synthetic data reflecting varying sequencing depths and cell states. Furthermore, scDesign3 is capable of producing spatial transcriptomic data based on cell type proportions derived from single-cell sequencing, simulating realistic ATAC-seq datasets at both count and read levels, and generating multi-omics datasets by integrating separate omics datasets like RNA expression or DNA methylation. 

In summary, statistical modeling provides a highly interpretable framework for data generation and analysis, and its integration with DL is emerging as a significant trend.

GlouNdGAN\cite{zinati2024groundgan}  is a causal model-based data generation method that allows the generation of realistic simulated data aligned with the underlying principles of GRN. The architecture of GlouNdGAN consists of five sub-networks: Causal Controller, Target Generator, Critic, Labeler, and Anti-labeler. The Causal Controller generates expression values for transcription factors (TFs), while the Target Generators produce expression values for target genes based on the causal GRN framework. The Critic estimates the Wasserstein distance between the generated data and the real data to ensure that the target gene expression is causally related to TF expression. The Labeler predicts TF expression based on generated and real target gene expression, while the Anti-labeler estimates TF expression solely from generated target gene expression. This model pre-trains the TF expression generation module and subsequently generates expression values for other genes via the Target Generator. GlouNdGAN allows researchers to simulate interference by manipulating TF expression values during the inference phase, enabling an accurate comparison of gene expression before and after interference while maintaining constant parameters. Additionally, by performing mutation experiments on TF expression for certain cell types, the researchers observed alterations in the characteristics of the generated cells, thus validating the function of TFs and their relationship to phenotypic labels. This capability positions GlouNdGAN as an ideal tool for in-situ interference experiments.

As discussed above, simulation data generation serves as a valuable tool for elucidating biological mechanisms in contexts where high-quality data is lacking. It allows the creation of diverse datasets with controllable parameters and facilitates the evaluation of model performance.

\begin{figure*}[ht]
  \centering
  \includegraphics[width=\linewidth]{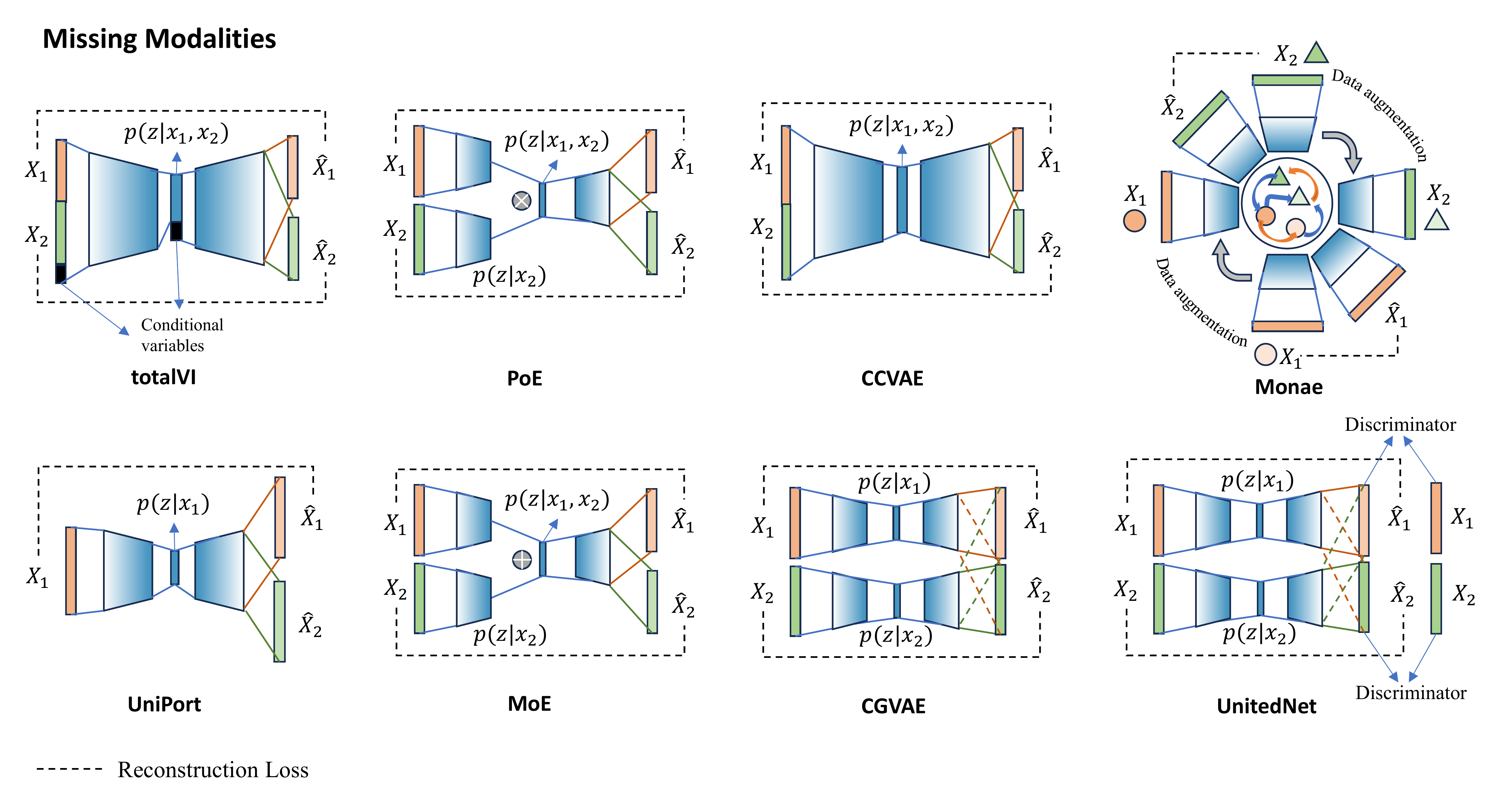}
  \caption{\textbf{The challenges and typical approaches for data scarcity.} DL-based methods mainly rely on VAE architectures, using either single-modality or dual-modality joint embeddings for feature modeling to learn a shared latent space. We plot the basic architectures of the following VAE-based methods: totalVI\cite{gayoso2021joint}, UniPort\cite{cao2022unified}, POE\cite{wu2018multimodal}, MOE\cite{shi2019variational}, CCVAE\cite{svahn2022ccvae}, CGVAE\cite{liu2018constrained}. Recently, additional strategies have been incorporated into generative models, such as introducing regularized discriminators (UnitedNet\cite{tang2023explainable}) and employing data augmentation strategies (Monae\cite{tang2024modal}).}
  \label{fig4}
\end{figure*}
\subsubsection{Missing modalities}

Although genetic information is transferred from DNA to RNA and then to proteins, each modality captures distinct biological information, making it impossible for one modality to substitute for another. It has been demonstrated that multimodal analysis enhances the overall understanding of cellular heterogeneity. However, multi-channel sequencing typically incurs higher costs compared to single-channel sequencing. DL-based solutions commonly rely on VAE architectures that use either single-modality or multi-modality joint embeddings for shared latent space modeling\cite{makrodimitris2024depth}. The difference lies in how the latent variables are modeled (Fig. \ref{fig4}). TotalVI\cite{gayoso2021joint} is trained on the joint embeddings of the two modalities with separate reconstruction. UniPort\cite{cao2022unified} trains a single-modality embedding to reconstruct two different modalities, compelling the encoder to learn features that are predictive of both. In the  Product of Experts (POE) model\cite{wu2018multimodal}, the joint latent variable is derived as a product of each modality's. Unlike POE, Mixture of Experts (MOE)\cite{shi2019variational} employs the sum of the joint latent variables for data reconstruction. Constrained Graph Variational Autoencoders (CGVAE) \cite{liu2018constrained} learns feature embeddings for each modality individually while applying constraints to ensure that each modality can reconstruct both itself and the other modalities. Based on the benchmark results, Makrodimitris, S. et al.\cite{makrodimitris2024depth} concluded that different joint embeddings can be used for different downstream tasks.

Scenarios of modality completion are often related to data sparsity. Monae\cite{tang2024modal} employs data imputation to perform data denoising and modality completion simultaneously, constrained by a cross-modal prediction loss. In the feature extraction phase, a graph encoder-decoder reconstruction process extracts embedding features from multiple modalities, and contrastive learning is applied to minimize the spatial distance between embeddings of the same cell type. Therefore, when discriminative information from one modality is lacking, the latent space embeddings of other modalities can be leveraged for reconstruction. UnitedNet\cite{tang2023explainable} combines multimodal ensemble with cross-modal prediction in a multi-task learning framework, trained with cross-modal prediction loss alongside generator and discriminator losses. 

\begin{figure}
  \centering
  \includegraphics[width=\linewidth]{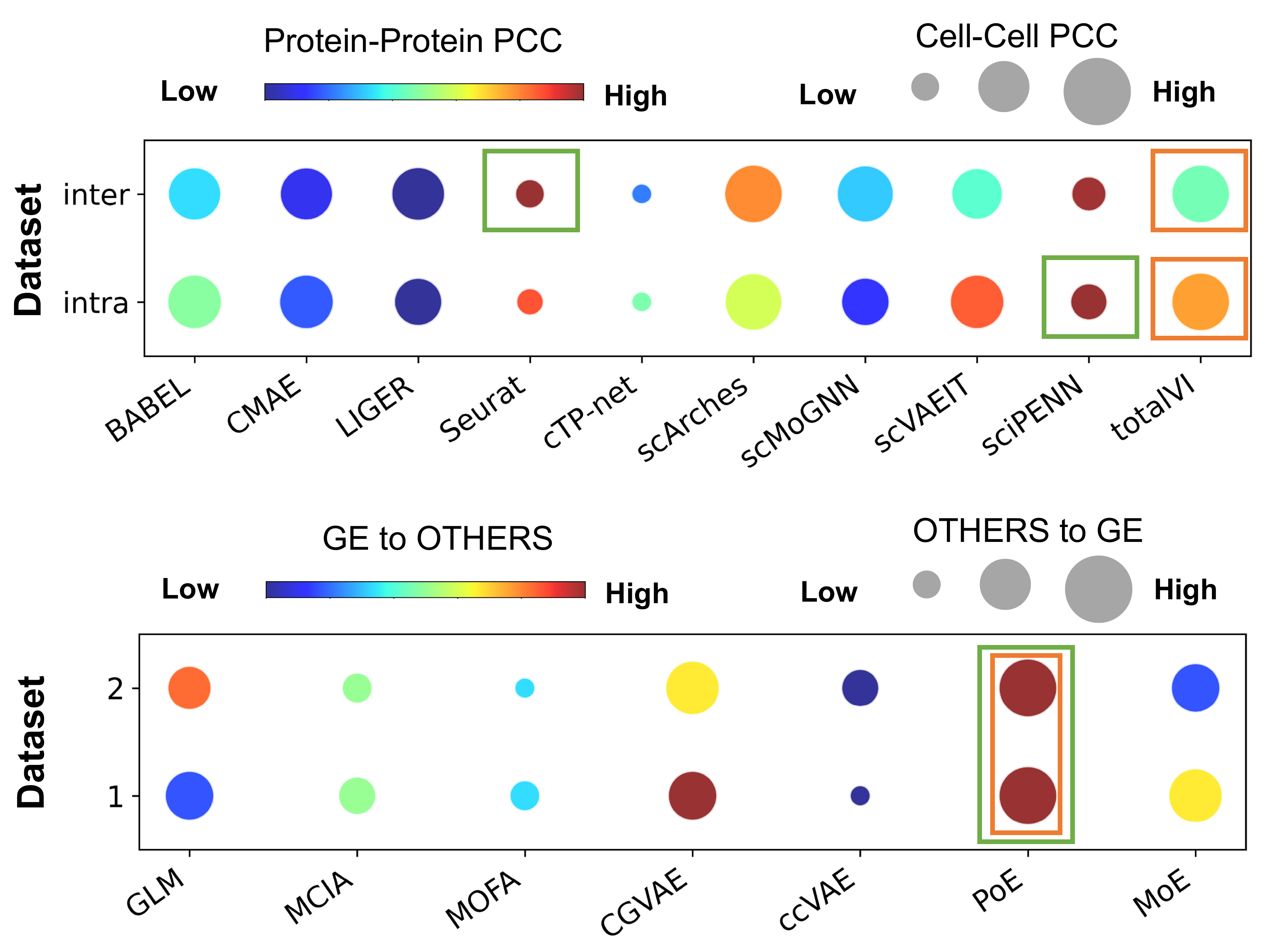}
  \caption{\textbf{Revisualize the benchmark results for modality prediction from four benchmark datasets\cite{hu2024benchmarking,makrodimitris2024depth}.} The orange rectangle represents the largest point size, while the green rectangle indicates the points with the highest color value, closest to red on the color bar. In Benchmark 1 (Dataset "inter" and Dataset "intra", the size of the bubbles represents the abundance of a protein (or chromatin accessibility) across two cells, while the color indicates the Pearson correlation coefficients (PCCs) between pairs of proteins. In Benchmark 2 (Dataset 1 and Dataset 2), 'GE to OTHERS' refers to the average accuracy of translation from GE to other modalities, and vice versa.}
  \label{benchmark_missingmodality}
\end{figure}

We have collected 17 methods, including BABEL\cite{wu2021babel}, CMAE\cite{yang2021multi}, LIGER\cite{welch2019single}, Seurat\cite{stuart2019comprehensive}, cTP-net\cite{zhou2020surface}, scArches\cite{lotfollahi2022mapping}, scMoGNN\cite{lan2023efficient}, scVAEIT\cite{du2022robust}, sciPENN\cite{lakkis2022multi}, totalVI\cite{gayoso2021joint}, Generalized Linear Model (GLM), MCIA\cite{meng2014multivariate}, MOFA\cite{argelaguet2020mofa+}, CGVAE\cite{liu2018constrained}, ccVAE\cite{svahn2022ccvae}, PoE\cite{wu2018multimodal}, MoE\cite{shi2019variational}, to evaluate their modality prediction performance on four benchmark datasets\cite{hu2024benchmarking,makrodimitris2024depth}. Among all the methods, totalVI shows highest cell-cell PCC on predicted modalities and PoE shows better performance than other VAE-based models (Fig.\ref{benchmark_missingmodality}).

\subsection{Data correlation}
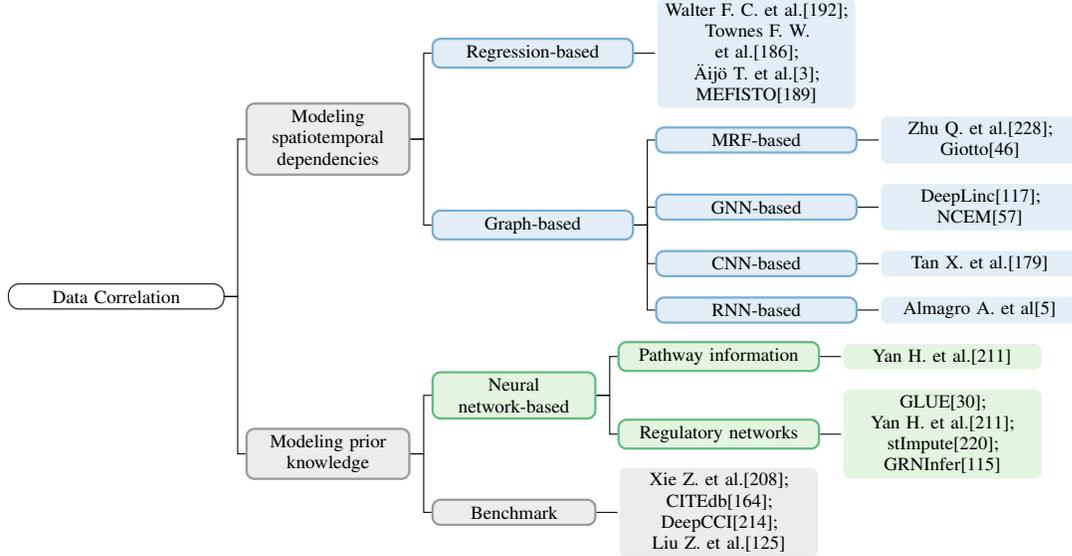
\begin{figure*}
    \scriptsize
    \hspace*{-30pt}
    \centering
    \begin{forest}
        for tree={
        forked edges,
        grow'=0,
        draw,
        rounded corners,
        node options={align=center,},
        text width=2.7cm,
        s sep=6pt,
        calign=edge midpoint,
        text=black,
        },
        [
        Data Correlation,
        [
        Modeling~\\ spatiotemporal~\\ dependencies, for tree={top_class}
        [
        Regression-based={fill=green!45,gpa},
        gpa_wide
        [
        Walter F. C. et al.\cite{walter2023fishfactor};\\
        Townes F. W. et al.\cite{townes2023nonnegative};\\
        Äijö T. et al.\cite{aijo2018temporal};\\
        MEFISTO\cite{velten2022identifying},
        gpa_work
        ]
        ]
        [
        Graph-based,
        for tree={fill=green!45,gpa},
        [
        MRF-based,
        gpa_wide
        [
        Zhu Q. et al.\cite{zhu2018identification};\\
        Giotto\cite{dries2021giotto},
        gpa_work
        ]
        ]
        [
        GNN-based,
        gpa_wide
        [
        DeepLinc\cite{li2022novo};\\
        NCEM\cite{fischer2023modeling},
        gpa_work
        ]
        ]
        [
        CNN-based,
        gpa_wide
        [
        Tan X. et al.\cite{tan2020spacell},
        gpa_work
        ]
        ]
        [
        RNN-based,
        gpa_wide
        [
        Almagro A. et al\cite{almagro2017deeploc},
        gpa_work
        ]
        ]
        ]
        ]
        [
        Modeling prior knowledge,
        for tree={ top_class}
        [
        Neural network-based,
        for tree={fill=green!45,pretraining},
        [
        Pathway information,
        pretraining_wide
        [
        Yan H. et al.\cite{yan2024prior},
        pretraining_work
        ]
        ]
        [
        Regulatory networks,
        pretraining_wide
        [
        GLUE\cite{cao2022multi};\\
        Yan H. et al.\cite{yan2024prior};\\
        stImpute\cite{zeng2024imputing};\\
        GRNInfer\cite{li2022integrating},
        pretraining_work
        ]
        ]
        ]
        [
        Benchmark={fill=gray!45,top_class}
        topclass_wide
        [
        Xie Z. et al.\cite{xie2023comparison};\\
        CITEdb\cite{shan2022citedb};\\
        DeepCCI\cite{yang2023deepcci};\\
        Liu Z. et al.\cite{liu2022evaluation},
        topclass_wide
        ]
        ]
        ]
        ]
    \end{forest}
    \caption{\textbf{The structure of section "Data correlation" and related methods.} The tree chart outlines the challenges related to data correlation, emphasizing challenges such as modeling spatiotemporal dependencies and prior knowledge.}
    \label{fig:correlation_structure}
\end{figure*}

Understanding the relationship between biological systems and external factors is crucial to gain deeper insights into cellular dynamics and the interactions between cells and their environment. Modeling data correlation involves capturing these complex interactions and dependencies, which are affected by both spatial and temporal variations, as well as biological prior knowledge. The overall structure of this section is shown in Fig. \ref{fig:correlation_structure}.

\subsubsection{Modeling spatiotemporal dependencies}

Modeling temporal and spatial dependencies is crucial for the analysis of single-cell and spatial transcriptomics data, as numerous biological processes exhibit dynamic spatiotemporal correlations. Examples include cell differentiation during development\cite{bar2012studying}, the spatial organization of cells within tissues\cite{larsson2021spatially}, disease progression pathways\cite{seferbekova2023spatial}, and variations in immune responses over time and space. Capturing these features can reveal dynamical shifts in cell states, cell-cell interactions, and complex tissue or disease structures. Spatiotemporal omics data encompass longitudinal molecular profiles from patients, molecular profiles across developmental stages, and continuous spatiotemporal omics maps. However, making comparisons across varying scales, biological samples, or conditions remains a challenging task. For example, establishing statistical correlations between samples requires the alignment of temporal and spatial coordinates across individuals or biological scales. Current approaches mainly rely on regression-based and graph-based models\cite{velten2023principles}.

Among regression-based models, Gaussian process-based probabilistic models are widely used\cite{walter2023fishfactor,townes2023nonnegative,aijo2018temporal}. These models are effective in capturing trends of continuous variability for both time series and spatially distributed data. MEFISTO\cite{velten2022identifying} leverages the Gaussian process to model latent factors by incorporating temporal and spatial information, as well as grouped data, into the covariates within the Gaussian kernel. The covariance function consists of two components: one that describes relationships across different groups (e.g., sample sets or experimental conditions), and another that captures smooth variations such as spatial locations or time points. This dual structure allows Gaussian processes (GP) to account for both inter-group heterogeneity and intra-group covariate variation. By ensuring that samples located closer together in the covariate space share more similar latent factors, the Gaussian process effectively models the continuity of relations between data points.  

Another approach for jointly modeling omics latent features $z$ involves the use of graph models. Markov random fields (MRF) are undirected graphical models that assume the distribution of each node depends only on the labels of its neighboring nodes. Compared to non-parametric regression models like GP, MRF offer greater computational efficiency, as they do not require inference of the complete covariance structure across all samples. Qian Zhu et al.\cite{zhu2018identification} proposed a Markovian property for spatial patterns, which constrains the correlations between neighboring nodes. By assuming that labels of neighboring cells, including gene expression states or cell types, exhibit a degree of similarity, the joint probability distribution over the spatial domain can be factorized into a product of local neighborhood probability distributions. The probability distribution for the label associated with the current node $s_i$ is jointly modeled using both its neighboring nodes' labels $C_{Ni}$ and its own gene expression $x_i$: 
\begin{equation}
    P\left(c_i \mid s_i, x_i, c_{N_i}\right)=\frac{1}{Z} P\left(x_i \mid c_i, s_i\right) P\left(c_i \mid s_i, c_{N_i}\right)
\end{equation}

Giotto\cite{dries2021giotto} utilizes MRF with conditional probability distributions, such as Gaussian or Poisson, to enhance spatial clustering. This approach effectively captures smooth and continuous expression changes, thereby facilitating the identification of spatially structured cell populations. 

\begin{figure*}[ht]
  \centering
  \includegraphics[width=\linewidth]{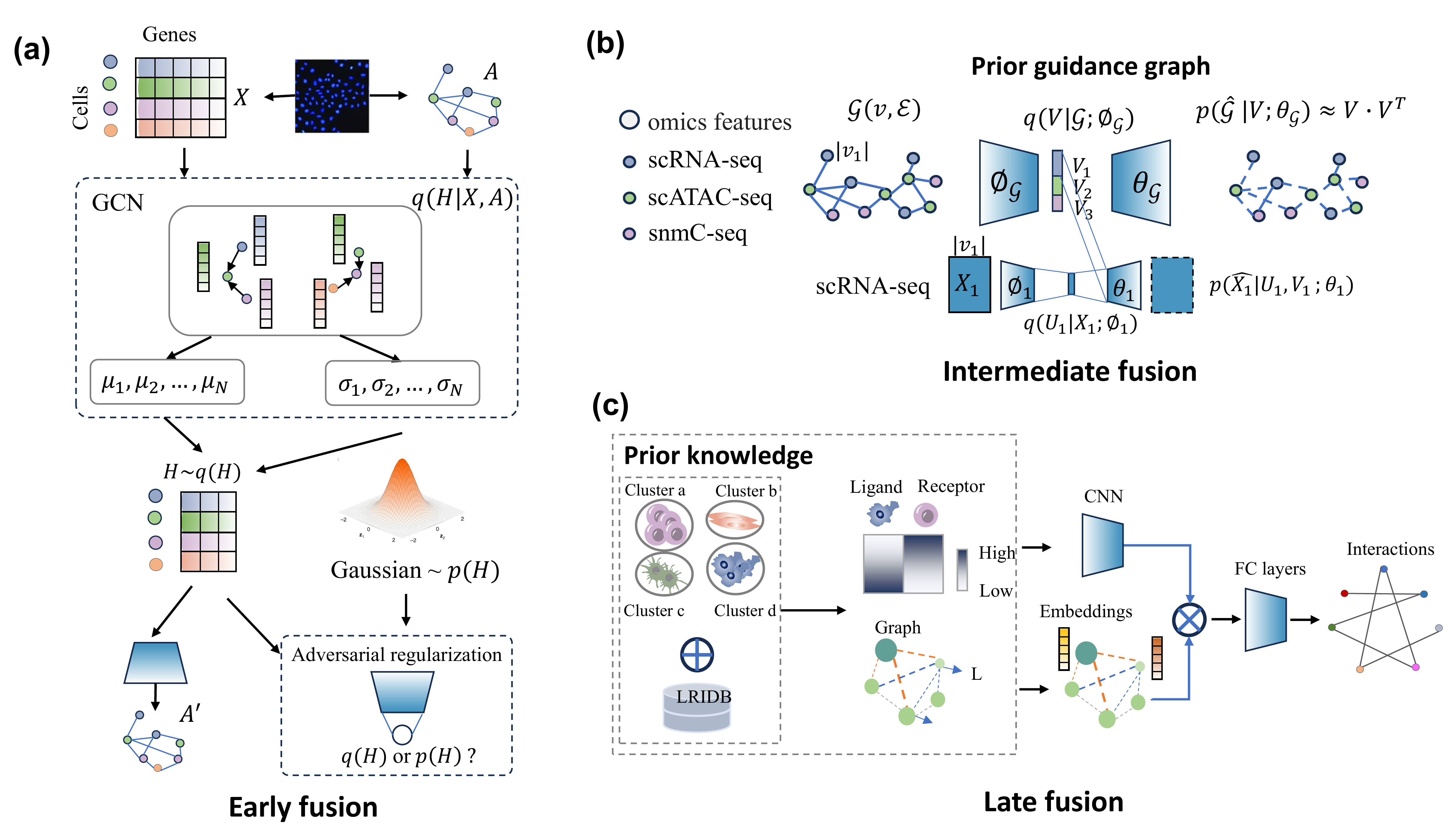}
  \caption{\textbf{The challenges and typical approaches for data correlation.} DL methods for integrating prior knowledge and modeling data correlation typically use three fusion strategies: early, intermediate, and late fusion. (a) For early fusion, DeepLinc\cite{li2022novo} constructs a cell adjacency graph based on the physical distance between cells and learns graph embedding features. (b) For intermediate fusion, GLUE\cite{cao2022multi} fuses nodes representation features (genes or accessible chromatin regions) from each modality. (c) For late fusion, DeepCCI\cite{yang2023deepcci} uses the LRIDB database to identify receptors and predicts interactions by combining ResNet and GCN outputs.}
  \label{fig5}
\end{figure*}

DL-based graph frameworks are increasingly employed to explore cell-cell interactions, including recurrent neural network (RNN), CNN, and GNN. 
In GNNs, cells are represented as nodes, with edges denoting potential interactions, such as those between ligand-receptor pairs. This approach effectively integrates spatial data and gene expression profiles to reveal interaction patterns. For instance, DeepLinc\cite{li2022novo} posits that neighboring cells are more likely to interact than cells further apart (Fig. \ref{fig5} (a)). It constructs a cell adjacency graph based on the physical distance between cells and learns embedding features that reflect the likelihood of interactions by aggregating information from each cell along with its neighbors. Using variational graph autoencoders (VGAEs) and adversarial networks, DeepLinc employs unsupervised learning techniques to uncover intrinsic associations between the cell adjacency graph and gene expression profiles, thereby reconstructing interaction networks. NCEM \cite{fischer2023modeling} utilizes GNN to reconstruct gene expression vectors from cell type labels and niche composition, which are represented through graph-level predictors and adjacency matrices derived from spatial proximity. While NCEM incorporates a linear model for mathematical interpretability, experimental results demonstrate that its nonlinear variant significantly outperforms the linear one. Different from GNNs, CNNs\cite{tan2020spacell} aggregate features from neighbouring regions in images through convolution operations. RNNs\cite{almagro2017deeploc}, on the other hand, propagate information from adjacent spatial points using recurrent connections. 

Overall, DL frameworks exhibit considerable flexibility in analyzing spatiotemporal omics data. 

\begin{figure}
  \centering
  \includegraphics[width=\linewidth]{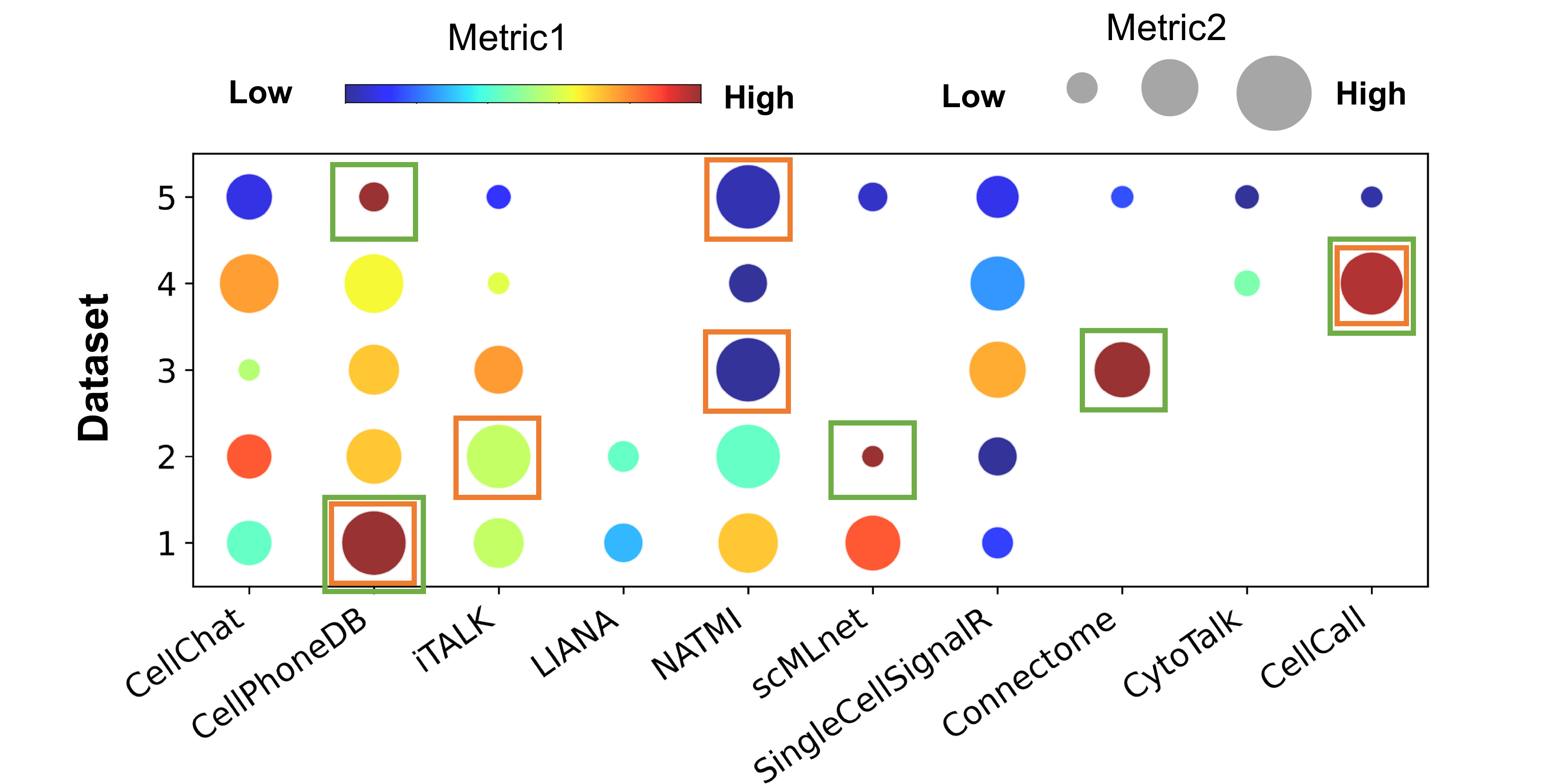}
  \caption{\textbf{Revisualize the benchmark results for cell-cell interactions from five benchmark datasets\cite{xie2023comparison, shan2022citedb,yang2023deepcci,liu2022evaluation}.}  "Metric 1 and Metric 2 are both accuracy metrics for cell-cell interactions. In Benchmark 1 (Datasets 1 and 2), 'Metric 1' refers to precision, and 'Metric 2' refers to F1 score. In Benchmark 2 (Dataset 3), 'Metric 1' refers to the sum of communication scores, and 'Metric 2' refers to the count of active LR pairs. In Benchmark 3 (Dataset 4), 'Metric 1' refers to AUC, and 'Metric 2' refers to precision. In Benchmark 4 (Dataset 5), 'Metric 1' refers to the distance enrichment score, and 'Metric 2' refers to F1 score. The orange rectangle represents the largest point size ("Metric2"), while the green rectangle indicates the points with the highest color value ("Metric1"), closest to red on the color bar.}
  \label{benchmark_cci}
\end{figure}

\subsubsection{Modeling prior knowledge}
Single-cell data is characterized by sparse features, multi-source heterogeneity, and lack of high-quality labels, making it unreliable to draw experimental conclusions solely from the observed data. However, the incorporation of prior biological knowledge such as gene regulatory networks, cell type characteristics, and developmental trajectories, can significantly enhance the accuracy and interpretability of the analysis. Nonetheless, effectively integrating prior knowledge while avoiding potential biases and overfitting remains a challenging task.

Prior knowledge mainly refers to pathway information and regulatory networks\cite{yan2024prior} obtained from databases. Pathway information concerns to molecular interactions and biochemical reactions that drive specific biological processes, such as signal transduction, metabolism, and cellular activity. This information aids in elucidating how cells respond to external stimuli or internal changes. In the context of single-cell and multi-omics analyses, pathway information is used to infer gene-gene interactions, characterize cell types, and determine cell states. It is typically sourced from well-established databases such as KEGG\cite{kanehisa2002kegg}, Reactome\cite{fabregat2018reactome}, and WikiPathways\cite{agrawal2024wikipathways}. Regulatory networks involve the interactions among genes, transcription factors, proteins, and other biomolecules that regulate gene expression and cellular function. These networks are commonly represented as graph where nodes denote biomolecules (e.g., genes or proteins) while edges indicate regulatory relationships (e.g., activation or inhibition). Databases such as STRING\cite{szklarczyk2021string} and GeneMANIA\cite{warde2010genemania} provide valuable insights into protein-protein interactions and gene-gene interactions, respectively. Regulatory networks facilitate the understanding of the mechanisms governing gene expression regulation, the identification of key regulatory factors, and the revelation of cell-specific patterns in gene expression.

GLUE\cite{cao2022multi} integrates multi-omics data through a guidance graph, where nodes represent features from various modalities, such as genes in scRNA data and accessible chromatin regions in ATAC-seq data(Fig. \ref{fig5} (b)). Graphs establish connections between ATAC peaks and RNA genes based on overlapping gene bodies or promoter regions. A variational posterior is employed to reconstruct the guidance map and its latent space is used as a prior for multi-omics data reconstruction. The decoder computes an inner product of feature and cell embeddings to ensure consistent embedding directions across different modalities. Hongxi Yan et al.\cite{yan2024prior} aggregate gene features within the same biological pathway to obtain pathway-level features for predictive modeling. They utilize the KEGG database together with an ensemble gradient method to identify key pathways, which significantly enhances model interpretability.

Some methods use prior knowledge from existing databases to initialise edge features in gene-gene interaction networks. For instance, DeepCCI\cite{yang2023deepcci} uses the constructed LRIDB database to define receptors and establishes interaction networks between cell clusters based on known ligand-receptor (L–R) pairs, predicting interactions by a combination of ResNet and graph convolutional network (GCN) outputs (Fig. \ref{fig5} (c)). stImpute\cite{zeng2024imputing} leverages the ESM-2 protein language model to embed proteins and constructs a network of gene relationships using cosine similarity. GRNInfer\cite{li2022integrating} incorporates gene regulatory relationships from RegNetwork\cite{liu2015regnetwork} as prior information to construct a gene graph network.

Furthermore, we have collected 10 methods, including CellChat\cite{jin2021inference}, CellPhoneDB\cite{efremova2020cellphonedb}, iTALK\cite{wang2019italk}, LIANA\cite{dimitrov2022comparison}, NATMI\cite{hou2020predicting}, scMLnet\cite{cheng2021inferring}, SingleCellSignalR\cite{cabello2020singlecellsignalr}, Connectome\cite{raredon2022computation}, CytoTalk\cite{hu2021cytotalk}, and CellCall\cite{zhang2021cellcall}. These methods leverage existing L-R pair knowledge to infer cell-cell communication, and we evaluate their cell-cell interaction prediction performance on five benchmark datasets\cite{xie2023comparison, shan2022citedb,yang2023deepcci,liu2022evaluation}. Among all the methods, CellPhoneDB ranks among the top across all benchmark datasets, demonstrating the robustness of extracting cellular context (Fig. \ref{benchmark_cci}).
\section{Conclusion and Future Perspectives}
\subsection{Innovative AI Method for Single-Cell and spatial transcriptomics data analysis} 
The rapid expansion in the size, depth, and complexity of single-cell and spatial transcriptomics data requires the development of algorithms capable of effectively capturing complex gene expression patterns and spatial distributions.

Recently, foundational models have emerged as a focus of single-cell omics. By leveraging self-supervised pre-training on large unlabeled scRNA-seq datasets, these models capture complex features and patterns, producing unified representations that can be fine-tuned for specific downstream tasks\cite{yang2022scbert,theodoris2023transfer,hao2024large,cui2024scgpt}. For instance, SCimilarity \cite{heimberg2023scalable} enables rapid querying of cell states for cell type annotation. scMulan\cite{bian2024scmulan} converts single-cell transcriptomic data along with rich metadata (e.g., cell type, spatial context, and temporal aspects) into "cell sentences" (c-sentences), achieving superior performance in tasks like zero-shot cell annotation and batch correction. scGPT\cite{cui2024scgpt}, which is based on the GPT architecture, employs self-supervised pretraining with condition tokens to model gene interactions within cells, incorporating cell type labels for tasks such as cell type prediction, and applies a "binning" technique to ensure semantic alignment across diverse datasets.

Future advancements in large language models, such as OpenAI’s O1\footnote{\href{https://openai.com/o1/}{https://openai.com/o1/}}, and agent-based methods, are expected to further enhance single-cell and spatial transcriptomics analysis. O1 leverages large-scale reinforcement learning algorithms to achieve chain-of-thought (COT) reasoning, thereby improving inference accuracy. Agent-based approaches, such as ReAct\cite{yao2023react}, integrate real-time observations to guide decision-making, facilitating more efficient and proactive error correction. These models offer considerable potential for providing interpretable inferences across a variety of downstream tasks in single-cell analysis.

However, in contrast to parametric modeling approaches, end-to-end networks often operate as "black boxes," providing limited interpretability of the inference processes. Moreover, these networks are typically trained on specific datasets and lack dynamic updates, which constrains their generalizability and robustness on unseen or uncertain data. For scientific scenarios, it is crucial to strike a balance between interpretability and accuracy. As indicated by previous studies, it may be feasible to enhance interpretability by establishing connections between prior models and observed outcomes through interpretable parametric rules.

\subsection{Benchmark Datasets and Evaluation Metrics}

Relevant benchmarks have been established for various stages of single-cell sequencing data analysis workflows, including imputation\cite{dai2022scimc}, cell identification\cite{abdelaal2019comparison}, clustering\cite{wang2022comparison}, gene regulatory networks\cite{badia2023gene}, cell-cell interactions\cite{wang2022systematic}, and multi-omics data integration\cite{athaya2023multimodal}. However, several studies have mentioned that the datasets and evaluation metrics employed in these benchmarks do not accurately reflect the strengths and weaknesses of contemporary algorithms\cite{pratapa2020benchmarking}. Moreover, as large-scale sequencing data continues to evolve, algorithms that previously demonstrated strong performance may no longer be applicable in different application contexts\cite{khan2023reusability}. These datasets may exhibit increased heterogeneity due to samples derived from diverse sequencing platforms or continuous-time and continuous-space settings.

Current evaluation metrics mainly emphasize performance metrics such as accuracy, AUC, and RMSE, with a limited focus on biological relevance. To ensure the biological validity of model predictions, systematic validation through in vitro experiments is essential. For example, trends in gene expression, molecular properties, or cellular behavior can be compared to experimental results to confirm biological significance. Therefore, it is important to establish benchmarks that provide a more comprehensive and objective assessment of the generalizability and applicability of algorithms. Data simulations that generate benchmark datasets based on well-defined rules can provide a diverse array of labeled data for evaluation. This approach has already been applied to tasks such as cell identity recognition and modelling of gene regulatory networks\cite{song2024scdesign3,zinati2024groundgan}, highlighting its potential as a robust tool for evaluating model performance.



\subsection{Application of DL in practical scenarios}

Here, we summarize the applications of single-cell and spatial transcriptomics in biology, medicine, and clinical practice, providing an overview of the background for DL applications in these fields.

In biology, single-cell and spatial transcriptomics focus on embryonic\cite{proks2024deep}, tissue\cite{dohmen2022identifying}, and organ development\cite{biancalani2021deep}. These techniques facilitate the identification and classification of various cell types and lineages, while providing insights into the evolution of cell populations throughout organogenesis.

In precision medicine, the analysis of single-cell transcriptomic data is critical for investigating disease heterogeneity\cite{halawani2023deep}, identifying distinct subclones within diseases\cite{qin2024statistical}, discovering critical disease biomarkers\cite{alamin2024single}, characterizing interactions between normal and diseased cells\cite{yang2024deciphering}, elucidating relevant signaling pathways\cite{ji2023single}, and predicting resistance\cite{wegmann2024single}. Single-cell transcriptomics facilitates the construction of comprehensive cellular maps that significantly enhance the discovery of novel biomarkers and therapeutic targets\cite{van2023applications}. In additon, scRNA-seq has demonstrated significant potential for improving patient outcomes and accelerating the development of personalized therapies\cite{tang2023sparx, ianevski2024single}.

In clinical applications, scRNA-seq plays a crucial role in characterizing patient-specific features\cite{sorin2023single}. It assists in the identification of biomarkers for patient stratification\cite{niu2023identification}, elucidates the underlying mechanisms of drug action and resistance\cite{qi2023trends}, supports the development of personalized treatment strategies, and enables monitoring of drug response and disease progression\cite{liu2024integrated}.

\section{}
\fbox{
\begin{minipage}{0.9\columnwidth}
\textbf{Key Points}  
\vspace{0.5em}  
    \begin{itemize}
        \item This review discusses four major challenges and related deep learning approaches in single-cell and spatial transcriptomics data analysis.

        \item This review curates 21 datasets from 9 benchmarks covering 58 computational methods and compares their performance on their respective modeling tasks.

        \item This review outlines three future research directions regarding data, methods, and applications for single-cell and spatial omics data analysis. 
    \end{itemize}
\end{minipage}
}


\section{Competing interests}
No competing interest is declared.

\section{Author contributions statement}

Shuang Ge and Zhixiang Ren collected and reviewed literature. Shuang Ge, Shuqing Sun, Qiang Cheng and Zhixiang Ren drafted the manuscript. All authors read and approved the final manuscript.

{
\bibliographystyle{plain}
\bibliography{references}

\begin{thebibliography}{100}

\bibitem{abdelaal2019comparison}
Tamim Abdelaal, Lieke Michielsen, Davy Cats, Dylan Hoogduin, Hailiang Mei, Marcel~JT Reinders, and Ahmed Mahfouz.
\newblock A comparison of automatic cell identification methods for single-cell rna sequencing data.
\newblock {\em Genome biology}, 20:1--19, 2019.

\bibitem{agrawal2024wikipathways}
Ayushi Agrawal, Hasan Balc{\i}, Kristina Hanspers, Susan~L Coort, Marvin Martens, Denise~N Slenter, Friederike Ehrhart, Daniela Digles, Andra Waagmeester, Isabel Wassink, et~al.
\newblock Wikipathways 2024: next generation pathway database.
\newblock {\em Nucleic acids research}, 52(D1):D679--D689, 2024.

\bibitem{aijo2018temporal}
Tarmo {\"A}ij{\"o}, Christian~L M{\"u}ller, and Richard Bonneau.
\newblock Temporal probabilistic modeling of bacterial compositions derived from 16s rrna sequencing.
\newblock {\em Bioinformatics}, 34(3):372--380, 2018.

\bibitem{alamin2024single}
Md~Alamin, Most Humaira~Sultana, Isaac~Adeyemi Babarinde, AKM Azad, Mohammad~Ali Moni, and Haiming Xu.
\newblock Single-cell rna-seq data analysis reveals functionally relevant biomarkers of early brain development and their regulatory footprints in human embryonic stem cells (hescs).
\newblock {\em Briefings in Bioinformatics}, 25(3):bbae230, 2024.

\bibitem{almagro2017deeploc}
Jos{\'e}~Juan Almagro~Armenteros, Casper~Kaae S{\o}nderby, S{\o}ren~Kaae S{\o}nderby, Henrik Nielsen, and Ole Winther.
\newblock Deeploc: prediction of protein subcellular localization using deep learning.
\newblock {\em Bioinformatics}, 33(21):3387--3395, 2017.

\bibitem{alon2021expansion}
Shahar Alon, Daniel~R Goodwin, Anubhav Sinha, Asmamaw~T Wassie, Fei Chen, Evan~R Daugharthy, Yosuke Bando, Atsushi Kajita, Andrew~G Xue, Karl Marrett, et~al.
\newblock Expansion sequencing: Spatially precise in situ transcriptomics in intact biological systems.
\newblock {\em Science}, 371(6528):eaax2656, 2021.

\bibitem{angermueller2016parallel}
Christof Angermueller, Stephen~J Clark, Heather~J Lee, Iain~C Macaulay, Mabel~J Teng, Tim~Xiaoming Hu, Felix Krueger, S{\'e}bastien~A Smallwood, Chris~P Ponting, Thierry Voet, et~al.
\newblock Parallel single-cell sequencing links transcriptional and epigenetic heterogeneity.
\newblock {\em Nature methods}, 13(3):229--232, 2016.

\bibitem{nature_methods_2013}
Anonymous.
\newblock Method of the year 2013.
\newblock {\em Nature Methods}, 11:1, 2014.
\newblock Published: 30 December 2013, Issue Date: January 2014.

\bibitem{argelaguet2020mofa+}
Ricard Argelaguet, Damien Arnol, Danila Bredikhin, Yonatan Deloro, Britta Velten, John~C Marioni, and Oliver Stegle.
\newblock Mofa+: a statistical framework for comprehensive integration of multi-modal single-cell data.
\newblock {\em Genome biology}, 21:1--17, 2020.

\bibitem{arisdakessian2019deepimpute}
C{\'e}dric Arisdakessian, Olivier Poirion, Breck Yunits, Xun Zhu, and Lana~X Garmire.
\newblock Deepimpute: an accurate, fast, and scalable deep neural network method to impute single-cell rna-seq data.
\newblock {\em Genome biology}, 20:1--14, 2019.

\bibitem{ashuach2023multivi}
Tal Ashuach, Mariano~I Gabitto, Rohan~V Koodli, Giuseppe-Antonio Saldi, Michael~I Jordan, and Nir Yosef.
\newblock Multivi: deep generative model for the integration of multimodal data.
\newblock {\em Nature Methods}, 20(8):1222--1231, 2023.

\bibitem{athaya2023multimodal}
Tasbiraha Athaya, Rony~Chowdhury Ripan, Xiaoman Li, and Haiyan Hu.
\newblock Multimodal deep learning approaches for single-cell multi-omics data integration.
\newblock {\em Briefings in Bioinformatics}, 24(5):bbad313, 2023.

\bibitem{badia2023gene}
Pau Badia-i Mompel, Lorna Wessels, Sophia M{\"u}ller-Dott, R{\'e}mi Trimbour, Ricardo~O Ramirez~Flores, Ricard Argelaguet, and Julio Saez-Rodriguez.
\newblock Gene regulatory network inference in the era of single-cell multi-omics.
\newblock {\em Nature Reviews Genetics}, 24(11):739--754, 2023.

\bibitem{bai2024sae}
Liang Bai, Boya Ji, and Shulin Wang.
\newblock Sae-impute: imputation for single-cell data via subspace regression and auto-encoders.
\newblock {\em BMC bioinformatics}, 25(1):317, 2024.

\bibitem{bao2022deep}
Siqi Bao, Ke~Li, Congcong Yan, Zicheng Zhang, Jia Qu, and Meng Zhou.
\newblock Deep learning-based advances and applications for single-cell rna-sequencing data analysis.
\newblock {\em Briefings in Bioinformatics}, 23(1):bbab473, 2022.

\bibitem{bar2012studying}
Ziv Bar-Joseph, Anthony Gitter, and Itamar Simon.
\newblock Studying and modelling dynamic biological processes using time-series gene expression data.
\newblock {\em Nature Reviews Genetics}, 13(8):552--564, 2012.

\bibitem{barrett2012ncbi}
Tanya Barrett, Stephen~E Wilhite, Pierre Ledoux, Carlos Evangelista, Irene~F Kim, Maxim Tomashevsky, Kimberly~A Marshall, Katherine~H Phillippy, Patti~M Sherman, Michelle Holko, et~al.
\newblock Ncbi geo: archive for functional genomics data sets—update.
\newblock {\em Nucleic acids research}, 41(D1):D991--D995, 2012.

\bibitem{becht2019dimensionality}
Etienne Becht, Leland McInnes, John Healy, Charles-Antoine Dutertre, Immanuel~WH Kwok, Lai~Guan Ng, Florent Ginhoux, and Evan~W Newell.
\newblock Dimensionality reduction for visualizing single-cell data using umap.
\newblock {\em Nature biotechnology}, 37(1):38--44, 2019.

\bibitem{bennett2020predicting}
WF~Drew Bennett, Stewart He, Camille~L Bilodeau, Derek Jones, Delin Sun, Hyojin Kim, Jonathan~E Allen, Felice~C Lightstone, and Helgi~I Ing{\'o}lfsson.
\newblock Predicting small molecule transfer free energies by combining molecular dynamics simulations and deep learning.
\newblock {\em Journal of Chemical Information and Modeling}, 60(11):5375--5381, 2020.

\bibitem{bian2024scmulan}
Haiyang Bian, Yixin Chen, Xiaomin Dong, Chen Li, Minsheng Hao, Sijie Chen, Jinyi Hu, Maosong Sun, Lei Wei, and Xuegong Zhang.
\newblock scmulan: a multitask generative pre-trained language model for single-cell analysis.
\newblock In {\em International Conference on Research in Computational Molecular Biology}, pages 479--482. Springer, 2024.

\bibitem{biancalani2021deep}
Tommaso Biancalani, Gabriele Scalia, Lorenzo Buffoni, Raghav Avasthi, Ziqing Lu, Aman Sanger, Neriman Tokcan, Charles~R Vanderburg, {\AA}sa Segerstolpe, Meng Zhang, et~al.
\newblock Deep learning and alignment of spatially resolved single-cell transcriptomes with tangram.
\newblock {\em Nature methods}, 18(11):1352--1362, 2021.

\bibitem{boileau2020exploring}
Philippe Boileau, Nima~S Hejazi, and Sandrine Dudoit.
\newblock Exploring high-dimensional biological data with sparse contrastive principal component analysis.
\newblock {\em Bioinformatics}, 36(11):3422--3430, 2020.

\bibitem{brendel2022application}
Matthew Brendel, Chang Su, Zilong Bai, Hao Zhang, Olivier Elemento, and Fei Wang.
\newblock Application of deep learning on single-cell rna sequencing data analysis: a review.
\newblock {\em Genomics, Proteomics and Bioinformatics}, 20(5):814--835, 2022.

\bibitem{cabello2020singlecellsignalr}
Simon Cabello-Aguilar, M{\'e}lissa Alame, Fabien Kon-Sun-Tack, Caroline Fau, Matthieu Lacroix, and Jacques Colinge.
\newblock Singlecellsignalr: inference of intercellular networks from single-cell transcriptomics.
\newblock {\em Nucleic acids research}, 48(10):e55--e55, 2020.

\bibitem{cable2022robust}
Dylan~M Cable, Evan Murray, Luli~S Zou, Aleksandrina Goeva, Evan~Z Macosko, Fei Chen, and Rafael~A Irizarry.
\newblock Robust decomposition of cell type mixtures in spatial transcriptomics.
\newblock {\em Nature biotechnology}, 40(4):517--526, 2022.

\bibitem{cao2024decoder}
Jiao Cao, Zhong Zheng, Di~Sun, Xin Chen, Rui Cheng, Tianpeng Lv, Yu~An, Junhua Zheng, Jia Song, Lingling Wu, et~al.
\newblock Decoder-seq enhances mrna capture efficiency in spatial rna sequencing.
\newblock {\em Nature Biotechnology}, pages 1--12, 2024.

\bibitem{cao2018joint}
Junyue Cao, Darren~A Cusanovich, Vijay Ramani, Delasa Aghamirzaie, Hannah~A Pliner, Andrew~J Hill, Riza~M Daza, Jose~L McFaline-Figueroa, Jonathan~S Packer, Lena Christiansen, et~al.
\newblock Joint profiling of chromatin accessibility and gene expression in thousands of single cells.
\newblock {\em Science}, 361(6409):1380--1385, 2018.

\bibitem{cao2022unified}
Kai Cao, Qiyu Gong, Yiguang Hong, and Lin Wan.
\newblock A unified computational framework for single-cell data integration with optimal transport.
\newblock {\em Nature Communications}, 13(1):7419, 2022.

\bibitem{cao2021benchmark}
Yue Cao, Pengyi Yang, and Jean Yee~Hwa Yang.
\newblock A benchmark study of simulation methods for single-cell rna sequencing data.
\newblock {\em Nature communications}, 12(1):6911, 2021.

\bibitem{cao2022multi}
Zhi-Jie Cao and Ge~Gao.
\newblock Multi-omics single-cell data integration and regulatory inference with graph-linked embedding.
\newblock {\em Nature Biotechnology}, 40(10):1458--1466, 2022.

\bibitem{catacutan2024machine}
Denise~B Catacutan, Jeremie Alexander, Autumn Arnold, and Jonathan~M Stokes.
\newblock Machine learning in preclinical drug discovery.
\newblock {\em Nature Chemical Biology}, 20(8):960--973, 2024.

\bibitem{chen2022spatiotemporal}
Ao~Chen, Sha Liao, Mengnan Cheng, Kailong Ma, Liang Wu, Yiwei Lai, Xiaojie Qiu, Jin Yang, Jiangshan Xu, Shijie Hao, et~al.
\newblock Spatiotemporal transcriptomic atlas of mouse organogenesis using dna nanoball-patterned arrays.
\newblock {\em Cell}, 185(10):1777--1792, 2022.

\bibitem{chen2017spatial}
Jun Chen, Shengbao Suo, Patrick~PL Tam, Jing-Dong~J Han, Guangdun Peng, and Naihe Jing.
\newblock Spatial transcriptomic analysis of cryosectioned tissue samples with geo-seq.
\newblock {\em Nature protocols}, 12(3):566--580, 2017.

\bibitem{chen2022heca}
Sijie Chen, Yanting Luo, Haoxiang Gao, Fanhong Li, Yixin Chen, Jiaqi Li, Renke You, Minsheng Hao, Haiyang Bian, Xi~Xi, et~al.
\newblock heca: the cell-centric assembly of a cell atlas.
\newblock {\em Iscience}, 25(5), 2022.

\bibitem{chen2019high}
Song Chen, Blue~B Lake, and Kun Zhang.
\newblock High-throughput sequencing of the transcriptome and chromatin accessibility in the same cell.
\newblock {\em Nature biotechnology}, 37(12):1452--1457, 2019.

\bibitem{cheng2021inferring}
Jinyu Cheng, Ji~Zhang, Zhongdao Wu, and Xiaoqiang Sun.
\newblock Inferring microenvironmental regulation of gene expression from single-cell rna sequencing data using scmlnet with an application to covid-19.
\newblock {\em Briefings in bioinformatics}, 22(2):988--1005, 2021.

\bibitem{cho2021microscopic}
Chun-Seok Cho, Jingyue Xi, Yichen Si, Sung-Rye Park, Jer-En Hsu, Myungjin Kim, Goo Jun, Hyun~Min Kang, and Jun~Hee Lee.
\newblock Microscopic examination of spatial transcriptome using seq-scope.
\newblock {\em Cell}, 184(13):3559--3572, 2021.

\bibitem{chowdhury2022single}
Ratul Chowdhury, Nazim Bouatta, Surojit Biswas, Christina Floristean, Anant Kharkar, Koushik Roy, Charlotte Rochereau, Gustaf Ahdritz, Joanna Zhang, George~M Church, et~al.
\newblock Single-sequence protein structure prediction using a language model and deep learning.
\newblock {\em Nature Biotechnology}, 40(11):1617--1623, 2022.

\bibitem{clark2018scnmt}
Stephen~J Clark, Ricard Argelaguet, Chantriolnt-Andreas Kapourani, Thomas~M Stubbs, Heather~J Lee, Celia Alda-Catalinas, Felix Krueger, Guido Sanguinetti, Gavin Kelsey, John~C Marioni, et~al.
\newblock scnmt-seq enables joint profiling of chromatin accessibility dna methylation and transcription in single cells.
\newblock {\em Nature communications}, 9(1):781, 2018.

\bibitem{cui2024scgpt}
Haotian Cui, Chloe Wang, Hassaan Maan, Kuan Pang, Fengning Luo, Nan Duan, and Bo~Wang.
\newblock scgpt: toward building a foundation model for single-cell multi-omics using generative ai.
\newblock {\em Nature Methods}, pages 1--11, 2024.

\bibitem{dai2022scimc}
Chichi Dai, Yi~Jiang, Chenglin Yin, Ran Su, Xiangxiang Zeng, Quan Zou, Kenta Nakai, and Leyi Wei.
\newblock scimc: a platform for benchmarking comparison and visualization analysis of scrna-seq data imputation methods.
\newblock {\em Nucleic Acids Research}, 50(9):4877--4899, 2022.

\bibitem{dey2015integrated}
Siddharth~S Dey, Lennart Kester, Bastiaan Spanjaard, Magda Bienko, and Alexander Van~Oudenaarden.
\newblock Integrated genome and transcriptome sequencing of the same cell.
\newblock {\em Nature biotechnology}, 33(3):285--289, 2015.

\bibitem{dimitrov2022comparison}
Daniel Dimitrov, D{\'e}nes T{\"u}rei, Martin Garrido-Rodriguez, Paul~L Burmedi, James~S Nagai, Charlotte Boys, Ricardo~O Ramirez~Flores, Hyojin Kim, Bence Szalai, Ivan~G Costa, et~al.
\newblock Comparison of methods and resources for cell-cell communication inference from single-cell rna-seq data.
\newblock {\em Nature communications}, 13(1):3224, 2022.

\bibitem{ding2018interpretable}
Jiarui Ding, Anne Condon, and Sohrab~P Shah.
\newblock Interpretable dimensionality reduction of single cell transcriptome data with deep generative models.
\newblock {\em Nature communications}, 9(1):2002, 2018.

\bibitem{dohmen2022identifying}
Jan Dohmen, Artem Baranovskii, Jonathan Ronen, Bora Uyar, Vedran Franke, and Altuna Akalin.
\newblock Identifying tumor cells at the single-cell level using machine learning.
\newblock {\em Genome Biology}, 23(1):123, 2022.

\bibitem{dries2021giotto}
Ruben Dries, Qian Zhu, Rui Dong, Chee-Huat~Linus Eng, Huipeng Li, Kan Liu, Yuntian Fu, Tianxiao Zhao, Arpan Sarkar, Feng Bao, et~al.
\newblock Giotto: a toolbox for integrative analysis and visualization of spatial expression data.
\newblock {\em Genome biology}, 22:1--31, 2021.

\bibitem{du2022robust}
Jin-Hong Du, Zhanrui Cai, and Kathryn Roeder.
\newblock Robust probabilistic modeling for single-cell multimodal mosaic integration and imputation via scvaeit.
\newblock {\em Proceedings of the National Academy of Sciences}, 119(49):e2214414119, 2022.

\bibitem{efremova2020cellphonedb}
Mirjana Efremova, Miquel Vento-Tormo, Sarah~A Teichmann, and Roser Vento-Tormo.
\newblock Cellphonedb: inferring cell--cell communication from combined expression of multi-subunit ligand--receptor complexes.
\newblock {\em Nature protocols}, 15(4):1484--1506, 2020.

\bibitem{eng2019transcriptome}
Chee-Huat~Linus Eng, Michael Lawson, Qian Zhu, Ruben Dries, Noushin Koulena, Yodai Takei, Jina Yun, Christopher Cronin, Christoph Karp, Guo-Cheng Yuan, et~al.
\newblock Transcriptome-scale super-resolved imaging in tissues by rna seqfish+.
\newblock {\em Nature}, 568(7751):235--239, 2019.

\bibitem{eraslan2019single}
G{\"o}kcen Eraslan, Lukas~M Simon, Maria Mircea, Nikola~S Mueller, and Fabian~J Theis.
\newblock Single-cell rna-seq denoising using a deep count autoencoder.
\newblock {\em Nature communications}, 10(1):390, 2019.

\bibitem{erfanian2023deep}
Nafiseh Erfanian, A~Ali Heydari, Adib~Miraki Feriz, Pablo Ia{\~n}ez, Afshin Derakhshani, Mohammad Ghasemigol, Mohsen Farahpour, Seyyed~Mohammad Razavi, Saeed Nasseri, Hossein Safarpour, et~al.
\newblock Deep learning applications in single-cell genomics and transcriptomics data analysis.
\newblock {\em Biomedicine \& Pharmacotherapy}, 165:115077, 2023.

\bibitem{fabregat2018reactome}
Antonio Fabregat, Steven Jupe, Lisa Matthews, Konstantinos Sidiropoulos, Marc Gillespie, Phani Garapati, Robin Haw, Bijay Jassal, Florian Korninger, Bruce May, et~al.
\newblock The reactome pathway knowledgebase.
\newblock {\em Nucleic acids research}, 46(D1):D649--D655, 2018.

\bibitem{fan2015combinatorial}
H~Christina Fan, Glenn~K Fu, and Stephen~PA Fodor.
\newblock Combinatorial labeling of single cells for gene expression cytometry.
\newblock {\em Science}, 347(6222):1258367, 2015.

\bibitem{fan2020single}
Jean Fan, Kamil Slowikowski, and Fan Zhang.
\newblock Single-cell transcriptomics in cancer: computational challenges and opportunities.
\newblock {\em Experimental \& Molecular Medicine}, 52(9):1452--1465, 2020.

\bibitem{fan2020spatialdb}
Zhen Fan, Runsheng Chen, and Xiaowei Chen.
\newblock Spatialdb: a database for spatially resolved transcriptomes.
\newblock {\em Nucleic acids research}, 48(D1):D233--D237, 2020.

\bibitem{fan2023spascer}
Zhiwei Fan, Yangyang Luo, Huifen Lu, Tiangang Wang, YuZhou Feng, Weiling Zhao, Pora Kim, and Xiaobo Zhou.
\newblock Spascer: spatial transcriptomics annotation at single-cell resolution.
\newblock {\em Nucleic Acids Research}, 51(D1):D1138--D1149, 2023.

\bibitem{fischer2023modeling}
David~S Fischer, Anna~C Schaar, and Fabian~J Theis.
\newblock Modeling intercellular communication in tissues using spatial graphs of cells.
\newblock {\em Nature Biotechnology}, 41(3):332--336, 2023.

\bibitem{flores2022deep}
Mario Flores, Zhentao Liu, Tinghe Zhang, Md~Musaddaqui Hasib, Yu-Chiao Chiu, Zhenqing Ye, Karla Paniagua, Sumin Jo, Jianqiu Zhang, Shou-Jiang Gao, et~al.
\newblock Deep learning tackles single-cell analysis—a survey of deep learning for scrna-seq analysis.
\newblock {\em Briefings in bioinformatics}, 23(1):bbab531, 2022.

\bibitem{franzen2019panglaodb}
Oscar Franz{\'e}n, Li-Ming Gan, and Johan~LM Bj{\"o}rkegren.
\newblock Panglaodb: a web server for exploration of mouse and human single-cell rna sequencing data.
\newblock {\em Database}, 2019:baz046, 2019.

\bibitem{gao2021iterative}
Chao Gao, Jialin Liu, April~R Kriebel, Sebastian Preissl, Chongyuan Luo, Rosa Castanon, Justin Sandoval, Angeline Rivkin, Joseph~R Nery, Margarita~M Behrens, et~al.
\newblock Iterative single-cell multi-omic integration using online learning.
\newblock {\em Nature biotechnology}, 39(8):1000--1007, 2021.

\bibitem{gawlikowski2023survey}
Jakob Gawlikowski, Cedrique Rovile~Njieutcheu Tassi, Mohsin Ali, Jongseok Lee, Matthias Humt, Jianxiang Feng, Anna Kruspe, Rudolph Triebel, Peter Jung, Ribana Roscher, et~al.
\newblock A survey of uncertainty in deep neural networks.
\newblock {\em Artificial Intelligence Review}, 56(Suppl 1):1513--1589, 2023.

\bibitem{gayoso2021joint}
Adam Gayoso, Zo{\"e} Steier, Romain Lopez, Jeffrey Regier, Kristopher~L Nazor, Aaron Streets, and Nir Yosef.
\newblock Joint probabilistic modeling of single-cell multi-omic data with totalvi.
\newblock {\em Nature methods}, 18(3):272--282, 2021.

\bibitem{gerlach2019combined}
Jan~P Gerlach, Jessie~AG van Buggenum, Sabine~EJ Tanis, Mark Hogeweg, Branco~MH Heuts, Mauro~J Muraro, Lisa Elze, Francesca Rivello, Agata Rakszewska, Alexander van Oudenaarden, et~al.
\newblock Combined quantification of intracellular (phospho-) proteins and transcriptomics from fixed single cells.
\newblock {\em Scientific reports}, 9(1):1469, 2019.

\bibitem{gierahn2017seq}
Todd~M Gierahn, Marc~H Wadsworth, Travis~K Hughes, Bryan~D Bryson, Andrew Butler, Rahul Satija, Sarah Fortune, J~Christopher Love, and Alex~K Shalek.
\newblock Seq-well: portable, low-cost rna sequencing of single cells at high throughput.
\newblock {\em Nature methods}, 14(4):395--398, 2017.

\bibitem{giladi2020dissecting}
Amir Giladi, Merav Cohen, Chiara Medaglia, Yael Baran, Baoguo Li, Mor Zada, Pierre Bost, Ronnie Blecher-Gonen, Tomer-Meir Salame, Johannes~U Mayer, et~al.
\newblock Dissecting cellular crosstalk by sequencing physically interacting cells.
\newblock {\em Nature biotechnology}, 38(5):629--637, 2020.

\bibitem{gong2021cobolt}
Boying Gong, Yun Zhou, and Elizabeth Purdom.
\newblock Cobolt: integrative analysis of multimodal single-cell sequencing data.
\newblock {\em Genome biology}, 22:1--21, 2021.

\bibitem{gong2018drimpute}
Wuming Gong, Il-Youp Kwak, Pruthvi Pota, Naoko Koyano-Nakagawa, and Daniel~J Garry.
\newblock Drimpute: imputing dropout events in single cell rna sequencing data.
\newblock {\em BMC bioinformatics}, 19:1--10, 2018.

\bibitem{goodfellow2020generative}
Ian Goodfellow, Jean Pouget-Abadie, Mehdi Mirza, Bing Xu, David Warde-Farley, Sherjil Ozair, Aaron Courville, and Yoshua Bengio.
\newblock Generative adversarial networks.
\newblock {\em Communications of the ACM}, 63(11):139--144, 2020.

\bibitem{gossett2010label}
Daniel~R Gossett, Westbrook~M Weaver, Albert~J Mach, Soojung~Claire Hur, Henry Tat~Kwong Tse, Wonhee Lee, Hamed Amini, and Dino Di~Carlo.
\newblock Label-free cell separation and sorting in microfluidic systems.
\newblock {\em Analytical and bioanalytical chemistry}, 397:3249--3267, 2010.

\bibitem{gunady2019scgain}
Mohamed~K Gunady, Jayaram Kancherla, H{\'e}ctor~Corrada Bravo, and Soheil Feizi.
\newblock scgain: single cell rna-seq data imputation using generative adversarial networks.
\newblock {\em BioRxiv}, page 837302, 2019.

\bibitem{hagemann2020single}
Michael Hagemann-Jensen, Christoph Ziegenhain, Ping Chen, Daniel Ramsk{\"o}ld, Gert-Jan Hendriks, Anton~JM Larsson, Omid~R Faridani, and Rickard Sandberg.
\newblock Single-cell rna counting at allele and isoform resolution using smart-seq3.
\newblock {\em Nature biotechnology}, 38(6):708--714, 2020.

\bibitem{haghverdi2018batch}
Laleh Haghverdi, Aaron~TL Lun, Michael~D Morgan, and John~C Marioni.
\newblock Batch effects in single-cell rna-sequencing data are corrected by matching mutual nearest neighbors.
\newblock {\em Nature biotechnology}, 36(5):421--427, 2018.

\bibitem{hahaut2022fast}
Vincent Hahaut, Dinko Pavlinic, Walter Carbone, Sven Schuierer, Pierre Balmer, Mathieu Quinodoz, Magdalena Renner, Guglielmo Roma, Cameron~S Cowan, and Simone Picelli.
\newblock Fast and highly sensitive full-length single-cell rna sequencing using flash-seq.
\newblock {\em Nature biotechnology}, 40(10):1447--1451, 2022.

\bibitem{halawani2023deep}
Raid Halawani, Michael Buchert, and Yi-Ping~Phoebe Chen.
\newblock Deep learning exploration of single-cell and spatially resolved cancer transcriptomics to unravel tumour heterogeneity.
\newblock {\em Computers in Biology and Medicine}, 164:107274, 2023.

\bibitem{han2022self}
Wenkai Han, Yuqi Cheng, Jiayang Chen, Huawen Zhong, Zhihang Hu, Siyuan Chen, Licheng Zong, Liang Hong, Ting-Fung Chan, Irwin King, et~al.
\newblock Self-supervised contrastive learning for integrative single cell rna-seq data analysis.
\newblock {\em Briefings in Bioinformatics}, 23(5):bbac377, 2022.

\bibitem{han2018mapping}
Xiaoping Han, Renying Wang, Yincong Zhou, Lijiang Fei, Huiyu Sun, Shujing Lai, Assieh Saadatpour, Ziming Zhou, Haide Chen, Fang Ye, et~al.
\newblock Mapping the mouse cell atlas by microwell-seq.
\newblock {\em Cell}, 172(5):1091--1107, 2018.

\bibitem{hao2024large}
Minsheng Hao, Jing Gong, Xin Zeng, Chiming Liu, Yucheng Guo, Xingyi Cheng, Taifeng Wang, Jianzhu Ma, Xuegong Zhang, and Le~Song.
\newblock Large-scale foundation model on single-cell transcriptomics.
\newblock {\em Nature Methods}, pages 1--11, 2024.

\bibitem{he2020integrating}
Bryan He, Ludvig Bergenstr{\aa}hle, Linnea Stenbeck, Abubakar Abid, Alma Andersson, {\AA}ke Borg, Jonas Maaskola, Joakim Lundeberg, and James Zou.
\newblock Integrating spatial gene expression and breast tumour morphology via deep learning.
\newblock {\em Nature biomedical engineering}, 4(8):827--834, 2020.

\bibitem{heimberg2023scalable}
Graham Heimberg, Tony Kuo, Daryle DePianto, Tobias Heigl, Nathaniel Diamant, Omar Salem, Gabriele Scalia, Tommaso Biancalani, Shannon Turley, Jason Rock, et~al.
\newblock Scalable querying of human cell atlases via a foundational model reveals commonalities across fibrosis-associated macrophages.
\newblock {\em BioRxiv}, pages 2023--07, 2023.

\bibitem{herzenberg1976fluorescence}
Leonard~A Herzenberg, Richard~G Sweet, and Leonore~A Herzenberg.
\newblock Fluorescence-activated cell sorting.
\newblock {\em Scientific American}, 234(3):108--118, 1976.

\bibitem{hie2019efficient}
Brian Hie, Bryan Bryson, and Bonnie Berger.
\newblock Efficient integration of heterogeneous single-cell transcriptomes using scanorama.
\newblock {\em Nature biotechnology}, 37(6):685--691, 2019.

\bibitem{hinton2006reducing}
Geoffrey~E Hinton and Ruslan~R Salakhutdinov.
\newblock Reducing the dimensionality of data with neural networks.
\newblock {\em science}, 313(5786):504--507, 2006.

\bibitem{hou2020predicting}
Rui Hou, Elena Denisenko, Huan~Ting Ong, Jordan~A Ramilowski, and Alistair~RR Forrest.
\newblock Predicting cell-to-cell communication networks using natmi.
\newblock {\em Nature communications}, 11(1):5011, 2020.

\bibitem{hou2016single}
Yu~Hou, Huahu Guo, Chen Cao, Xianlong Li, Boqiang Hu, Ping Zhu, Xinglong Wu, Lu~Wen, Fuchou Tang, Yanyi Huang, et~al.
\newblock Single-cell triple omics sequencing reveals genetic, epigenetic, and transcriptomic heterogeneity in hepatocellular carcinomas.
\newblock {\em Cell research}, 26(3):304--319, 2016.

\bibitem{hu2022versatile}
Jialu Hu, Yuanke Zhong, and Xuequn Shang.
\newblock A versatile and scalable single-cell data integration algorithm based on domain-adversarial and variational approximation.
\newblock {\em Briefings in Bioinformatics}, 23(1):bbab400, 2022.

\bibitem{hu2024benchmarking}
Yunfei Hu, Manfei Xie, Yikang Li, Mingxing Rao, Wenjun Shen, Can Luo, Haoran Qin, Jihoon Baek, and Xin~Maizie Zhou.
\newblock Benchmarking clustering, alignment, and integration methods for spatial transcriptomics.
\newblock {\em Genome Biology}, 25(1):212, 2024.

\bibitem{hu2021cytotalk}
Yuxuan Hu, Tao Peng, Lin Gao, and Kai Tan.
\newblock Cytotalk: De novo construction of signal transduction networks using single-cell transcriptomic data.
\newblock {\em Science Advances}, 7(16):eabf1356, 2021.

\bibitem{huang2018saver}
Mo~Huang, Jingshu Wang, Eduardo Torre, Hannah Dueck, Sydney Shaffer, Roberto Bonasio, John~I Murray, Arjun Raj, Mingyao Li, and Nancy~R Zhang.
\newblock Saver: gene expression recovery for single-cell rna sequencing.
\newblock {\em Nature methods}, 15(7):539--542, 2018.

\bibitem{hwang2024big}
Hyeonseo Hwang, Hyeonseong Jeon, Nagyeong Yeo, and Daehyun Baek.
\newblock Big data and deep learning for rna biology.
\newblock {\em Experimental \& Molecular Medicine}, pages 1--29, 2024.

\bibitem{ianevski2024single}
Aleksandr Ianevski, Kristen Nader, Kyriaki Driva, Wojciech Senkowski, Daria Bulanova, Lidia Moyano-Galceran, Tanja Ruokoranta, Heikki Kuusanm{\"a}ki, Nemo Ikonen, Philipp Sergeev, et~al.
\newblock Single-cell transcriptomes identify patient-tailored therapies for selective co-inhibition of cancer clones.
\newblock {\em Nature Communications}, 15(1):8579, 2024.

\bibitem{ji2023single}
Xuanrui Ji, Quanwei Pei, Junpei Zhang, Pengqi Lin, Bin Li, Hongpeng Yin, Jingmei Sun, Dezhan Su, Xiufen Qu, and Dechun Yin.
\newblock Single-cell sequencing combined with machine learning reveals the mechanism of interaction between epilepsy and stress cardiomyopathy.
\newblock {\em Frontiers in immunology}, 14:1078731, 2023.

\bibitem{jin2020sctssr}
Ke~Jin, Le~Ou-Yang, Xing-Ming Zhao, Hong Yan, and Xiao-Fei Zhang.
\newblock sctssr: gene expression recovery for single-cell rna sequencing using two-side sparse self-representation.
\newblock {\em Bioinformatics}, 36(10):3131--3138, 2020.

\bibitem{jin2021inference}
Suoqin Jin, Christian~F Guerrero-Juarez, Lihua Zhang, Ivan Chang, Raul Ramos, Chen-Hsiang Kuan, Peggy Myung, Maksim~V Plikus, and Qing Nie.
\newblock Inference and analysis of cell-cell communication using cellchat.
\newblock {\em Nature communications}, 12(1):1088, 2021.

\bibitem{jin2020scai}
Suoqin Jin, Lihua Zhang, and Qing Nie.
\newblock scai: an unsupervised approach for the integrative analysis of parallel single-cell transcriptomic and epigenomic profiles.
\newblock {\em Genome biology}, 21:1--19, 2020.

\bibitem{junker2014genome}
Jan~Philipp Junker, Emily~S Noel, Victor Guryev, Kevin~A Peterson, Gopi Shah, Jan Huisken, Andrew~P McMahon, Eugene Berezikov, Jeroen Bakkers, and Alexander van Oudenaarden.
\newblock Genome-wide rna tomography in the zebrafish embryo.
\newblock {\em Cell}, 159(3):662--675, 2014.

\bibitem{kanehisa2002kegg}
Minoru Kanehisa.
\newblock The kegg database.
\newblock In {\em ‘In silico’simulation of biological processes: Novartis Foundation Symposium 247}, volume 247, pages 91--103. Wiley Online Library, 2002.

\bibitem{keller2019unravelling}
Laura Keller and Klaus Pantel.
\newblock Unravelling tumour heterogeneity by single-cell profiling of circulating tumour cells.
\newblock {\em Nature Reviews Cancer}, 19(10):553--567, 2019.

\bibitem{khan2023scaegan}
Sumeer~Ahmad Khan, Robert Lehmann, Xabier Martinez-de Morentin, Alberto Maillo, Vincenzo Lagani, Narsis~A Kiani, David Gomez-Cabrero, and Jesper Tegner.
\newblock scaegan: Unification of single-cell genomics data by adversarial learning of latent space correspondences.
\newblock {\em Plos one}, 18(2):e0281315, 2023.

\bibitem{khan2023reusability}
Sumeer~Ahmad Khan, Alberto Maillo, Vincenzo Lagani, Robert Lehmann, Narsis~A Kiani, David Gomez-Cabrero, and Jesper Tegner.
\newblock Reusability report: Learning the transcriptional grammar in single-cell rna-sequencing data using transformers.
\newblock {\em Nature Machine Intelligence}, 5(12):1437--1446, 2023.

\bibitem{kim2023similarity}
Gwangwoo Kim and Hyonho Chun.
\newblock Similarity-assisted variational autoencoder for nonlinear dimension reduction with application to single-cell rna sequencing data.
\newblock {\em BMC bioinformatics}, 24(1):432, 2023.

\bibitem{kim2024seq}
Yongsung Kim, Weiqiu Cheng, Chun-Seok Cho, Yongha Hwang, Yichen Si, Anna Park, Mitchell Schrank, Jer-En Hsu, Angelo Anacleto, Jingyue Xi, et~al.
\newblock Seq-scope: repurposing illumina sequencing flow cells for high-resolution spatial transcriptomics.
\newblock {\em Nature Protocols}, pages 1--47, 2024.

\bibitem{kingma2013auto}
Diederik~P Kingma.
\newblock Auto-encoding variational bayes.
\newblock {\em arXiv preprint arXiv:1312.6114}, 2013.

\bibitem{klein2015droplet}
Allon~M Klein, Linas Mazutis, Ilke Akartuna, Naren Tallapragada, Adrian Veres, Victor Li, Leonid Peshkin, David~A Weitz, and Marc~W Kirschner.
\newblock Droplet barcoding for single-cell transcriptomics applied to embryonic stem cells.
\newblock {\em Cell}, 161(5):1187--1201, 2015.

\bibitem{kobak2019art}
Dmitry Kobak and Philipp Berens.
\newblock The art of using t-sne for single-cell transcriptomics.
\newblock {\em Nature communications}, 10(1):5416, 2019.

\bibitem{korsunsky2019fast}
Ilya Korsunsky, Nghia Millard, Jean Fan, Kamil Slowikowski, Fan Zhang, Kevin Wei, Yuriy Baglaenko, Michael Brenner, Po-ru Loh, and Soumya Raychaudhuri.
\newblock Fast, sensitive and accurate integration of single-cell data with harmony.
\newblock {\em Nature methods}, 16(12):1289--1296, 2019.

\bibitem{lahnemann2020eleven}
David L{\"a}hnemann, Johannes K{\"o}ster, Ewa Szczurek, Davis~J McCarthy, Stephanie~C Hicks, Mark~D Robinson, Catalina~A Vallejos, Kieran~R Campbell, Niko Beerenwinkel, Ahmed Mahfouz, et~al.
\newblock Eleven grand challenges in single-cell data science.
\newblock {\em Genome biology}, 21:1--35, 2020.

\bibitem{lakkis2022multi}
Justin Lakkis, Amelia Schroeder, Kenong Su, Michelle~YY Lee, Alexander~C Bashore, Muredach~P Reilly, and Mingyao Li.
\newblock A multi-use deep learning method for cite-seq and single-cell rna-seq data integration with cell surface protein prediction and imputation.
\newblock {\em Nature machine intelligence}, 4(11):940--952, 2022.

\bibitem{lan2023efficient}
Meng Lan, Shixiong Zhang, and Lin Gao.
\newblock Efficient generation of paired single-cell multiomics profiles by deep learning.
\newblock {\em Advanced Science}, 10(21):2301169, 2023.

\bibitem{larsson2021spatially}
Ludvig Larsson, Jonas Fris{\'e}n, and Joakim Lundeberg.
\newblock Spatially resolved transcriptomics adds a new dimension to genomics.
\newblock {\em Nature methods}, 18(1):15--18, 2021.

\bibitem{lecun1989backpropagation}
Yann LeCun, Bernhard Boser, John~S Denker, Donnie Henderson, Richard~E Howard, Wayne Hubbard, and Lawrence~D Jackel.
\newblock Backpropagation applied to handwritten zip code recognition.
\newblock {\em Neural computation}, 1(4):541--551, 1989.

\bibitem{lee2022fastrna}
Hanbin Lee and Buhm Han.
\newblock Fastrna: An efficient solution for pca of single-cell rna-sequencing data based on a batch-accounting count model.
\newblock {\em The American Journal of Human Genetics}, 109(11):1974--1985, 2022.

\bibitem{li2024tissue}
Bohan Li, Feng Bao, Yimin Hou, Fengji Li, Hongjue Li, Yue Deng, and Qionghai Dai.
\newblock Tissue characterization at an enhanced resolution across spatial omics platforms with deep generative model.
\newblock {\em Nature Communications}, 15(1):6541, 2024.

\bibitem{li2022deep}
Gaoyang Li, Shaliu Fu, Shuguang Wang, Chenyu Zhu, Bin Duan, Chen Tang, Xiaohan Chen, Guohui Chuai, Ping Wang, and Qi~Liu.
\newblock A deep generative model for multi-view profiling of single-cell rna-seq and atac-seq data.
\newblock {\em Genome biology}, 23(1):20, 2022.

\bibitem{li2023single}
Hui Li, Kuangwen Hsieh, Pak~Kin Wong, Kathleen~E Mach, Joseph~C Liao, and Tza-Huei Wang.
\newblock Single-cell pathogen diagnostics for combating antibiotic resistance.
\newblock {\em Nature Reviews Methods Primers}, 3(1):6, 2023.

\bibitem{li2022integrating}
Jiawei Li, Fan Yang, Fang Wang, Yu~Rong, Peilin Zhao, Shizhan Chen, Jianhua Yao, Jijun Tang, and Fei Guo.
\newblock Integrating prior knowledge with graph encoder for gene regulatory inference from single-cell rna-seq data.
\newblock In {\em 2022 IEEE International Conference on Bioinformatics and Biomedicine (BIBM)}, pages 102--107. IEEE, 2022.

\bibitem{li2022disco}
Mengwei Li, Xiaomeng Zhang, Kok~Siong Ang, Jingjing Ling, Raman Sethi, Nicole Yee~Shin Lee, Florent Ginhoux, and Jinmiao Chen.
\newblock Disco: a database of deeply integrated human single-cell omics data.
\newblock {\em Nucleic acids research}, 50(D1):D596--D602, 2022.

\bibitem{li2022novo}
Runze Li and Xuerui Yang.
\newblock De novo reconstruction of cell interaction landscapes from single-cell spatial transcriptome data with deeplinc.
\newblock {\em Genome Biology}, 23(1):124, 2022.

\bibitem{li2018accurate}
Wei~Vivian Li and Jingyi~Jessica Li.
\newblock An accurate and robust imputation method scimpute for single-cell rna-seq data.
\newblock {\em Nature communications}, 9(1):997, 2018.

\bibitem{li2015embl}
Weizhong Li, Andrew Cowley, Mahmut Uludag, Tamer Gur, Hamish McWilliam, Silvano Squizzato, Young~Mi Park, Nicola Buso, and Rodrigo Lopez.
\newblock The embl-ebi bioinformatics web and programmatic tools framework.
\newblock {\em Nucleic acids research}, 43(W1):W580--W584, 2015.

\bibitem{li2022soar}
Y~Li, S~Dennis, MR~Hutch, Y~Ding, Y~Zhou, Y~Li, M~Pillai, S~Ghotbaldini, MA~Garcia, MS~Broad, et~al.
\newblock Soar elucidates disease mechanisms and empowers drug discovery through spatial transcriptomics.
\newblock {\em bioRxiv}, 2022.

\bibitem{linderman2019fast}
George~C Linderman, Manas Rachh, Jeremy~G Hoskins, Stefan Steinerberger, and Yuval Kluger.
\newblock Fast interpolation-based t-sne for improved visualization of single-cell rna-seq data.
\newblock {\em Nature methods}, 16(3):243--245, 2019.

\bibitem{linderman2022zero}
George~C Linderman, Jun Zhao, Manolis Roulis, Piotr Bielecki, Richard~A Flavell, Boaz Nadler, and Yuval Kluger.
\newblock Zero-preserving imputation of single-cell rna-seq data.
\newblock {\em Nature communications}, 13(1):192, 2022.

\bibitem{liu2018constrained}
Qi~Liu, Miltiadis Allamanis, Marc Brockschmidt, and Alexander Gaunt.
\newblock Constrained graph variational autoencoders for molecule design.
\newblock {\em Advances in neural information processing systems}, 31, 2018.

\bibitem{liu2024integrated}
Yang Liu, Hanlin Li, Tianyu Zeng, Yang Wang, Hongqi Zhang, Ying Wan, Zheng Shi, Renzhi Cao, and Hua Tang.
\newblock Integrated bulk and single-cell transcriptomes reveal pyroptotic signature in prognosis and therapeutic options of hepatocellular carcinoma by combining deep learning.
\newblock {\em Briefings in Bioinformatics}, 25(1):bbad487, 2024.

\bibitem{liu2022evaluation}
Zhaoyang Liu, Dongqing Sun, and Chenfei Wang.
\newblock Evaluation of cell-cell interaction methods by integrating single-cell rna sequencing data with spatial information.
\newblock {\em Genome Biology}, 23(1):218, 2022.

\bibitem{liu2015regnetwork}
Zhi-Ping Liu, Canglin Wu, Hongyu Miao, and Hulin Wu.
\newblock Regnetwork: an integrated database of transcriptional and post-transcriptional regulatory networks in human and mouse.
\newblock {\em Database}, 2015:bav095, 2015.

\bibitem{long2024deciphering}
Yahui Long, Kok~Siong Ang, Raman Sethi, Sha Liao, Yang Heng, Lynn van Olst, Shuchen Ye, Chengwei Zhong, Hang Xu, Di~Zhang, et~al.
\newblock Deciphering spatial domains from spatial multi-omics with spatialglue.
\newblock {\em Nature Methods}, pages 1--10, 2024.

\bibitem{lopez2018deep}
Romain Lopez, Jeffrey Regier, Michael~B Cole, Michael~I Jordan, and Nir Yosef.
\newblock Deep generative modeling for single-cell transcriptomics.
\newblock {\em Nature methods}, 15(12):1053--1058, 2018.

\bibitem{lotfollahi2022mapping}
Mohammad Lotfollahi, Mohsen Naghipourfar, Malte~D Luecken, Matin Khajavi, Maren B{\"u}ttner, Marco Wagenstetter, {\v{Z}}iga Avsec, Adam Gayoso, Nir Yosef, Marta Interlandi, et~al.
\newblock Mapping single-cell data to reference atlases by transfer learning.
\newblock {\em Nature biotechnology}, 40(1):121--130, 2022.

\bibitem{lovatt2014transcriptome}
Ditte Lovatt, Brittani~K Ruble, Jaehee Lee, Hannah Dueck, Tae~Kyung Kim, Stephen Fisher, Chantal Francis, Jennifer~M Spaethling, John~A Wolf, M~Sean Grady, et~al.
\newblock Transcriptome in vivo analysis (tiva) of spatially defined single cells in live tissue.
\newblock {\em Nature methods}, 11(2):190--196, 2014.

\bibitem{luecken2022benchmarking}
Malte~D Luecken, Maren B{\"u}ttner, Kridsadakorn Chaichoompu, Anna Danese, Marta Interlandi, Michaela~F M{\"u}ller, Daniel~C Strobl, Luke Zappia, Martin Dugas, Maria Colom{\'e}-Tatch{\'e}, et~al.
\newblock Benchmarking atlas-level data integration in single-cell genomics.
\newblock {\em Nature methods}, 19(1):41--50, 2022.

\bibitem{lv2024multi}
Tongxuan Lv, Yong Zhang, Junlin Liu, Qiang Kang, and Lin Liu.
\newblock Multi-omics integration for both single-cell and spatially resolved data based on dual-path graph attention auto-encoder.
\newblock {\em Briefings in Bioinformatics}, 25(5):bbae450, 2024.

\bibitem{ma2022deep}
Qin Ma and Dong Xu.
\newblock Deep learning shapes single-cell data analysis.
\newblock {\em Nature Reviews Molecular Cell Biology}, 23(5):303--304, 2022.

\bibitem{ma2020chromatin}
Sai Ma, Bing Zhang, Lindsay~M LaFave, Andrew~S Earl, Zachary Chiang, Yan Hu, Jiarui Ding, Alison Brack, Vinay~K Kartha, Tristan Tay, et~al.
\newblock Chromatin potential identified by shared single-cell profiling of rna and chromatin.
\newblock {\em Cell}, 183(4):1103--1116, 2020.

\bibitem{mackiewicz1993principal}
Andrzej Ma{\'c}kiewicz and Waldemar Ratajczak.
\newblock Principal components analysis (pca).
\newblock {\em Computers \& Geosciences}, 19(3):303--342, 1993.

\bibitem{macosko2015highly}
Evan~Z Macosko, Anindita Basu, Rahul Satija, James Nemesh, Karthik Shekhar, Melissa Goldman, Itay Tirosh, Allison~R Bialas, Nolan Kamitaki, Emily~M Martersteck, et~al.
\newblock Highly parallel genome-wide expression profiling of individual cells using nanoliter droplets.
\newblock {\em Cell}, 161(5):1202--1214, 2015.

\bibitem{makrodimitris2024depth}
Stavros Makrodimitris, Bram Pronk, Tamim Abdelaal, and Marcel Reinders.
\newblock An in-depth comparison of linear and non-linear joint embedding methods for bulk and single-cell multi-omics.
\newblock {\em Briefings in Bioinformatics}, 25(1):bbad416, 2024.

\bibitem{marconato2024spatialdata}
Luca Marconato, Giovanni Palla, Kevin~A Yamauchi, Isaac Virshup, Elyas Heidari, Tim Treis, Wouter-Michiel Vierdag, Marcella Toth, Sonja Stockhaus, Rahul~B Shrestha, et~al.
\newblock Spatialdata: an open and universal data framework for spatial omics.
\newblock {\em Nature Methods}, pages 1--5, 2024.

\bibitem{marx2021method}
Vivien Marx.
\newblock Method of the year: spatially resolved transcriptomics.
\newblock {\em Nature Methods}, 18:9--14, 2021.
\newblock Published: 06 January 2021, Issue Date: January 2021.

\bibitem{medaglia2017spatial}
Chiara Medaglia, Amir Giladi, Liat Stoler-Barak, Marco De~Giovanni, Tomer~Meir Salame, Adi Biram, Eyal David, Hanjie Li, Matteo Iannacone, Ziv Shulman, et~al.
\newblock Spatial reconstruction of immune niches by combining photoactivatable reporters and scrna-seq.
\newblock {\em Science}, 358(6370):1622--1626, 2017.

\bibitem{megill2021cellxgene}
Colin Megill, Bruce Martin, Charlotte Weaver, Sidney Bell, Lia Prins, Seve Badajoz, Brian McCandless, Angela~Oliveira Pisco, Marcus Kinsella, Fiona Griffin, et~al.
\newblock Cellxgene: a performant, scalable exploration platform for high dimensional sparse matrices.
\newblock {\em bioRxiv}, pages 2021--04, 2021.

\bibitem{meng2014multivariate}
Chen Meng, Bernhard Kuster, Aed{\'\i}n~C Culhane, and Amin~Moghaddas Gholami.
\newblock A multivariate approach to the integration of multi-omics datasets.
\newblock {\em BMC bioinformatics}, 15:1--13, 2014.

\bibitem{miltenyi1990high}
Stefan Miltenyi, Werner M{\"u}ller, Walter Weichel, and Andreas Radbruch.
\newblock High gradient magnetic cell separation with macs.
\newblock {\em Cytometry: The Journal of the International Society for Analytical Cytology}, 11(2):231--238, 1990.

\bibitem{minoura2021mixture}
Kodai Minoura, Ko~Abe, Hyunha Nam, Hiroyoshi Nishikawa, and Teppei Shimamura.
\newblock A mixture-of-experts deep generative model for integrated analysis of single-cell multiomics data.
\newblock {\em Cell reports methods}, 1(5), 2021.

\bibitem{molho2024deep}
Dylan Molho, Jiayuan Ding, Wenzhuo Tang, Zhaoheng Li, Hongzhi Wen, Yixin Wang, Julian Venegas, Wei Jin, Renming Liu, Runze Su, et~al.
\newblock Deep learning in single-cell analysis.
\newblock {\em ACM Transactions on Intelligent Systems and Technology}, 15(3):1--62, 2024.

\bibitem{moreno2022expression}
Pablo Moreno, Silvie Fexova, Nancy George, Jonathan~R Manning, Zhichiao Miao, Suhaib Mohammed, Alfonso Mu{\~n}oz-Pomer, Anja Fullgrabe, Yalan Bi, Natassja Bush, et~al.
\newblock Expression atlas update: gene and protein expression in multiple species.
\newblock {\em Nucleic acids research}, 50(D1):D129--D140, 2022.

\bibitem{myung2024deep}
Yoochan Myung, Alex~GC de~S{\'a}, and David~B Ascher.
\newblock Deep-pk: deep learning for small molecule pharmacokinetic and toxicity prediction.
\newblock {\em Nucleic Acids Research}, page gkae254, 2024.

\bibitem{ng2011sparse}
Andrew Ng et~al.
\newblock Sparse autoencoder.
\newblock {\em CS294A Lecture notes}, 72(2011):1--19, 2011.

\bibitem{niu2023identification}
Xiaoguang Niu, Man Xu, Jian Zhu, Shaowei Zhang, and Yanning Yang.
\newblock Identification of the immune-associated characteristics and predictive biomarkers of keratoconus based on single-cell rna-sequencing and bulk rna-sequencing.
\newblock {\em Frontiers in Immunology}, 14:1220646, 2023.

\bibitem{pearce2021deep}
Robin Pearce and Yang Zhang.
\newblock Deep learning techniques have significantly impacted protein structure prediction and protein design.
\newblock {\em Current opinion in structural biology}, 68:194--207, 2021.

\bibitem{pratapa2020benchmarking}
Aditya Pratapa, Amogh~P Jalihal, Jeffrey~N Law, Aditya Bharadwaj, and TM~Murali.
\newblock Benchmarking algorithms for gene regulatory network inference from single-cell transcriptomic data.
\newblock {\em Nature methods}, 17(2):147--154, 2020.

\bibitem{proks2024deep}
Martin Proks, Nazmus Salehin, and Joshua~M Brickman.
\newblock Deep learning-based models for preimplantation mouse and human embryos based on single-cell rna sequencing.
\newblock {\em Nature Methods}, pages 1--10, 2024.

\bibitem{qi2023trends}
Ren Qi and Quan Zou.
\newblock Trends and potential of machine learning and deep learning in drug study at single-cell level.
\newblock {\em Research}, 6:0050, 2023.

\bibitem{qin2024statistical}
Fei Qin, Guoshuai Cai, Christopher~I Amos, and Feifei Xiao.
\newblock A statistical learning method for simultaneous copy number estimation and subclone clustering with single-cell sequencing data.
\newblock {\em Genome Research}, 34(1):85--93, 2024.

\bibitem{raredon2022computation}
Micha Sam~Brickman Raredon, Junchen Yang, James Garritano, Meng Wang, Dan Kushnir, Jonas~Christian Schupp, Taylor~S Adams, Allison~M Greaney, Katherine~L Leiby, Naftali Kaminski, et~al.
\newblock Computation and visualization of cell--cell signaling topologies in single-cell systems data using connectome.
\newblock {\em Scientific Reports}, 12(1):4187, 2022.

\bibitem{regev2017human}
Aviv Regev, Sarah~A Teichmann, Eric~S Lander, Ido Amit, Christophe Benoist, Ewan Birney, Bernd Bodenmiller, Peter Campbell, Piero Carninci, Menna Clatworthy, et~al.
\newblock The human cell atlas.
\newblock {\em elife}, 6:e27041, 2017.

\bibitem{rodriguez2019unravelling}
Alba Rodriguez-Meira, Gemma Buck, Sally-Ann Clark, Benjamin~J Povinelli, Veronica Alcolea, Eleni Louka, Simon McGowan, Angela Hamblin, Nikolaos Sousos, Nikolaos Barkas, et~al.
\newblock Unravelling intratumoral heterogeneity through high-sensitivity single-cell mutational analysis and parallel rna sequencing.
\newblock {\em Molecular cell}, 73(6):1292--1305, 2019.

\bibitem{salmen2022high}
Fredrik Salmen, Joachim De~Jonghe, Tomasz~S Kaminski, Anna Alemany, Guillermo~E Parada, Joe Verity-Legg, Ayaka Yanagida, Timo~N Kohler, Nicholas Battich, Floris van~den Brekel, et~al.
\newblock High-throughput total rna sequencing in single cells using vasa-seq.
\newblock {\em Nature Biotechnology}, 40(12):1780--1793, 2022.

\bibitem{satija2015spatial}
Rahul Satija, Jeffrey~A Farrell, David Gennert, Alexander~F Schier, and Aviv Regev.
\newblock Spatial reconstruction of single-cell gene expression data.
\newblock {\em Nature biotechnology}, 33(5):495--502, 2015.

\bibitem{scarselli2008graph}
Franco Scarselli, Marco Gori, Ah~Chung Tsoi, Markus Hagenbuchner, and Gabriele Monfardini.
\newblock The graph neural network model.
\newblock {\em IEEE transactions on neural networks}, 20(1):61--80, 2008.

\bibitem{schede2021spatial}
Halima~Hannah Schede, Christian~G Schneider, Johanna Stergiadou, Lars~E Borm, Anurag Ranjak, Tracy~M Yamawaki, Fabrice~PA David, Peter L{\"o}nnerberg, Maria~Antonietta Tosches, Simone Codeluppi, et~al.
\newblock Spatial tissue profiling by imaging-free molecular tomography.
\newblock {\em Nature Biotechnology}, 39(8):968--977, 2021.

\bibitem{schott2024open}
Marie Schott, Daniel Le{\'o}n-Peri{\~n}{\'a}n, Elena Splendiani, Leon Strenger, Jan~Robin Licha, Tancredi~Massimo Pentimalli, Simon Schallenberg, Jonathan Alles, Sarah~Samut Tagliaferro, Anastasiya Boltengagen, et~al.
\newblock Open-st: High-resolution spatial transcriptomics in 3d.
\newblock {\em Cell}, 187(15):3953--3972, 2024.

\bibitem{seferbekova2023spatial}
Zaira Seferbekova, Artem Lomakin, Lucy~R Yates, and Moritz Gerstung.
\newblock Spatial biology of cancer evolution.
\newblock {\em Nature Reviews Genetics}, 24(5):295--313, 2023.

\bibitem{shan2022citedb}
Nayang Shan, Yao Lu, Hao Guo, Dongyu Li, Jitong Jiang, Linlin Yan, Jiudong Gao, Yong Ren, Xingming Zhao, and Lin Hou.
\newblock Citedb: a manually curated database of cell--cell interactions in human.
\newblock {\em Bioinformatics}, 38(22):5144--5148, 2022.

\bibitem{shi2023husch}
Xiaoying Shi, Zhiguang Yu, Pengfei Ren, Xin Dong, Xuanxin Ding, Jiaming Song, Jing Zhang, Taiwen Li, and Chenfei Wang.
\newblock Husch: an integrated single-cell transcriptome atlas for human tissue gene expression visualization and analyses.
\newblock {\em Nucleic Acids Research}, 51(D1):D1029--D1037, 2023.

\bibitem{shi2019variational}
Yuge Shi, Brooks Paige, Philip Torr, et~al.
\newblock Variational mixture-of-experts autoencoders for multi-modal deep generative models.
\newblock {\em Advances in neural information processing systems}, 32, 2019.

\bibitem{sohn2015learning}
Kihyuk Sohn, Honglak Lee, and Xinchen Yan.
\newblock Learning structured output representation using deep conditional generative models.
\newblock {\em Advances in neural information processing systems}, 28, 2015.

\bibitem{song2024scdesign3}
Dongyuan Song, Qingyang Wang, Guanao Yan, Tianyang Liu, Tianyi Sun, and Jingyi~Jessica Li.
\newblock scdesign3 generates realistic in silico data for multimodal single-cell and spatial omics.
\newblock {\em Nature Biotechnology}, 42(2):247--252, 2024.

\bibitem{sorin2023single}
Mark Sorin, Morteza Rezanejad, Elham Karimi, Benoit Fiset, Lysanne Desharnais, Lucas~JM Perus, Simon Milette, Miranda~W Yu, Sarah~M Maritan, Samuel Dor{\'e}, et~al.
\newblock Single-cell spatial landscapes of the lung tumour immune microenvironment.
\newblock {\em Nature}, 614(7948):548--554, 2023.

\bibitem{staahl2016visualization}
Patrik~L St{\aa}hl, Fredrik Salm{\'e}n, Sanja Vickovic, Anna Lundmark, Jos{\'e}~Fern{\'a}ndez Navarro, Jens Magnusson, Stefania Giacomello, Michaela Asp, Jakub~O Westholm, Mikael Huss, et~al.
\newblock Visualization and analysis of gene expression in tissue sections by spatial transcriptomics.
\newblock {\em Science}, 353(6294):78--82, 2016.

\bibitem{stickels2021highly}
Robert~R Stickels, Evan Murray, Pawan Kumar, Jilong Li, Jamie~L Marshall, Daniela~J Di~Bella, Paola Arlotta, Evan~Z Macosko, and Fei Chen.
\newblock Highly sensitive spatial transcriptomics at near-cellular resolution with slide-seqv2.
\newblock {\em Nature biotechnology}, 39(3):313--319, 2021.

\bibitem{stoeckius2017simultaneous}
Marlon Stoeckius, Christoph Hafemeister, William Stephenson, Brian Houck-Loomis, Pratip~K Chattopadhyay, Harold Swerdlow, Rahul Satija, and Peter Smibert.
\newblock Simultaneous epitope and transcriptome measurement in single cells.
\newblock {\em Nature methods}, 14(9):865--868, 2017.

\bibitem{stuart2019comprehensive}
Tim Stuart, Andrew Butler, Paul Hoffman, Christoph Hafemeister, Efthymia Papalexi, William~M Mauck, Yuhan Hao, Marlon Stoeckius, Peter Smibert, and Rahul Satija.
\newblock Comprehensive integration of single-cell data.
\newblock {\em cell}, 177(7):1888--1902, 2019.

\bibitem{sun2024single}
Fengying Sun, Haoyan Li, Dongqing Sun, Shaliu Fu, Lei Gu, Xin Shao, Qinqin Wang, Xin Dong, Bin Duan, Feiyang Xing, et~al.
\newblock Single-cell omics: experimental workflow, data analyses and applications.
\newblock {\em Science China Life Sciences}, pages 1--98, 2024.

\bibitem{sun2020ensemble}
Xiaoxiao Sun, Yiwen Liu, and Lingling An.
\newblock Ensemble dimensionality reduction and feature gene extraction for single-cell rna-seq data.
\newblock {\em Nature Communications}, 11(1):5853, 2020.

\bibitem{svahn2022ccvae}
Caroline Svahn and Oleg Sysoev.
\newblock Ccvae: A variational autoencoder for handling censored covariates.
\newblock In {\em 2022 21st IEEE International Conference on Machine Learning and Applications (ICMLA)}, pages 709--714. IEEE, 2022.

\bibitem{szklarczyk2021string}
Damian Szklarczyk, Annika~L Gable, Katerina~C Nastou, David Lyon, Rebecca Kirsch, Sampo Pyysalo, Nadezhda~T Doncheva, Marc Legeay, Tao Fang, Peer Bork, et~al.
\newblock The string database in 2021: customizable protein--protein networks, and functional characterization of user-uploaded gene/measurement sets.
\newblock {\em Nucleic acids research}, 49(D1):D605--D612, 2021.

\bibitem{talwar2018autoimpute}
Divyanshu Talwar, Aanchal Mongia, Debarka Sengupta, and Angshul Majumdar.
\newblock Autoimpute: Autoencoder based imputation of single-cell rna-seq data.
\newblock {\em Scientific reports}, 8(1):16329, 2018.

\bibitem{tan2020spacell}
Xiao Tan, Andrew Su, Minh Tran, and Quan Nguyen.
\newblock Spacell: integrating tissue morphology and spatial gene expression to predict disease cells.
\newblock {\em Bioinformatics}, 36(7):2293--2294, 2020.

\bibitem{tang2023explainable}
Xin Tang, Jiawei Zhang, Yichun He, Xinhe Zhang, Zuwan Lin, Sebastian Partarrieu, Emma~Bou Hanna, Zhaolin Ren, Hao Shen, Yuhong Yang, et~al.
\newblock Explainable multi-task learning for multi-modality biological data analysis.
\newblock {\em Nature communications}, 14(1):2546, 2023.

\bibitem{tang2024modal}
Zhenchao Tang, Guanxing Chen, Shouzhi Chen, Jianhua Yao, Linlin You, and Calvin Yu-Chian Chen.
\newblock Modal-nexus auto-encoder for multi-modality cellular data integration and imputation.
\newblock {\em Nature Communications}, 15(1):9021, 2024.

\bibitem{tang2023sparx}
Ziyang Tang, Xiang Liu, Zuotian Li, Tonglin Zhang, Baijian Yang, Jing Su, and Qianqian Song.
\newblock Sparx: Elucidate single-cell spatial heterogeneity of drug responses for personalized treatment.
\newblock {\em Briefings in Bioinformatics}, 24(6):bbad338, 2023.

\bibitem{tarhan2023single}
Leyla Tarhan, Jon Bistline, Jean Chang, Bryan Galloway, Emily Hanna, and Eric Weitz.
\newblock Single cell portal: an interactive home for single-cell genomics data.
\newblock {\em BioRxiv}, 2023.

\bibitem{taud2018multilayer}
Hind Taud and Jean-Franccois Mas.
\newblock Multilayer perceptron (mlp).
\newblock {\em Geomatic approaches for modeling land change scenarios}, pages 451--455, 2018.

\bibitem{theodoris2023transfer}
Christina~V Theodoris, Ling Xiao, Anant Chopra, Mark~D Chaffin, Zeina~R Al~Sayed, Matthew~C Hill, Helene Mantineo, Elizabeth~M Brydon, Zexian Zeng, X~Shirley Liu, et~al.
\newblock Transfer learning enables predictions in network biology.
\newblock {\em Nature}, 618(7965):616--624, 2023.

\bibitem{townes2023nonnegative}
F~William Townes and Barbara~E Engelhardt.
\newblock Nonnegative spatial factorization applied to spatial genomics.
\newblock {\em Nature methods}, 20(2):229--238, 2023.

\bibitem{van2023applications}
Bram Van~de Sande, Joon~Sang Lee, Euphemia Mutasa-Gottgens, Bart Naughton, Wendi Bacon, Jonathan Manning, Yong Wang, Jack Pollard, Melissa Mendez, Jon Hill, et~al.
\newblock Applications of single-cell rna sequencing in drug discovery and development.
\newblock {\em Nature Reviews Drug Discovery}, 22(6):496--520, 2023.

\bibitem{van2018recovering}
David Van~Dijk, Roshan Sharma, Juozas Nainys, Kristina Yim, Pooja Kathail, Ambrose~J Carr, Cassandra Burdziak, Kevin~R Moon, Christine~L Chaffer, Diwakar Pattabiraman, et~al.
\newblock Recovering gene interactions from single-cell data using data diffusion.
\newblock {\em Cell}, 174(3):716--729, 2018.

\bibitem{velten2022identifying}
Britta Velten, Jana~M Braunger, Ricard Argelaguet, Damien Arnol, Jakob Wirbel, Danila Bredikhin, Georg Zeller, and Oliver Stegle.
\newblock Identifying temporal and spatial patterns of variation from multimodal data using mefisto.
\newblock {\em Nature methods}, 19(2):179--186, 2022.

\bibitem{velten2023principles}
Britta Velten and Oliver Stegle.
\newblock Principles and challenges of modeling temporal and spatial omics data.
\newblock {\em Nature Methods}, 20(10):1462--1474, 2023.

\bibitem{vickovic2019high}
Sanja Vickovic, G{\"o}kcen Eraslan, Fredrik Salm{\'e}n, Johanna Klughammer, Linnea Stenbeck, Denis Schapiro, Tarmo {\"A}ij{\"o}, Richard Bonneau, Ludvig Bergenstr{\aa}hle, Jos{\'e}~Fernand{\'e}z Navarro, et~al.
\newblock High-definition spatial transcriptomics for in situ tissue profiling.
\newblock {\em Nature methods}, 16(10):987--990, 2019.

\bibitem{walter2023fishfactor}
Florin~C Walter, Oliver Stegle, and Britta Velten.
\newblock Fishfactor: a probabilistic factor model for spatial transcriptomics data with subcellular resolution.
\newblock {\em Bioinformatics}, 39(5):btad183, 2023.

\bibitem{wang2022comparison}
Jiacheng Wang, Quan Zou, and Chen Lin.
\newblock A comparison of deep learning-based pre-processing and clustering approaches for single-cell rna sequencing data.
\newblock {\em Briefings in Bioinformatics}, 23(1):bbab345, 2022.

\bibitem{wang2020single}
Liang Wang, Steven Mo, Xin Li, Yingzhi He, and Jing Yang.
\newblock Single-cell rna-seq reveals the immune escape and drug resistance mechanisms of mantle cell lymphoma.
\newblock {\em Cancer biology \& medicine}, 17(3):726, 2020.

\bibitem{wang2022systematic}
Saidi Wang, Hansi Zheng, James~S Choi, Jae~K Lee, Xiaoman Li, and Haiyan Hu.
\newblock A systematic evaluation of the computational tools for ligand-receptor-based cell--cell interaction inference.
\newblock {\em Briefings in functional genomics}, 21(5):339--356, 2022.

\bibitem{wang2024single}
Teng Wang, Junhua Peng, Jiaqi Fan, Ni~Tang, Rui Hua, Xueliang Zhou, Zhihao Wang, Longfei Wang, Yanling Bai, Xiaowan Quan, et~al.
\newblock Single-cell multi-omics profiling of human preimplantation embryos identifies cytoskeletal defects during embryonic arrest.
\newblock {\em Nature Cell Biology}, 26(2):263--277, 2024.

\bibitem{wang2019bermuda}
Tongxin Wang, Travis~S Johnson, Wei Shao, Zixiao Lu, Bryan~R Helm, Jie Zhang, and Kun Huang.
\newblock Bermuda: a novel deep transfer learning method for single-cell rna sequencing batch correction reveals hidden high-resolution cellular subtypes.
\newblock {\em Genome biology}, 20:1--15, 2019.

\bibitem{wang2018three}
Xiao Wang, William~E Allen, Matthew~A Wright, Emily~L Sylwestrak, Nikolay Samusik, Sam Vesuna, Kathryn Evans, Cindy Liu, Charu Ramakrishnan, Jia Liu, et~al.
\newblock Three-dimensional intact-tissue sequencing of single-cell transcriptional states.
\newblock {\em Science}, 361(6400):eaat5691, 2018.

\bibitem{wang2021single}
Yang Wang, Peng Yuan, Zhiqiang Yan, Ming Yang, Ying Huo, Yanli Nie, Xiaohui Zhu, Jie Qiao, and Liying Yan.
\newblock Single-cell multiomics sequencing reveals the functional regulatory landscape of early embryos.
\newblock {\em Nature communications}, 12(1):1247, 2021.

\bibitem{wang2021joint}
Yinqiao Wang, Lu~Chen, Jaemin Jo, and Yunhai Wang.
\newblock Joint t-sne for comparable projections of multiple high-dimensional datasets.
\newblock {\em IEEE Transactions on Visualization and Computer Graphics}, 28(1):623--632, 2021.

\bibitem{wang2019italk}
Yuanxin Wang, Ruiping Wang, Shaojun Zhang, Shumei Song, Changying Jiang, Guangchun Han, Michael Wang, Jaffer Ajani, Andy Futreal, and Linghua Wang.
\newblock italk: an r package to characterize and illustrate intercellular communication.
\newblock {\em BioRxiv}, page 507871, 2019.

\bibitem{warde2010genemania}
David Warde-Farley, Sylva~L Donaldson, Ovi Comes, Khalid Zuberi, Rashad Badrawi, Pauline Chao, Max Franz, Chris Grouios, Farzana Kazi, Christian~Tannus Lopes, et~al.
\newblock The genemania prediction server: biological network integration for gene prioritization and predicting gene function.
\newblock {\em Nucleic acids research}, 38(suppl\_2):W214--W220, 2010.

\bibitem{wegmann2024single}
Rebekka Wegmann, Ximena Bonilla, Ruben Casanova, St{\'e}phane Chevrier, Ricardo Coelho, Cinzia Esposito, Joanna Ficek-Pascual, Sandra Goetze, Gabriele Gut, Francis Jacob, et~al.
\newblock Single-cell landscape of innate and acquired drug resistance in acute myeloid leukemia.
\newblock {\em Nature Communications}, 15(1):9402, 2024.

\bibitem{welch2019single}
Joshua~D Welch, Velina Kozareva, Ashley Ferreira, Charles Vanderburg, Carly Martin, and Evan~Z Macosko.
\newblock Single-cell multi-omic integration compares and contrasts features of brain cell identity.
\newblock {\em Cell}, 177(7):1873--1887, 2019.

\bibitem{wu2021babel}
Kevin~E Wu, Kathryn~E Yost, Howard~Y Chang, and James Zou.
\newblock Babel enables cross-modality translation between multiomic profiles at single-cell resolution.
\newblock {\em Proceedings of the National Academy of Sciences}, 118(15):e2023070118, 2021.

\bibitem{wu2018multimodal}
Mike Wu and Noah Goodman.
\newblock Multimodal generative models for scalable weakly-supervised learning.
\newblock {\em Advances in neural information processing systems}, 31, 2018.

\bibitem{xia2019spatial}
Chenglong Xia, Jean Fan, George Emanuel, Junjie Hao, and Xiaowei Zhuang.
\newblock Spatial transcriptome profiling by merfish reveals subcellular rna compartmentalization and cell cycle-dependent gene expression.
\newblock {\em Proceedings of the National Academy of Sciences}, 116(39):19490--19499, 2019.

\bibitem{xie2023comparison}
Zihong Xie, Xuri Li, and Antonio Mora.
\newblock A comparison of cell-cell interaction prediction tools based on scrna-seq data.
\newblock {\em Biomolecules}, 13(8):1211, 2023.

\bibitem{xu2020scigans}
Yungang Xu, Zhigang Zhang, Lei You, Jiajia Liu, Zhiwei Fan, and Xiaobo Zhou.
\newblock scigans: single-cell rna-seq imputation using generative adversarial networks.
\newblock {\em Nucleic acids research}, 48(15):e85--e85, 2020.

\bibitem{xu2024stomicsdb}
Zhicheng Xu, Weiwen Wang, Tao Yang, Ling Li, Xizheng Ma, Jing Chen, Jieyu Wang, Yan Huang, Joshua Gould, Huifang Lu, et~al.
\newblock Stomicsdb: a comprehensive database for spatial transcriptomics data sharing, analysis and visualization.
\newblock {\em Nucleic acids research}, 52(D1):D1053--D1061, 2024.

\bibitem{yan2024prior}
Hongxi Yan, Dawei Weng, Dongguo Li, Yu~Gu, Wenji Ma, and Qingjie Liu.
\newblock Prior knowledge-guided multilevel graph neural network for tumor risk prediction and interpretation via multi-omics data integration.
\newblock {\em Briefings in Bioinformatics}, 25(3):bbae184, 2024.

\bibitem{yang2022scbert}
Fan Yang, Wenchuan Wang, Fang Wang, Yuan Fang, Duyu Tang, Junzhou Huang, Hui Lu, and Jianhua Yao.
\newblock scbert as a large-scale pretrained deep language model for cell type annotation of single-cell rna-seq data.
\newblock {\em Nature Machine Intelligence}, 4(10):852--866, 2022.

\bibitem{yang2021multi}
Karren~Dai Yang, Anastasiya Belyaeva, Saradha Venkatachalapathy, Karthik Damodaran, Abigail Katcoff, Adityanarayanan Radhakrishnan, GV~Shivashankar, and Caroline Uhler.
\newblock Multi-domain translation between single-cell imaging and sequencing data using autoencoders.
\newblock {\em Nature communications}, 12(1):31, 2021.

\bibitem{yang2023deepcci}
Wenyi Yang, Pingping Wang, Meng Luo, Yideng Cai, Chang Xu, Guangfu Xue, Xiyun Jin, Rui Cheng, Jinhao Que, Fenglan Pang, et~al.
\newblock Deepcci: a deep learning framework for identifying cell--cell interactions from single-cell rna sequencing data.
\newblock {\em Bioinformatics}, 39(10):btad596, 2023.

\bibitem{yang2024deciphering}
Wenyi Yang, Pingping Wang, Shouping Xu, Tao Wang, Meng Luo, Yideng Cai, Chang Xu, Guangfu Xue, Jinhao Que, Qian Ding, et~al.
\newblock Deciphering cell--cell communication at single-cell resolution for spatial transcriptomics with subgraph-based graph attention network.
\newblock {\em Nature Communications}, 15(1):7101, 2024.

\bibitem{yao2023react}
Shunyu Yao, Jeffrey Zhao, Dian Yu, Nan Du, Izhak Shafran, Karthik Narasimhan, and Yuan Cao.
\newblock React: synergizing reasoning and acting in language models (2022).
\newblock {\em arXiv preprint arXiv:2210.03629}, 2023.

\bibitem{yuan2023sodb}
Zhiyuan Yuan, Wentao Pan, Xuan Zhao, Fangyuan Zhao, Zhimeng Xu, Xiu Li, Yi~Zhao, Michael~Q Zhang, and Jianhua Yao.
\newblock Sodb facilitates comprehensive exploration of spatial omics data.
\newblock {\em Nature Methods}, 20(3):387--399, 2023.

\bibitem{zeng2022cancerscem}
Jingyao Zeng, Yadong Zhang, Yunfei Shang, Jialin Mai, Shuo Shi, Mingming Lu, Congfan Bu, Zhewen Zhang, Zaichao Zhang, Yang Li, et~al.
\newblock Cancerscem: a database of single-cell expression map across various human cancers.
\newblock {\em Nucleic acids research}, 50(D1):D1147--D1155, 2022.

\bibitem{zeng2023understanding}
Qun Zeng, Mira Mousa, Aisha~Shigna Nadukkandy, Lies Franssens, Halima Alnaqbi, Fatima~Yousif Alshamsi, Habiba~Al Safar, and Peter Carmeliet.
\newblock Understanding tumour endothelial cell heterogeneity and function from single-cell omics.
\newblock {\em Nature Reviews Cancer}, 23(8):544--564, 2023.

\bibitem{zeng2024imputing}
Yuansong Zeng, Yujie Song, Chengyang Zhang, Haoxuan Li, Yongkang Zhao, Weijiang Yu, Shiqi Zhang, Hongyu Zhang, Zhiming Dai, and Yuedong Yang.
\newblock Imputing spatial transcriptomics through gene network constructed from protein language model.
\newblock {\em Communications Biology}, 7(1):1271, 2024.

\bibitem{zhang2023single}
Shu Zhang, Wen Fang, Siqi Zhou, Dongming Zhu, Ruidong Chen, Xin Gao, Zhuojin Li, Yao Fu, Yixuan Zhang, Fa~Yang, et~al.
\newblock Single cell transcriptomic analyses implicate an immunosuppressive tumor microenvironment in pancreatic cancer liver metastasis.
\newblock {\em Nature communications}, 14(1):5123, 2023.

\bibitem{zhang2021cellcall}
Yang Zhang, Tianyuan Liu, Xuesong Hu, Mei Wang, Jing Wang, Bohao Zou, Puwen Tan, Tianyu Cui, Yiying Dou, Lin Ning, et~al.
\newblock Cellcall: integrating paired ligand--receptor and transcription factor activities for cell--cell communication.
\newblock {\em Nucleic acids research}, 49(15):8520--8534, 2021.

\bibitem{zheng2017massively}
Grace~XY Zheng, Jessica~M Terry, Phillip Belgrader, Paul Ryvkin, Zachary~W Bent, Ryan Wilson, Solongo~B Ziraldo, Tobias~D Wheeler, Geoff~P McDermott, Junjie Zhu, et~al.
\newblock Massively parallel digital transcriptional profiling of single cells.
\newblock {\em Nature communications}, 8(1):14049, 2017.

\bibitem{zheng2023aquila}
Yimin Zheng, Yitian Chen, Xianting Ding, Koon~Ho Wong, and Edwin Cheung.
\newblock Aquila: a spatial omics database and analysis platform.
\newblock {\em Nucleic Acids Research}, 51(D1):D827--D834, 2023.

\bibitem{zhou2024sorc}
Weiwei Zhou, Minghai Su, Tiantongfei Jiang, Qingyi Yang, Qisen Sun, Kang Xu, Jingyi Shi, Changbo Yang, Na~Ding, Yongsheng Li, et~al.
\newblock Sorc: an integrated spatial omics resource in cancer.
\newblock {\em Nucleic Acids Research}, 52(D1):D1429--D1437, 2024.

\bibitem{zhou2023integrating}
Xiang Zhou, Kangning Dong, and Shihua Zhang.
\newblock Integrating spatial transcriptomics data across different conditions, technologies and developmental stages.
\newblock {\em Nature Computational Science}, 3(10):894--906, 2023.

\bibitem{zhou2020surface}
Zilu Zhou, Chengzhong Ye, Jingshu Wang, and Nancy~R Zhang.
\newblock Surface protein imputation from single cell transcriptomes by deep neural networks.
\newblock {\em Nature communications}, 11(1):651, 2020.

\bibitem{zhu2018identification}
Qian Zhu, Sheel Shah, Ruben Dries, Long Cai, and Guo-Cheng Yuan.
\newblock Identification of spatially associated subpopulations by combining scrnaseq and sequential fluorescence in situ hybridization data.
\newblock {\em Nature biotechnology}, 36(12):1183--1190, 2018.

\bibitem{zinati2024groundgan}
Yazdan Zinati, Abdulrahman Takiddeen, and Amin Emad.
\newblock Groundgan: Grn-guided simulation of single-cell rna-seq data using causal generative adversarial networks.
\newblock {\em Nature Communications}, 15(1):4055, 2024.

\bibitem{zuo2021deep}
Chunman Zuo and Luonan Chen.
\newblock Deep-joint-learning analysis model of single cell transcriptome and open chromatin accessibility data.
\newblock {\em Briefings in Bioinformatics}, 22(4):bbaa287, 2021.

\end{thebibliography}
}
\end{document}